\documentclass[epj]{svjour}

\usepackage[english]{babel}
\usepackage{indentfirst} 
\usepackage[latin1,utf8]{inputenc}
\usepackage{lmodern}
\usepackage[T1]{fontenc}
\usepackage{setspace}
\usepackage{hyperref}
\usepackage[french]{varioref}
\usepackage{wrapfig}
\usepackage{vmargin}
\usepackage[]{subfig}
\usepackage{multirow}
\usepackage{bbm}
\usepackage{slashed}
\usepackage{braket}
\usepackage{enumitem}
\usepackage{mathtools, cuted}
\usepackage{float}
\usepackage{mdframed}
\usepackage{cite}

\usepackage{cellspace}
\newcommand{\Mark}[1]{\textsuperscript{#1}}

\usepackage[dvipsnames]{xcolor}

\newcommand{\rchi}[2]{\raisebox{\depth}{$\chi$}}

\newcommand{\lambdabar}{\mbox{\makebox[-0.7ex][l]{$\lambda$} \raisebox{0.5ex}[0pt][0pt]{\textbf{--}}}}

\usepackage{amsmath,amsfonts,amssymb}

\makeatletter 
\newcommand{\Autoref}[1]{\@first@ref#1,@} 
\def\@throw@dot#1.#2@{#1}
\def\@set@refname#1{
    \edef\@tmp{\getrefbykeydefault{#1}{anchor}{}}%
    \def\@refname{\@nameuse{\expandafter\@throw@dot\@tmp.@autorefname}s}
}
\def\@first@ref#1,#2{%
  \ifx#2@\autoref{#1}\let\@nextref\@gobble
  \else%
    \@set@refname{#1}
    \@refname~\ref{#1}
    \let\@nextref\@next@ref
  \fi%
  \@nextref#2%
}
\def\@next@ref#1,#2{%
   \ifx#2@ and~\ref{#1}\let\@nextref\@gobble
   \else, \ref{#1}
   \fi%
   \@nextref#2%
}

\setmarginsrb{1.5cm}{1.2cm}{1.5cm}{1.2cm}{1cm}{1cm}{1cm}{2cm}

\usepackage{titletoc}

\begin{document}
\onecolumn
\renewcommand{\sectionautorefname}{sect.}
\renewcommand{\figureautorefname}{fig.}
\renewcommand{\equationautorefname}{eq.}
\renewcommand{\tableautorefname}{table}
\renewcommand{\appendixautorefname}{appendix}

 \title{Parametrization of cross sections for elementary hadronic collisions involving strange particles}
 \author{J. Hirtz\inst{1,2}
 \and J.-C. David\inst{1}
 \and A. Boudard\inst{1}
 \and J. Cugnon\inst{3}
 \and S. Leray\inst{1}
 \and I. Leya\inst{2}
 \and D. Mancusi\inst{4}
 }
 \institute{IRFU, CEA, Université Paris-Saclay, F-91191, Gif-sur-Yvette, France \and Space Research and Planetary Sciences, Physics Institute, University of Bern, Sidlerstrasse 5, 3012 Bern, Switzerland \and AGO department, University of Liège, allée du 6 août 17, bât. B5, B-4000 Liège 1, Belgium \and Den-Service d'étude des réacteurs et de mathématiques appliquées (SERMA), CEA, Université Paris-Saclay, F-91191, Gif-sur-Yvette, France}

\abstract{The production of strange particles (kaons, hyperons) and hypernuclei in light charged particle induced reactions in the energy range of a few GeV (2-15 GeV) has become a topic of active research in several facilities (\textit{e.g.}, HypHI and PANDA at GSI and/or FAIR (Germany), JLab (USA), and JPARC (Japan)). This energy range represents the low-energy limit of the string models (degree of freedom: quark and gluon) or the high-energy limit of the so-called spallation models (degree of freedom: hadrons). A well known spallation model is INCL, the Liège intranuclear cascade model (combined with a de-excitation model to complete the reaction). INCL, known to give good results up to 2-3 GeV, was recently upgraded by the implementation of multiple pion emission to extend the energy range of applicability up to roughly 15 GeV. The next step, to account also for strange particle production, both for refining the high energy domain and making it usable when strangeness appears, requires the following main ingredients: i) the relevant elementary cross sections (production, scattering, and absorption) and ii) the characteristics of the associated final states. Some of those ingredients are already known and, sometimes, are already used in models of the same type (\textit{e.g.}, Bertini, GiBUU), but this paper aims at reviewing the situation by compiling, updating, and comparing the necessary elementary information which are independent of the model used.}

\titlerunning{Parametrization of cross sections for elementary hadronic collisions involving strange particles}
\maketitle

\section{Introduction}

The modelling of nuclear reactions involving a light projectile and an atomic nucleus from a few tens of MeV to a few GeV is important for a large variety of applications, ranging from nuclear waste transmutation, to spacecraft shielding, through hadron therapy. This type of reactions is called spallation reactions. Technically, it is assumed that a proper description of spallation reactions starts at about 100-200 MeV. However, special attention to the low energy domain showed that results down to a few tens of MeV could be as good as those obtained via models dedicated to the description of low energy nuclear reactions \cite{Dav08}. Spallation reactions are usually described by two steps. The first step is called the intranuclear cascade (INC), because the incident projectile gives rise to a cascade of hadronic reactions within the nucleus with emission of energetic particles leading to a remaining excited nucleus. The second step is the deexcitation of the nucleus via evaporation, fission, fermi-breakup, or multifragmentation. During the last twenty years great improvements have been achieved modelling those reactions, often driven by projects on spallation neutron sources (shielding of neutron beams or transmutation of nuclear waste). In 2010, IAEA tested the reliability for most of the models used worldwide \cite{Ler11}. Various observables enabled to scrutinize the qualities and shortcomings of the models.

The INCL (Liège Intranuclear Cascade) model, which is developed by the authors of this paper, was recognized as one of the best spallation code up to 2-3 GeV according to the IAEA 2010 benchmark. We decided then to improve and extend our model \cite{incl}. Among the different topics one can cite the improvement of low-energy cluster-induced reactions \cite{Dav13}, the few-nucleon removal study \cite{Man15,jose}, and the extension to high energies (up to 15 GeV) \cite{Ped11,Ped12}. For extending the model to high energies we introduced the main new channel, which is multiple pion emission in $NN$ and $\pi N$ interactions. This type of emission is based on the hypothesis that the produced baryonic resonances have so short lifetimes that their decay, in several pions, occurs before they interact with another particle in the nucleus. In addition, the overlap of their large widths makes difficult the choice of a specific resonance. Finally, the very good results obtained when comparing the new model predictions to experimental data and other models confirmed that the main features can be described on this manner. However, other particles, especially strange particles, can be produced to a lesser extent when the energy goes up. Even if they only play a minor role during the cascade, strange particle production contributes a few percent of the nucleon-nucleon inelastic cross section for energies from 2 GeV to 15 GeV; therefore, taking them into account could improve the modelling. The improvements and implementations will also bring new possibilities, which are important for simulating specific experiment involving, for instance, kaon emission. Comparisons with experimental data may also probe the nuclear medium effects. In addition, hypernuclei, whose interest grows with new facilities and experiments (\textit{e.g.}, HypHI and PANDA at GSI and/or FAIR (Germany), JLab (USA), J-PARC (Japan)), can also be studied. We want to stress the particular interest of possibility of studying hypernuclei with the extended INCL model. Beyond having a general high predictive power, the combination of the INCL model with the de-excitation Abla model is probably the most suitable tool to study the propagation of baryons in a nuclear medium, as testified by the IAEA intercomparison mentioned above.

The needed ingredients to account for strange particles (limited in this paper to Kaons, antiKaons, Lambda, Sigmas) are their characteristics, reaction cross sections involving strange particles in the initial and/or final state, angular distributions, momentum and charge repartition of the particles in the final state. This paper describes the ingredients and especially the parametrizations of the reaction cross sections involving strange particles. These ingredients are independent of the code considered and can be used in any other code. It is worth mentioning that hyperon and kaon production from a nucleus are already modelled in several codes, \textit{e.g.}, GiBUU \cite{gibuu}, JAM \cite{jam}, LAQGSM \cite{Mas08}, INCL2.0\cite{joseph,deneye}, and Bertini \cite{bertini}. Numerous scenarios exist to treat the production of strange particles. Some models split the energy range in two parts: a low-energy part with a center-of-mass energy roughly below 3-4 GeV and a high-energy part. The low-energy part is described either by resonances or directly by their decay products. However, the cross sections are then often treated differently as it is done in INCL; there are often given in resonant and non-resonant terms. For the high energy part the LUND string model \cite{And83} is usually used. Some other models, like Bertini and INCL, which both focus on the energy domain considered here, \textit{i.e.}, below 15 GeV, consider directly the decay products of the resonances and they rest on experimental data, calculation results (\textit{e.g.}, from string models), and approximations. Therefore, some information already exists. However, we investigated new parametrizations by using all available materials (experimental data, hypotheses, and models) and here we use the opportunity to report our best knowledge of the thus determined cross sections and to improve some parametrizations. Our goal is also to provide a rather comprehensive set of cross sections and angular distributions in an as simple and accurate  as possible shape, that can be used by other model builders and/or end-users. In addition, our work  attempts to  a systematic and coherent elaboration of fitted cross sections,  largely based on symmetry and simple hadronic models, as explained in detail in this manuscript.

The paper starts with the list of particles and reactions considered. Then, the way the reaction cross sections have been parametrized is described in \autoref{III}. \hyperref[IV]{Sect.~\ref*{IV}} is devoted to the particles in the final states and more precisely to their emission angles and momenta. There we also describe the charge repartition. Since such information already partly exist in literature, comparisons of the earlier data with the new results obtained here are given in \autoref{V}. Finally, we draw some conclusions.

\section{Particles and reactions}
\label{II}

In a first step, only the non-resonant particles with one unit of strangeness were considered. Therefore Kaons ($K^0$ and $K^+$), antiKaons ($\overline{K}^0$ and $K^-$)(the difference between Kaons and antiKaons is relevant in this paper), Sigmas ($\Sigma^-$,$\Sigma^0$, and $\Sigma^+$), and the Lambda ($\Lambda$) were added, \textit{i.e.}, particles with a nuclear spin $J=0$ and $J=1/2$ and with a strangeness $-1$ for baryons and $\pm 1$ for mesons.

The types of particles considered also define the types of reactions that must be considered. Doing so, we use their  relative importance, given by the experimental cross sections. Knowing that the main particles that evolve during the intranuclear cascade are nucleons and pions, we consider reactions contributing at least 1\% to the $NN$ and $\pi N$ total cross section and at least 10\% of the total cross section for $YN$($Y=\Lambda$ or $\Sigma$), $\overline{K}N $, and $KN$ reaction. The reactions take into account in this work are listed in \autoref{reac1}. This choice is based on available experimental data.

\begin{table}[ht]
\begin{center}
\begin{tabular}{ccl|ccl|ccl|ccl}
$NN$ & $ \rightarrow$ & $ N \Lambda K$ & $\pi N$ & $ \rightarrow$ & $\Lambda K $ & $N \overline{K}$ & $ \rightarrow$ & $N \overline{K}$ & $N K$ & $ \rightarrow$ & $N K $\\
     & $ \rightarrow$ & $ N \Sigma K $         & & $ \rightarrow$ & $\Sigma K $        &            & $ \rightarrow$ & $\Lambda \pi$    & & $\rightarrow $ & $N K \pi$\\  
     & $ \rightarrow$ & $ N \Lambda K \pi$     & & $ \rightarrow$ & $\Lambda K \pi$    &            & $ \rightarrow$ & $\Sigma \pi$     & & $\rightarrow $ & $N K \pi \pi$\\
     & $ \rightarrow$ & $ N \Sigma K \pi$      & & $ \rightarrow$ & $\Sigma K \pi$     &            & $ \rightarrow$ & $N \overline{K} \pi$ & $N \Lambda $ & $ \rightarrow$ & $ N \Lambda$\\
     & $ \rightarrow$ & $ N \Lambda K \pi\pi$  & & $ \rightarrow$ & $\Lambda K \pi\pi$ &            & $ \rightarrow$ & $\Lambda \pi \pi$ & &  $ \rightarrow  $ & $ N \Sigma$\\
     & $ \rightarrow$ & $ N \Sigma K \pi\pi$   & & $ \rightarrow$ & $\Lambda \pi \pi$  &            & $ \rightarrow$ & $\Sigma \pi \pi$ & $N \Sigma $ & $ \rightarrow$ & $ N \Lambda$\\
     & $ \rightarrow$ & $ NN K \overline{K}$   & & $ \rightarrow$ & $N K \overline{K}$ &            & $ \rightarrow$ & $N \overline{K} \pi \pi$ & & $\rightarrow$ & $ N \Sigma$

\end{tabular}
\caption{\label{reac1} List of considered reactions involving strangeness based on experimental data.}
\end{center}
\end{table}

In addition, we include two other types of reactions. The first one considers strangeness production via $\Delta N$ reactions. $\Delta$'s are less numerous than nucleons and $\pi$'s, but are nevertheless expected to contribute significantly to the strangeness production according to the study of Tsushima et \textit{al.}\cite{tsushima}. The second type is the strange production in reactions where many particles are produced in the final state but no measurements are available. Since their contributions increase significantly with increasing energy, a specific study was necessary to get the correct inclusive strangeness production cross section. \autoref{reac2} lists the channels for both types of reactions also taken into account.

\begin{table}
\begin{center}
\begin{tabular}{ccl|ccl}
$\Delta N$ & $\rightarrow$ & $N \Lambda K$           & $NN$    & $\rightarrow$  & $K + X$  \\
                 & $ \rightarrow$ & $N \Sigma K $             &             &                         &    \\  
                 & $ \rightarrow$ & $ \Delta \Lambda K$   &$\pi N$ & $ \rightarrow$ & $K + X$  \\
                 & $ \rightarrow$ & $ \Delta \Sigma K$      &             &                        &   \\
                & $ \rightarrow$ & $ NN K \overline{K}$    &              &                        &   \\
\end{tabular}
\caption{\label{reac2} List of the reactions involving strangeness and requiring information to be taken exclusively from models. Meaning of $X$ is explained in \autoref{III} and excludes the reactions cited in \autoref{reac1}}
\end{center}
\end{table}

In the reaction listed in \autoref{reac2}, Kaon production is equivalent to strangeness production, since it is the only particle with strangeness $+1$ , in the energy range under consideration in this paper, which can counterbalance the production of strangeness $-1$ of $\Lambda$, $\Sigma$ and $\overline{K}$ particles (strangeness is conserved in strong interaction processes).

Considering isospin, there are 488 channels, excluding the reactions $NN~\rightarrow~K~+~X$ and $\pi~N~\rightarrow~K~+~X$ of \autoref{reac2}, which must be characterized by their reaction cross sections (\autoref{III}) and their final state, \textit{i.e.}, charge repartitions, emission angle, and energy of the particles (\autoref{IV}).

\section{Reaction cross sections}
\label{III}

Among the ingredients needed to include new particles in an INC model, reaction cross sections are the most important. As far as possible they are taken from experimental data. However, measurements are not always performed on the entire energy range, rarely for all isospin channels, and are often inexistent when numerous particles exist in the final state. To overcome these limitations a step-by-step procedure has been developed to obtain parametrizations of the required cross sections (\Autoref{reac1,reac2}). First, an overview of the available experimental data has been performed. Second, two methods based on isospin symmetry allowed to extend our database by increasing the available information. Third, the still missing cross sections were determined using models and/or similar reactions with the help of plausible hypotheses. Finally, generic formulae, which can be applied to parametrize the cross sections, are given in the last subsection.

\subsection{Available experimental data}

The number of measured data for each reaction from \autoref{reac1} are given in \autoref{tab_data}. The energy range goes up to 32 GeV and the data are taken from Landolt-Börnstein~\cite{landolt} and two other papers~\cite{hires,sibir2}.

Since some of the published experimental data are rather old, our study offers the possibility to check and summarize our knowledge of the cross sections. We therefore give for each reaction the number of isospin channels, number of experimental data points, and the Gini coefficient.

The Gini coefficient\cite{gini} is a statistical tool used typically in economy to measure the dispersion of a system (usually the income distribution of the residents of a nation). The coefficient takes values between 0 (perfect repartition) and 1 (maximal inequality). The Gini coefficient for the discrete case is calculated as following:

\begin{table}[!ht]
\centering
\begin{tabular}{|Sc|Sc|Sc|Sc|}
\hline
Reaction & \begin{tabular}{c} \# of \\ channels \end{tabular} & \begin{tabular}{c} \# of \\ data \end{tabular}&\begin{tabular}{c} Gini \\ coefficient \end{tabular} \\
\hline
\hline
\multicolumn{4}{|Sc|}{$NN$ to}\\
\hline
 $N \Lambda K $	& 4 & 31 &	0.62 \\
 $N \Sigma K $	& 10 & 44 &	0.69 \\
 $N \Lambda K \pi$	& 10 & 29 & 0.63 \\
 $N \Sigma K \pi$	& 26 & 43 & 0.74 \\
 $N \Lambda K \pi\pi$	& 16 & 14 & 0.77 \\
 $N \Sigma K \pi\pi$	& 44 & 15 & 0.87 \\
 $NN K \overline{K}$	& 10 & 16 & 0.71 \\
\hline
\hline
\multicolumn{4}{|Sc|}{$\pi N$ to}\\
\hline
 $\Lambda K $		& 4 & 108 &	0.75 \\
 $\Sigma K $		& 10 & 158 &	0.74 \\
 $\Lambda K \pi$	& 10 & 68 &	0.72 \\
 $\Sigma K \pi$		& 26 & 148 &	0.77 \\
 $\Lambda K \pi\pi$	& 16 & 60 &	0.81 \\
 $\Sigma K \pi\pi$	& 44 & 63 &	0.86 \\
 $N K \overline{K}$	& 14 & 57 &	0.81 \\
\hline
\hline
\multicolumn{4}{|Sc|}{$\Lambda N$ to}\\
\hline
 $N \Lambda$	& 2 & 44 & 0.5 \\
 $N \Sigma$	& 4 & 11 &	0.75 \\
\hline
\hline
\multicolumn{4}{|Sc|}{$\Sigma N$ to}\\
\hline
 $N \Lambda$	& 4 & 11 & 0.75 \\
 $N \Sigma$	& 10 & 21 & 0.80 \\
\hline
\hline
\multicolumn{4}{|Sc|}{$\overline{K}N$ to}\\
\hline
 $N \overline{K}$	& 6 & 687 &	0.61\\
 $N \overline{K} \pi$	& 14 & 500 &	0.72\\
 $N \overline{K} \pi \pi$	& 22 & 124 &	0.87\\
 $\Lambda \pi$	& 4 & 349 &	0.52\\
 $\Sigma \pi$	& 10 & 685 &	0.59\\
 $\Lambda \pi \pi$	& 6 & 256 &	0.66\\
 $\Sigma \pi \pi$	& 16 & 496 &	0.73\\
\hline
\hline
\multicolumn{4}{|Sc|}{$KN$ to}\\
\hline
 $N K $	& 4 & 134 &	0.69\\
 $N K \pi $	& 14 & 223 &	0.62\\
 $N K \pi \pi$	& 22 & 123 &	0.89\\
\hline
\end{tabular}
\caption{Available experimental data points for reactions studied in this work}
\label{tab_data}
\end{table}
\begin{equation}
G = \frac{2 \sum\limits_{i=1}^{n} i  \ y_i}{n \sum\limits_{i=1}^{n} \ y_i} - \frac{n+1}{n},
\end{equation}
with $y_i$ the number of data in the i\Mark{th} channel arranged as $y_{i+1} \geq y_i$ (non-cumulative). This coefficient measures the repartition of data in the different isospin channels and, in our case (with a very high inequality and a high number of channels), correspond to the missing part of data for each reaction. For example, if $ G = 0.8$ approximatively 80\% of more data are needed to complete the 20\% of existing data and to make the entire database for each channel as precise as for the most precise channel.

\begin{figure}[ht]
\centering
\subfloat{\includegraphics[width=0.6\columnwidth]{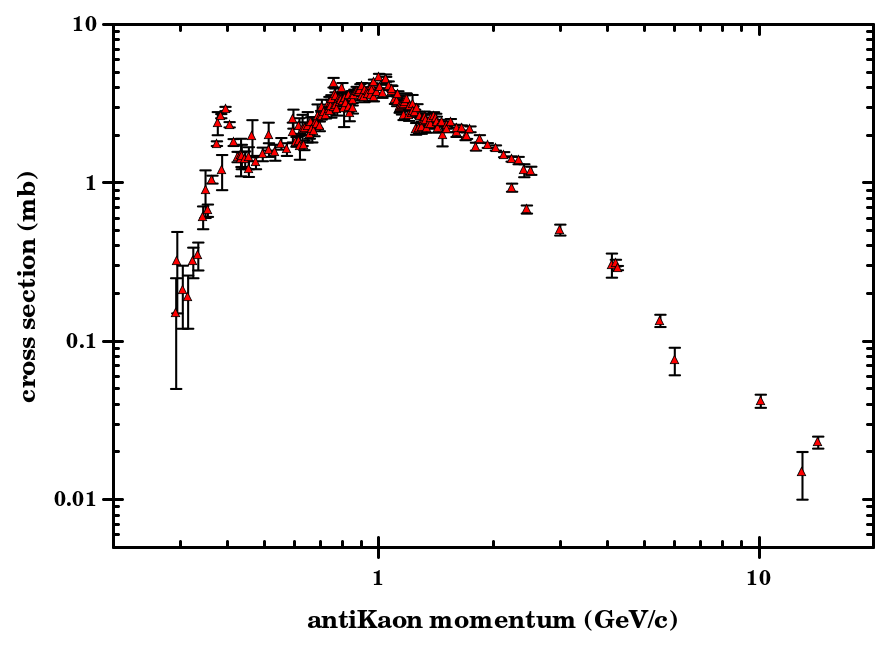}}
\caption{$K^- p \rightarrow \Lambda \pi^+ \pi^-$ reaction cross section as a function of the $K^-$ momentum. Data taken from \protect{\cite{landolt}}.}
\label{repartition}
\end{figure}

The \autoref{tab_data} shows the number of data depends strongly on the given reaction. For example, in average there are only 2-3 points per channel for the $NN$ collisions while for the $N \overline{K}$ reactions more than 35 data per channel are available. However, the Gini coefficients exhibit an important inhomogeneity ($G > 0.5$) with respect to the isospin channels. There is also a significant inhomogeneity, not given by the Gini coefficient, with respect to the energy range studied, with more data at the threshold and in the resonances region (see \autoref{repartition}). When available, these data nevertheless enable a reliable parametrization over the entire considered energy range (up to 15 GeV).

Using only experimental data the reaction cross sections were determined (sometimes partially) only for about 17\% of the channels listed in \autoref{reac1}. For the remaining 83\% various hypotheses were necessary, which are explained in some detail in the next subsections.

\subsection{The Bystricky procedure}
\label{bystricky}

The first method used to get information on missing isospin channels is based on the assumption of isospin symmetry, which is described in detail by Bystricky et \textit{al.} \cite{bystricky}. Their goal was to provide a phenomenological calculation tool for elastic and inelastic cross sections in the framework of isospin symmetry for the reactions $N \! N  \! \rightarrow  \! N \! N \! \pi$ and $N \! N  \! \rightarrow  \! N \! N \! \pi \!  \pi$.

The procedure, which is based on the isospin decomposition of systems, was used by Sophie Pedoux \cite{Ped11} to find missing cross sections in channels involving multiple pion production. The procedure  was applied up to the production of four pions and determined cross section were then implemented in a previous version of INCL.  Briefly, the initial state of two nucleons $\ket{NN}$ is projected on the final state decomposed into the nucleon final state $\bra{NN}$ and the pion final state $\bra{x \pi}$. The amplitude of the reaction is given by the following equation:
\begin{equation}
\label{ampli1}
\mathcal{M}(N \! N \rightarrow N \! N x \pi ) = \left(\bra{N \! N} \otimes \bra{x \pi} \right) M \ket{N \! N},
\end{equation}
with $M$ the reduced matrix element. The \autoref{ampli1} is subsequently decomposed using isospin projection:
\begin{equation}
\bra{I^{(1)} I^{(1)}_3 \ I^{(2)} I^{(2)}_3} M \ket{I^i I^i_3} = CG \ M_{I^{(1)}I^{(2)}I^i},
\end{equation}
with $CG$ the associated Clebsch-Gordan coefficient, $I^{(1)}$ and  $I^{(1)}_3$ the $NN$ system isospin and its projection, $I^{(2)}$ and  $I^{(2)}_3$ the $x\pi$ system isospin and its projection, $I^{(i)}$ and  $I^{(i)}_3$ the initial state isospin and its projection and $M_{I^{(1)}I^{(2)}I^i}$ the reduced matrix element for the isospin decomposition $I^i I^{(1)}I^{(2)}$. This equation can be written as the isospin decomposition on each multiplet system involved in the initial and final state contracted on the reduced matrix element.

Next, by integrating over all kinematic variables of the final state and summing over all permutations we obtain a decomposition of the cross section on isospin states, which is then compared with others to establish relations between the different cross sections.

This same procedure was then applied to reactions involving strange particles. In our case, \autoref{ampli1} can be written as the tensor product of the nucleon, pion, Kaon, antiKaon, Lambda, and Sigma systems of the initial and final state contracted on the reduced matrix element. With this \autoref{ampli1} becomes: 
\begin{small}
\begin{align}
\label{ampli2}
\mathcal{M} = \left( Initial~state \rightarrow x_N \! N \  x_{\pi} \! \pi \  x_Y \! Y \  x_K \! K \  x_{\overline{K}} \! \overline{K} \right) &= \left( \bra{x_N \! N} \otimes \bra{x_{\pi} \! \pi} \otimes \bra{x_Y \! Y} \otimes \bra{x_K \! K} \otimes \bra{x_{\overline{K}} \! \overline{K}} \right) M \ket{Initial~state} \nonumber \\
&= \left( \bra{system 1} \otimes \bra{system 2} \right) M \ket{Initial~state},
\end{align}
\end{small}
with $\bra{system 1}$ and $\bra{system 2}$ a contraction of the final multiplet systems in two arbitrary systems. Note that the final result does not depend on the choice of contraction.

The so obtained results are either simple equalities between individual cross sections, resulting form the Clebsch-Gordan coefficients associated with isospin symmetry, or equations between several cross sections resulting from the cross sections associated with a given total value of the isospin which can be expressed as sums of partial cross sections on various final charge states. Non trivial expressions of this kind are reported in \autoref{channel} in bold. As example, for the reaction $N \! \pi \! \rightarrow \! N \! K \! \overline{K}$ we get:
\begin{align}
\sigma(\pi^+ p\! \rightarrow p K^+ \! \overline{K}^0) &= \! \sigma(\pi^- n\! \rightarrow n K^0 \! K^-), \\
\nonumber \\
\sigma(\pi^- p \! \rightarrow n K^0 \! \overline{K}^0) + \sigma(\pi^- p \! \rightarrow n K^+ \! K^-) &+ \sigma(\pi^- p \! \rightarrow p K^0 \! K^-) + \sigma(\pi^+ p \! \rightarrow p K^+ \! \overline{K}^0) \nonumber \\
= 2 \sigma(\pi^0 p \! \rightarrow n K^+ \! \overline{K}^0) + 2 \sigma(\pi^0 p\!  &\rightarrow p K^0 \! \overline{K}^0) + 2 \sigma(\pi^0 p \! \rightarrow p K^+ \! K^-).
\label{sym_NpitoNKKb}
\end{align}

Errors arising from this procedure are introduced by the isospin invariance hypothesis and are estimated to be in the range of a few percent, which is approximately the mass differences between particles belonging to a same multiplet.

The Bystricky procedure allowed us to reduce the missing information on the reaction cross sections by approximately a factor 2, \textit{i.e.}, we increased the knowledge of the reaction cross sections by about a factor of 2. Thus, at this stage 35\% of the channels were parametrized, still 65\% are missing. For establishing a complete database another method, also based on isospin symmetry, was used (see next section).

\subsection{Hadron exchange model}
\label{HEM}

In order to complete the dataset, a procedure based on the hadron exchange model (HEM) was developed. The basic of the model is to apply the isospin symmetry at the Feynman diagram level, considering only diagrams at leading order, to obtain cross section ratios. This way, once again, unknown cross sections can be determined from known cross sections.

This procedure is an adaptation of the method used by Li \cite{li} and Sibirtsev \cite{sibirtsev}. In this method, complete Feynman diagrams are considered and not only the initial and final states as in the Bystricky procedure \cite{bystricky}. The method used by Li and Sibirtsev treats the case of pion and kaon exchange. Here, baryon exchange is also considered because of the type of studied cross sections. Initially, the hadron exchange model was developed with the idea to calculate explicitly a cross section and then using the isospin symmetry to determine easily other channel cross sections for a specific type of reaction. Here, the explicit calculation is replaced by a fit of experimental data. In the following, the method is explained and illustrated in an example.

Similar to the Bystricky method, the procedure determines in a first step relations between matrix elements and, in a second step, the cross section ratios by integrating over all kinematic variables of the squared matrix elements:
\begin{equation}
\label{HEM_1}
\sigma = \int |\mathcal{M}_{fi}|^2 d\Omega.
\end{equation}

To make things easier, the method used by Li and Sibirtsev neglects interferences between diagrams. They estimated that this hypothesis could change their result by about 30\%. In our case, first we consider only the ratios between cross sections and second we check, as far as possible, the results by comparing to experimental data or results arising from the Bystricky procedure. Doing so, the cross section of a specific isospin channel can be rewritten as the sum of all individual diagram contributions:
\begin{equation}
\label{HEM_2}
\sigma(channel) = \sum_i \int |\mathcal{M}_{X_i}(channel)|^2 d\Omega,
\end{equation}
with $\mathcal{M}_{X_i}(channel)$ the diagram amplitude of the isospin channel with the exchange particle $X_i$. In the reduced matrix element amplitude, there are three types of contribution: the initial and final fields, the propagators, and the vertices. Due to isospin symmetry, in the case of the same type of exchange particles, propagators and fields are identical. Therefore, the only difference between matrix elements comes from the vertices. However, the vertices have the same structure when the same particle types are involved. Consequently, these vertices are linked together by the isospin symmetry and this link can be obtained using Clebsch-Gordan coefficients. Note that Kaons and antiKaons have the same field and the same propagator because of the matter/antimatter symmetry.
Considering a specific vertex with two incoming particles and one outgoing particle, the contribution can be written as:
\begin{equation}
\label{HEM_3}
\bra{I^{out} I^{out}_3} \mathcal{V} \ket{I^{in(1)} I^{in(1)}_3, I^{in(2)} I^{in(2)}_3} = CG \ \mathcal{V}_{X,Y,Z},
\end{equation}
with $I^{out}$ and $I^{out}_3$ the outgoing particle isospin and its projection, $I^{in(i)}$ and $I^{in(i)}_3$ the isospin and the projection of the $i^{th}$ incoming particle, $\mathcal{V}$ the matrix element associated to the vertex, $CG$ the associated Clebsch-Gordan coefficient, and $\mathcal{V}_{X,Y,Z}$ the projected matrix element for the incoming and outgoing particles of type $X,Y,Z$. Since Clebsch-Gordan coefficients are scalar, diagrams with the same type of exchange particle are linked by a coefficient that is independent of energy. The matrix element of one diagram can be rewritten as:
\begin{equation}
\label{HEM_4}
\mathcal{M}_{X_i}(channel) = a_{X_i}(channel) \times \mathfrak{M}_{X_i},
\end{equation}
with $\mathfrak{M}_{X_i}$ the isospin-independent part of the matrix element and $a_{X_i}(channel)$ the product of all Clebsch-Gordan coefficients coming from each vertex (isospin-dependent part). A factor $n!$ appears in the case of n identical particles in the final state. The matrix element $\mathfrak{M}_{X_i}$ contains all the propagators, field contributions, and the structure of the vertices. The $a_{X_i}(channel)$ coefficient is a real scalar, which contains only the factor linking the different matrix elements. Using \autoref{HEM_4}, \autoref{HEM_2} can be rewritten as:
\begin{equation}
\label{HEM_5}
\sigma(channel) = \sum_i |a_{X_i}(channel)|^2 \int |\mathfrak{M}_{X_i}|^2 d\Omega. 
\end{equation}

Two cases must be distinguished. In the first case, all $|a_{X}(channel_j)/a_{X}(channel_k)|$ ratios are equal, independent of the diagram. In such a case, the cross section ratio of the two channels can easily be determined. In the second case with unequal ratios, extra information and hypotheses are required. In a first step, global relations obtained from the Bystricky procedure were systematically used as extra information. In a second step, hypotheses linking diagrams together or neglecting some diagrams are needed. Small coupling constants involved and/or small disintegration rates of the intermediate particles allow to leave out some diagrams. Note that all resonances (the $\Delta$ particle is not considered as a nucleon resonance from an isospin point of view: $J_\Delta \neq J_N$) are automatically considered because a given particle and its resonances have the same isospin and the same isospin projection. Therefore the $a_{X}$ coefficients are identical. Consequently, the sum over all diagram amplitudes with the same type of exchange particle can be treated as:

\begin{align}
\sum_{X^{(*)}_i} |a_{X_i} (channel)|^2 \int |\mathfrak{M}_{X_i}|^2 d\Omega & = |a_{\mathcal{X}}(channel)|^2 \sum_{X^{(*)}_i}  \int |\mathfrak{M}_{X_i}|^2 d\Omega \\
& = |a_{\mathcal{X}}(channel)|^2 \int |\mathfrak{M}_{\mathcal{X}}|^2 d\Omega, \nonumber
\end{align}
with ${X^{(*)}_i}$ the particle and its resonances $(*)$ and $\mathfrak{M}_{\mathcal{X}}$ the isospin independent general matrix element of the particle type $\mathcal{X}$ defined as:
\begin{equation}
|\mathfrak{M}_{\mathcal{X}}|^2 = \sum_{X^{(*)}_i} |\mathfrak{M}_{X_i}|^2.
\end{equation}

In order to illustrate the basic procedure, the way to solve the difficulties but also demonstrating the limits, we discuss an illustrative case based on the $\pi N~\rightarrow~\Sigma K$ reaction. Sadly, the hadron exchange model give no solution in this case but instead present the rare advantage to be relatively simple but to exhibit numerous problems, which appear often in more complex cases.

\begin{figure}[ht]
\centering
\subfloat{\includegraphics[scale=0.25]{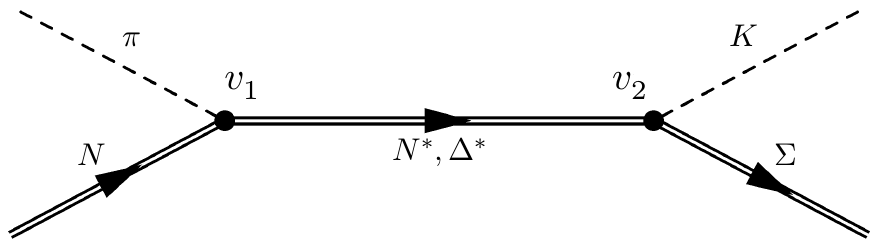}}\\
\subfloat{\includegraphics[scale=0.25]{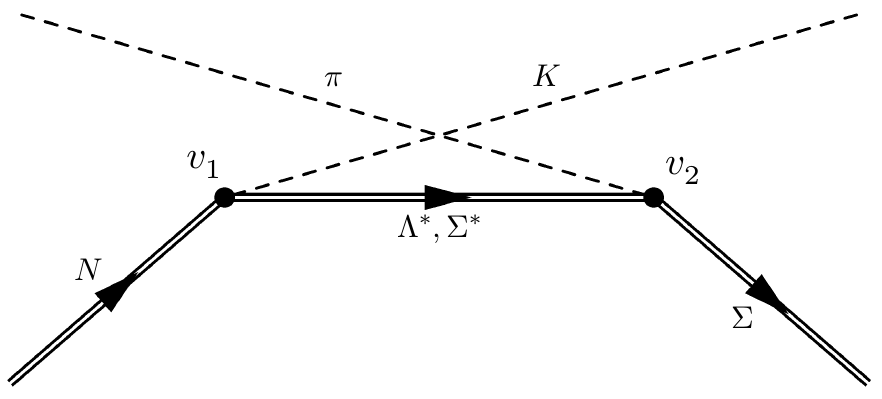}}\\
\subfloat{\includegraphics[scale=0.25]{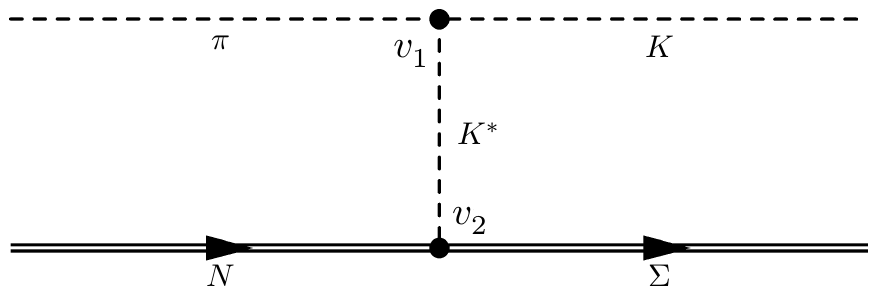}}
\caption{List of Feynman diagrams at leading order for the $\pi N \rightarrow \Sigma K$ reaction.}
\label{feyn}
\end{figure}

Five diagrams (three types), listed in \autoref{feyn}, are considered.

Using \autoref{HEM_5} the cross section is given by:
\begin{align}
\label{HEM_6}
\sigma (\pi N \rightarrow \Sigma K) = a_K^2 \int |\mathfrak{M}_K|^2 d\Omega & + a_{\Lambda}^2 \int |\mathfrak{M}_\Lambda|^2 d\Omega + a_{\Sigma}^2 \int |\mathfrak{M}_\Sigma|^2 d\Omega \nonumber \\
 & + a_{N}^2 \int |\mathfrak{M}_N|^2 d\Omega + a_{\Delta}^2 \int |\mathfrak{M}_\Delta|^2 d\Omega. 
\end{align}

In this example, there are two vertices in each diagram called $v_1^X$ and $v_2^X$ as shown in \autoref{feyn}. The $\Sigma$ exchange in the case $\pi^+p\rightarrow~\Sigma^+K^+$ is a $\Sigma^0$. Then, the projection on isospin eigenstates at $v_1^\Sigma$ is:
\begin{align}
P_r(v_1^\Sigma)( \pi^+ p \rightarrow \Sigma^+ K^+) & = (\bra{K^+} \otimes \bra{\Sigma^0})\mathcal{V} \ket{p} \nonumber \\
 & = \left( \left\langle \frac{1}{2} \frac{1}{2} \right| \otimes \bra{10} \right) \left|\frac{1}{2} \frac{1}{2} \right\rangle \mathcal{V}_{K\Sigma N} = \sqrt{\frac{1}{3}} \mathcal{V}_{K\Sigma N}.
\end{align}
\begin{table}[!ht]
\centering
\begin{tabular}{|c|Sccccc|}
\hline
 & $a_K^2$ & $a_\Lambda^2$ & $a_\Sigma^2$ & $a_N^2$ & $a_\Delta^2$\\
\hline
 $\pi^+ p \rightarrow \Sigma^+ K^+$	& 1 & 1 & 1/2 & 0 & 1 \\
 $\pi^0 p \rightarrow \Sigma^+ K^0$	& 1/2 & 0 & 1 & 1/2 & 2/9 \\
 $\pi^0 p \rightarrow \Sigma^0 K^+$	& 1/4 & 1 & 0 & 1/4 & 4/9 \\
 $\pi^- p \rightarrow \Sigma^0 K^0$	& 1/2 & 0 & 1 & 1/2 & 2/9\\
 $\pi^- p \rightarrow \Sigma^- K^+$	& 0 & 1 & 1/2 & 1 & 1/9\\
\hline
\end{tabular}
\caption{List of normalized $a_{X_i}(channel)$ squared coefficients for the reaction $\pi N \rightarrow \Sigma K$.}
\label{hdgh}
\end{table}

Doing the same calculation for each diagram, each channel, and each vertex gives the coefficients $a_{X_i}$ once a global normalization has been chosen. The counterweight of this normalization is hidden in the isospin-independ\-ent part of the matrix element. Here the choice is that the largest $a_{X_i}$ is equal to 1. All $a^2_{X_i}$ are given in \autoref{hdgh}. Only channels with an incoming proton are given here since channels with an incoming neutron can easily be deduced. It can be seen that the $a^2_{X_i}$ coefficients of the $\pi^0p\rightarrow~\Sigma^+K^0$ channel are equal to the ones of the $\pi^- p \rightarrow \Sigma^0 K^0$ channel. Therefore, we can infer:
\begin{equation}
\sigma(\pi^0 p \rightarrow \Sigma^+ K^0) = \sigma(\pi^- p \rightarrow \Sigma^0 K^0).
\end{equation}

Second, another interesting point is given by the following relations:
\begin{align}
2 a_N^2 = a_\Lambda^2 + 2 a_\Sigma^2 - 2 a_K^2,\\
9 a_\Delta^2 = 2 a_\Lambda^2 - 2 a_\Sigma^2 + 8 a_K^2.
\end{align}
Thus, if we define three new matrix elements:
\begin{align}
|\mathfrak{M}_1|^2 = |\mathfrak{M}_K|^2 - |\mathfrak{M}_N|^2 + \frac{8}{9}|\mathfrak{M}_\Delta|^2, \\
|\mathfrak{M}_2|^2 = |\mathfrak{M}_\Lambda|^2 + \frac{1}{2} |\mathfrak{M}_N|^2 + \frac{2}{9}|\mathfrak{M}_\Delta|^2, \\
|\mathfrak{M}_3|^2 = |\mathfrak{M}_\Sigma)|^2 + |\mathfrak{M}_N|^2 - \frac{2}{9}|\mathfrak{M}_\Delta|^2,
\end{align}
\autoref{HEM_6} becomes:
\begin{equation}
\sigma (\pi \! N \! \rightarrow \! \Sigma \! K) = a_K^2 \int |\mathfrak{M}_1|^2 d\Omega + a_{\Lambda}^2 \int |\mathfrak{M}_2|^2 d\Omega + a_{\Sigma}^2 \int |\mathfrak{M}_3|^2 d\Omega.
\end{equation}

The $|\mathfrak{M}_i|^2$ being unknown, extra hypotheses are needed to obtain other relations between the cross sections of the different channels. Their reliability will, however, directly affect the reliability of the final result.
The hypotheses for this show-case are: the experimental data exhibit some similarities between the known channel cross sections (3 channels in the 10 that which should be parametrized are reasonably well measured). It can be reasonably argued that:
\begin{equation}
\sigma (\pi^- p \rightarrow \Sigma^0 K^0) \approx \sigma(\pi^- p \rightarrow \Sigma^- K^+).
\end{equation}
That implies:
\begin{equation}
|\mathfrak{M}_1|^2 = 2 |\mathfrak{M}_2|^2 - |\mathfrak{M}_3|^2.
\end{equation} 

Finally, two more hypotheses are necessary to link the isospin channel cross sections of the reaction $\pi N~\rightarrow~\Sigma K$.  First $N$ and/or $\Delta$ exchanges were neglected, because the strange decay ratio is very weak for most of the resonances. Second,  the graphs with a $\Lambda$ exchange and a $\Sigma$ exchange are supposed to be equivalent, because of their similar nature. Doing so, it follows:
\begin{equation}
|\mathfrak{M}_K|^2 = |\mathfrak{M}_\Lambda|^2 = |\mathfrak{M}_\Sigma|^2.
\end{equation} 
We finally get:
\begin{align}
 \sigma(\pi^+ p \rightarrow \Sigma^+ K^+) & = \sigma(n \pi^- \rightarrow \Sigma^- K^0) \nonumber \\
= \frac{5}{3} \sigma(\pi^0 p \rightarrow \Sigma^+ K^0) & = \frac{5}{3} \sigma(\pi^- p \rightarrow \Sigma^0 K^0) \nonumber \\
= \frac{5}{3} \sigma(n \pi^+ \rightarrow \Sigma^0 K^+) & = \frac{5}{3} \sigma(n \pi^0 \rightarrow \Sigma^- K^+) \nonumber \\
= 2 \sigma(\pi^0 p \rightarrow \Sigma^0 K^+) & = 2 \sigma(n \pi^0 \rightarrow \Sigma^0 K^0) \nonumber \\
= \frac{5}{3} \sigma(\pi^- p \rightarrow \Sigma^- K^+) & = \frac{5}{3}\sigma(n \pi^+ \rightarrow \Sigma^+ K^0).
\end{align}

\begin{figure}[ht]
\centering
\subfloat{\includegraphics[width=0.6\columnwidth]{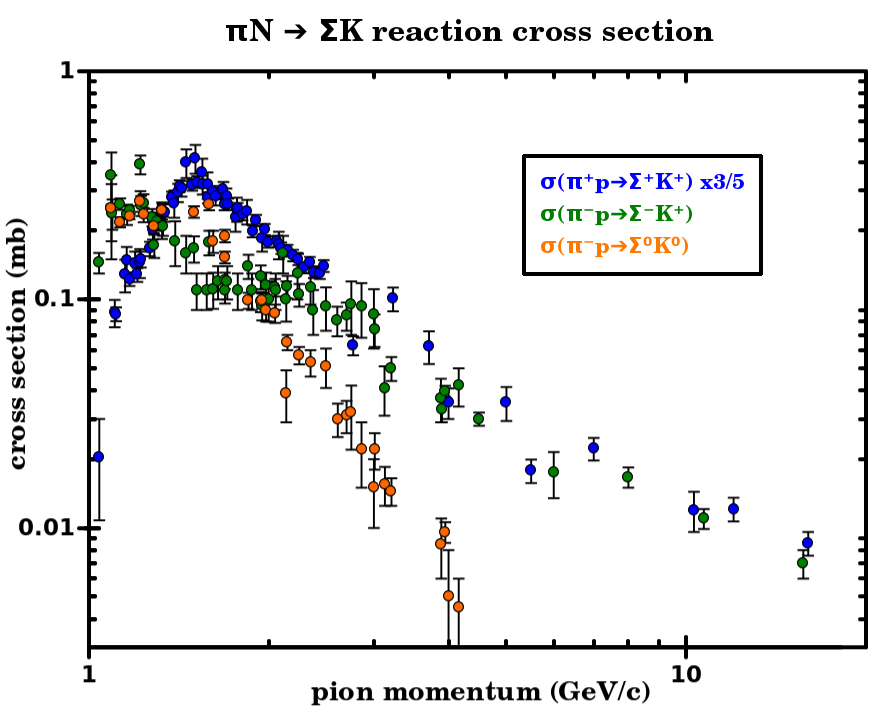}}
\caption{Experimental $\pi N \rightarrow \Sigma K$ cross sections}
\label{piNtoSK}
\end{figure}

After all necessary relations have been found, the result is always compared to the experimental data and/or the predictions by the Bystricky procedure, if available, in order to check if the hypotheses used are reasonable. Unfortunately, in this special case the result obtained by the HEM procedure is not very reasonable (see \autoref{piNtoSK}), likely due to unreliable hypotheses.

\begin{figure}[ht]
\centering
\subfloat{\includegraphics[width=0.7\columnwidth]{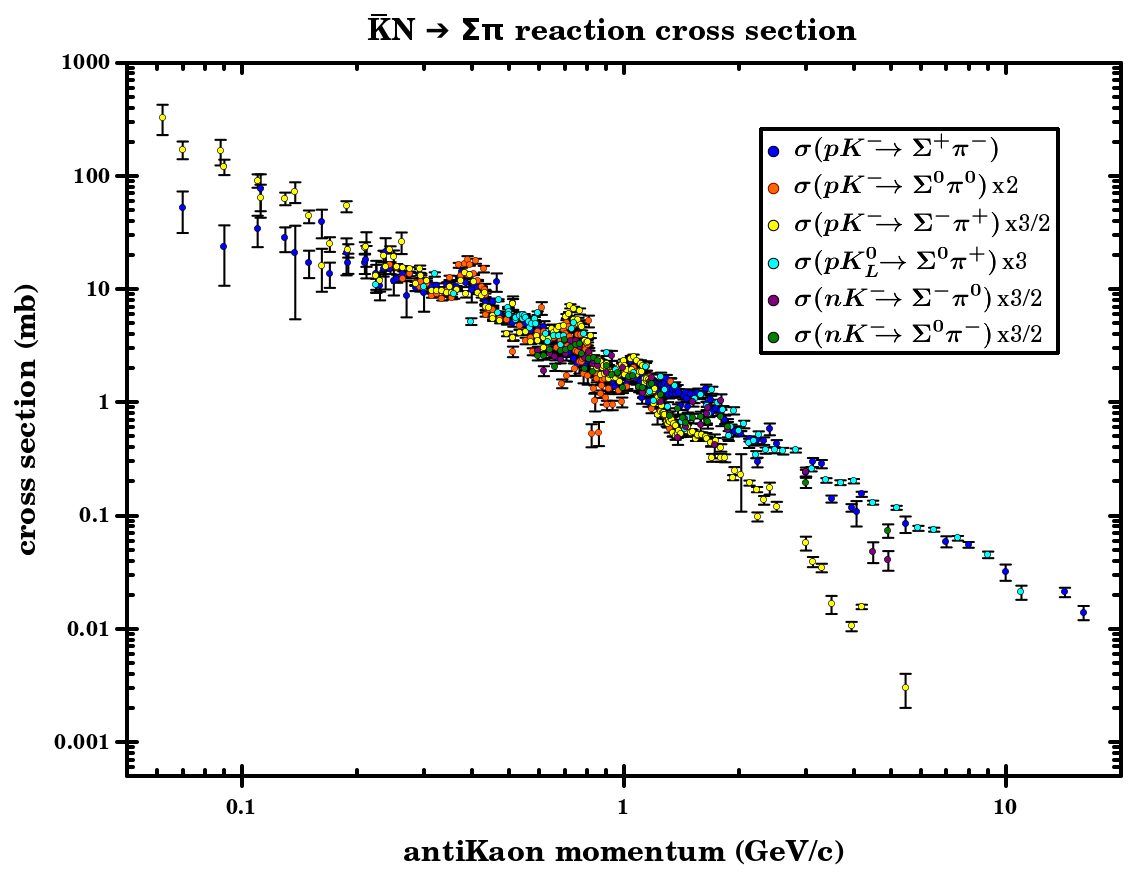}}
\caption{Experimental $\overline{K} N \rightarrow \Sigma \pi$ cross section showing the cross section of known channels normalized with the coefficients given by the HEM are equivalent.}
\label{KbNtoSpi}
\end{figure}

We anticipate that, the Bystricky procedure predictions associated to available experimental data are sufficient for parametrizing all $\pi N \rightarrow \Sigma K$ channels. Then, exclusive cross sections were fitted channel per channel for the ones with experimental data and the other cross sections are determined using the symmetries from the Bystricky procedure. However, in cases without enough experimental data, the relations obtained with sometimes questionable hypotheses must be kept. In general, the reliability of relations found using this method decreases with the increasing number of outgoing particles. This is due to the increasing number of Feynman diagrams, which should be taken into account and which then increases the number of hypothesis needed. An example of a case that works well even if the prediction does not match perfectly over the entire energy range with the experimental data for many channels is shown in \autoref{KbNtoSpi}.

The errors introduced by this method on the isospin average cross sections are estimated to be around 10\%-20\%, supposing that hypotheses are wisely chosen because, even if a specific isospin channel is under- or overestimated by a large factor, the Bystricky procedure provides relatively strong constraints on the isospin average cross sections. The list of all graphs considered and relations found are available in \autoref{channel}.

Thanks to the use of isospin symmetry in the hadron exchange model, combined with experimental data and the Bystricky procedure, around 72\% of the required information (\autoref{reac1}) can be obtained.

\subsection{Enlarging the data set}

Unfortunately, both methods, which are based on isospin symmetries in combination with experimental data, are not sufficient to provide a parametrization for all reactions listed in \autoref{reac1}\,. The missing cross sections were either obtained from models or from our best knowledge of {\it similar} reactions (notably based on reactions already studied in a previous version of INCL \cite{incl}).
The reactions of interest are: 
\begin{itemize}[label=$\bullet$]
\item $NN \rightarrow NN K \overline{K}$, 
\item $NN \rightarrow N \Lambda K \pi$, $NN \rightarrow N \Sigma K \pi$,
\item $NN \rightarrow N \Lambda K \pi\pi$,  $NN \rightarrow N \Sigma K \pi\pi$.
\end{itemize}

Parametrization of the $NN \rightarrow NN K \overline{K}$ reaction cross section parametrization is taken from \cite{sibirtsev}(Eq. 21).

For the other four reactions we assume similarities with the already included reactions $\sigma(NN \rightarrow NN \pi)$ and $\sigma(NN \rightarrow NN \pi\pi)$, taking into account the center of mass energy ($\sqrt s$ in MeV in the following equations). Actually, in these cases, the changes in the shape of the cross sections, when adding a pion in the final state, is supposed to be the same as if a hyperon and a kaon replace a nucleon and a pion.
\begin{equation}
\sigma_{NN \rightarrow N \Lambda K \pi}(\sqrt s) = 3\ \sigma_{NN \rightarrow N \Lambda K}(\sqrt s) \times \frac{\sigma_{NN \rightarrow NN \pi\pi}(\sqrt s - 540)}{\sigma_{NN \rightarrow NN \pi}(\sqrt s - 540)},
\label{NLKpi}
\end{equation}
\begin{equation}
\sigma_{NN \rightarrow N \Sigma K \pi}(\sqrt s) = 3\ \sigma_{NN \rightarrow N \Sigma K}(\sqrt s) \times \frac{\sigma_{NN \rightarrow NN \pi\pi}(\sqrt s - 620)}{\sigma_{NN \rightarrow NN \pi}(\sqrt s - 620)},
\label{NSKpi}
\end{equation}
\begin{equation}
\sigma_{NN \rightarrow N \Lambda K \pi\pi}(\sqrt s) = \sigma_{NN \rightarrow N \Lambda K \pi}(\sqrt s) \times \frac{\sigma_{NN \rightarrow NN \pi\pi}(\sqrt s - 675)}{\sigma_{NN \rightarrow NN \pi}(\sqrt s - 675)},
\end{equation}
\begin{equation}
\sigma_{NN \rightarrow N \Sigma K \pi\pi}(\sqrt s) = \sigma_{NN \rightarrow N \Sigma K \pi}(\sqrt s) \times \frac{\sigma_{NN \rightarrow NN \pi\pi}(\sqrt s - 755)}{\sigma_{NN \rightarrow NN \pi}(\sqrt s - 755)}.
\end{equation}

The factor 3 used in \autoref{NLKpi} and \autoref{NSKpi} is a normalization factor needed to fit the few available experimental data. The method was tested using the same type of reaction cross sections (strangeness produced or not) with the $\pi N$ initial state that are already relatively well described. It appears also  a factor of approximatively 3 between the cross section ratio $\sigma_{\pi N~\rightarrow~N \pi\pi\pi}/\sigma_{\pi N~\rightarrow~N \pi\pi}$ and the cross section ratio  $\sigma_{\pi N \rightarrow Y K \pi}/\sigma_{\pi N \rightarrow Y K}$ with the appropriately shifted center of mass energy. Note that this verification starts with the $\pi N~\rightarrow~N \pi \pi$ reaction, because the reaction $\pi N~\rightarrow~N \pi$ is an elastic reaction and therefore, is clearly not similar to $\pi N \rightarrow Y K$.

The charge repartition is determined by using the work done in \autoref{HEM} for $NN~\rightarrow~NN K \overline{K}$, $NN \rightarrow N \Lambda K \pi$, and $NN \rightarrow N \Sigma K \pi$ reactions. As discussed previously, the method based on the hadron exchange model is not used to calculate the total cross sections for those reactions (too many hypotheses needed), but it can be used to determine the charge repartition. The charge repartition for $NN \rightarrow N \Lambda K \pi\pi$ and $NN \rightarrow N \Sigma K \pi\pi$ were determined using an approach by Iljinov et \textit{al.}\cite{iljinov}, simplified to take into account only the combinatorics of the final state as it was done in the Bertini model \cite{bertini}. The method determines the ratio of channel cross sections from a same reaction based only on the particle multiplicities in the final state as:
\begin{equation}
\frac{\sigma \left( A+B\rightarrow \hspace{-2mm} \sum\limits_{i=n,p,\pi^+,...} \hspace{-2mm} x_i i \right) }{\sigma \left( A'+B'\rightarrow \hspace{-2mm} \sum\limits_{j=n,p,\pi^+,...} \hspace{-2mm} x_j j \right) } = \frac{\prod\limits_{i=n,p,\pi^+,...} x_i!}{\prod\limits_{j=n,p,\pi^+,...}^{} x_j!}
\end{equation}
with $x_i$ the number of particle $i$ in the final state.

In addition, and as mentioned in \autoref{II} and \autoref{reac2}, two additional reaction types must be taken into account: strangeness production reactions with numerous particles in final states and $\Delta$-induced strange production reactions.

\begin{figure}
\centering
\includegraphics[width=0.5\columnwidth]{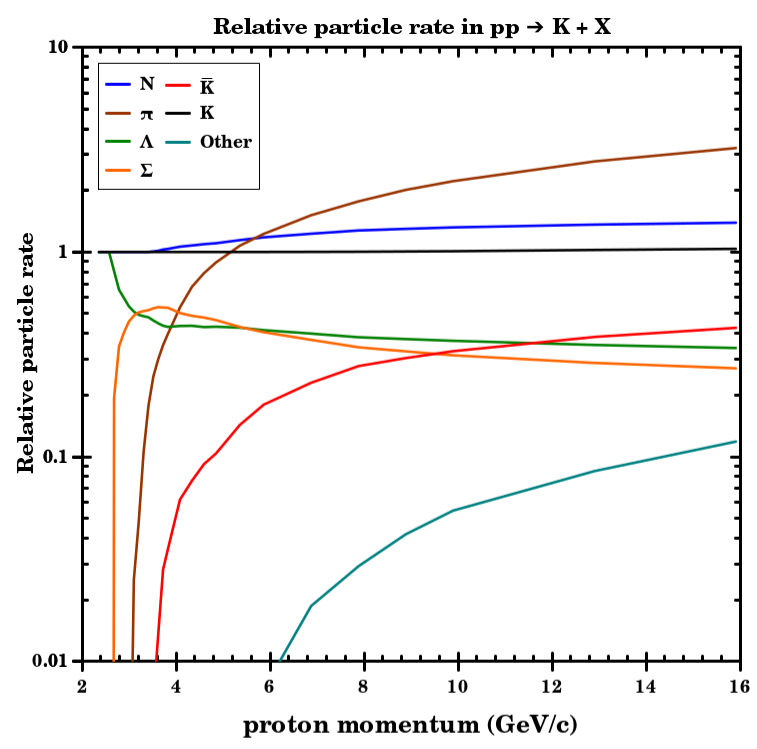}
\caption{Particle rate per reaction in $pp \rightarrow K + X$ reactions in the Fritiof model \protect{\cite{fritiof}} as a function of the incident proton momentum.}
\label{fritiof}
\end{figure}

For increasing energy, kaon production is associated with an increasing number of particles in the final state and, consequently, the reactions listed in \autoref{reac1} are not sufficient to account for kaon production. Actually, the additional particles are mostly pions as demonstrated by the Fritiof model \cite{fritiof} (see \autoref{fritiof}). Therefore, regarding the high-energy reactions $NN \rightarrow K + X$ and $\pi N \rightarrow K + X$, inclusive parametrizations of the cross sections are determined from experimental measurement and individual cross sections can be generated by trying to reproduce as good as possible the particle multiplicities given by the Fritiof model \cite{fritiof} using a random generator.

The parametrization for $\Delta$-induced strangeness production cross sections listed in \autoref{reac2} are taken from \cite{tsushima}, except for the reaction $\Delta N \rightarrow NN K \overline{K}$, which is discussed below and given in \autoref{param}. Since the estimates given by \cite{tsushima} for the cross sections related to $\Delta N$ collisions are very large compared to the cross sections related to $NN$ collisions with the same final states (factor $\sim$10), it was decided to take the isospin average cross section $\sigma(\Delta N\rightarrow~NNK\overline{K})$ as 10 times the isospin average cross section $\sigma(NN\rightarrow~NNK\overline{K})$.

Even if the number of $\Delta$ particles present in the nuclear volume during the collision is significantly lower than the number of pions and nucleons, $\Delta$-induced reaction are expected to contribute significantly to the strangeness production. Indeed, the cross sections calculated by Tsushima et \textit{al.} \cite{tsushima} for $\Delta$-induced reactions are much larger than those measured for pion-induced or nucleon-induced reactions. However, for these parametrizations, they used hypotheses, which are not obviously good for the entire energy range studied in this work and the experimental data in $NN\rightarrow~NYK$ calculated with the same hypotheses are not always well reproduced (see \cite{tsushima}, Fig.7). Considering the rather large uncertainties associated to these theoretical cross sections, this kind of reaction is supposed to be the largest source of error on strangeness production in our code.

The charge repartition was determined based on information obtained from the Bystricky procedure and the Hadron Exchange Model.

\subsection{Parametrizations}
\label{fitt}

Different generic formula were used to parametrize the reaction cross sections. The reactions considered are of two types: elastic and inelastic. This section presents our choice of fit functions. We give below the generic formula and in \autoref{param} the parametrizations for all reactions in the whole energy domain considered (momentum in laboratory frame of reference below $15~GeV$).

The elastic scattering cross sections become extremely large when the incoming particle momentum goes down to zero. Upper limits are placed at low energies to avoid cross section divergences. The limits have no consequences on the final result if placed high enough, because the cross sections are only used to determine which reaction will contribute. The elastic cross sections appear relatively complex in the energy range studied here to be defined by a singular function. As a result of which the energy ranges studied were split into several parts in order to get better parametrizations of the cross sections. The following functions were used:
\begin{equation}
\sigma(p_{Lab}) = a + b \ e^{-c \ p_{Lab}},
\end{equation}
\begin{equation}
\sigma(p_{Lab}) = a + b \ p_{Lab}^{-c}.
\end{equation}

Note that this kind of reaction is often resonant; the resonances are fitted by adding bumps of Gaussian shape on the underlying background.

The quasi-elastic reactions, which are $N K~\rightarrow~N' K'$, $N \overline{K}~\rightarrow~N' \overline{K}'$, and $N \Sigma~\rightarrow~N' \Sigma'$, are especially problematic at low energies with respect to the assumption of isospin symmetry because of the existence or absence of reaction thresholds. This asymmetry is taken into account by a cross section shift, which ``breaks'' the isospin symmetry hypothesis for both reactions.

The inelastic cross sections are the most important for the physics studied here. A lot of different formulae were tested. The following function, which is similar to formulae found in literature, gives good results for most reactions. We used the basic formula over the entire energy range even for those reactions where only few data concentrated in a narrow energy range exist.
\begin{equation}
\label{zea}
\sigma(p_{Lab}) = a \frac{(p_{Lab} - p_0)^b}{(p_{Lab} + p_0)^c \ p_{Lab}^d},
\end{equation}
with $p_0$ the threshold momentum and $a$, $b$, $c$, and $d$ positive fitting parameters.
In a few cases, Gaussian functions are added in order to fit resonances.

\section{Characteristics of the final states}
\label{IV}

After fixing the type of reaction, the final state must be determined. Doing so, charge and momentum must be assigned to each particle in the final state.

In most cases, charge repartition is determined using isospin symmetry and the hadron exchange model, which both predict relations between the isospin channel cross sections. The ratios are given in \autoref{channel}. We then randomly chose the charge repartition using the ratios determined before. For the reaction $NN \rightarrow N Y K \pi\pi$, the Bystricky procedure and the hadron exchange model discussed in \autoref{III} are not able to provide any ratio. Therefore, the simplified Iljinov et \textit{al.} approach \cite{iljinov} is used.

The other information needed to define the final state is the three-momentum of outgoing particles. In INCL, there are two different options to determine the kinematics of outgoing particles: the first one is to provide an angular distribution based on experimental measurements. The second one is to use a phase space generator, which is isotropic for the simplest cases or more sophisticated for more complex cases (Kopylov \cite{kopylov} or Raubold-Lynch \cite{james}). Typically, no experimental data are available and therefore, phase space generators are used. Nevertheless, studies providing Legendre coefficient have been carried out for $\overline{K}N$\cite{DA6,DA20,DA21,DA212,DA24,DA34,DA85,DA90,DA93,DA96,DA105,DAa,DAb,DAc} and $\pi N$\cite{piN1,piN2,piN3,piN4,piN5,piN6,piN7} elastic and quasi-elastic reactions. The results are used to provide angular distributions for $\overline{K}N$ and $\pi N$ reactions. Details are given and summarized in \autoref{ref_AD} .

\begin{table}[!ht]
\centering
\begin{tabular}{|c|c|c|}
\hline
 $\Delta p(MeV/c)$ & Reaction & Refs \\
\hline
225 - 2374 & $K^-p \rightarrow K^-p$ & \cite{DAb,DA96,DA21,DA34,DA105,DA212,DA93,DAa,DA6}\\
235 - 1355 & $K^-p \rightarrow \overline{K}^0 n$ & \cite{DAb,DA21,DA34,DAc,DA90,DA105}\\
436 - 1843 & $K^-p \rightarrow \Lambda \pi^0$ & \cite{DA21,DA24,DA85,DA90,DA105,DA20}\\
436 - 865 & $K^-p \rightarrow \Sigma^0 \pi^0$ & \cite{DA21,DA24,DA85}\\
436 - 1843 & $K^-p \rightarrow \Sigma^\pm \pi^\mp$ & \cite{DA21,DA90,DA105,DA24}\\
930 - 2375 & $\pi^- p \rightarrow K^0 \Lambda^0$ & \cite{piN1,piN2,piN3}\\
1040 - 2375 & $\pi^- p \rightarrow K^0 \Sigma^0$ & \cite{piN4,piN5}\\
1105 - 2473 & $\pi^+ p \rightarrow K^+ \Sigma^-$ & \cite{piN6,piN7}\\
\hline
\end{tabular}
\caption{List of reactions where the angular distributions were studied experimentally. Momentum range and references are given.}
\label{ref_AD}
\end{table}

The angular distributions for a given energy are usually parametrized using Legendre polynomials as follows:

\begin{equation}
\label{Leg}
\frac{d\sigma (\sqrt{s},\Theta_{c.m.})}{d\Omega} = \lambdabar^2(\sqrt{s}) \sum_{l=0}^{n} A_l(\sqrt{s}) P_l(\cos\Theta_{c.m.}),
\end{equation}
with $\lambdabar$ the c.m. reduced wavelength, $A_l$ the $l^{th}$ Legendre coefficient, $\sqrt{s}$ the center of mass energy, $\Theta_{c.m.}$ the angle of the outgoing particle with its initial momentum in the center of mass reference frame, and $P_l$ the $l^{th}$~order Legendre polynomial.

The experimental papers treating the angular distributions provide often $A_l$ at different energies\cite{DA6,DA20,DA21,DA212,DA24,DA34,DA85,DA90,DA93,DA96,DA105,DAa,DAb,DAc,piN1,piN2,piN3,piN4,piN5,piN6}. If it is not the case, like in \cite{piN7}, Legendre coefficients were determined by us (c.f.  \autoref{Leg_Table}). However, the Legendre coefficients determined in experiments strongly depend on the experimental set-up, like the backward detection and the angular binning, and can therefore provide an angular distribution that is only valid in a partial angular range. Sometimes, aberrations like negative density probability also appear. In an intranuclear cascade model, a description of Legendre coefficients as a function of the energy is needed. Doing so, a direct (non-parametric) fit of the $A_l$ using all Legendre coefficients coming from the experiments were done. Using these fitted $A_l$, we observed that most of the negative density probability problems disappeared. When negative probability density problems persists, the density is set to zero. Thanks to the cross section parametrization (see \autoref{III}), only the $A_i(\sqrt s)/A_0(\sqrt s)$ fittings are needed. Below, we elaborate on the two methods used to define the $A_i(\sqrt s)/A_0(\sqrt s)$ ratios in the given energy range.

The first method used is the Nadaraya–Watson kernel regression \cite{nadaraya}. The parametrization of the ratios is obtained by determining the function $\hat{m_h}(x)$ given by:
\begin{equation}
\label{Kernel}
\hat{m_h}(x) = \frac{\sum_{i=1}^n K_h(x-x_i) . y_i}{\sum_{i=1}^n K_h(x-x_i)},
\end{equation}
with $(x_i,y_i)$ the set of $n$ data, $K_h$ is a kernel, here a Gaussian with a standard deviation defined so that their quartiles (viewed as probability densities) are at $\pm 0.25 h$. The denominator in \autoref{Kernel} is the normalization term. In our analyse, the bandwidth was chosen as $h$ = $25$, $50$, $100$, $150$ or $200~MeV/c$ either on the whole energy range or according to energy bins. The latter case is used when complex structures or narrow resonances appear, taking care of avoiding fitting non physical fluctuations.

The second method used is the smoothing spline regression~\cite{smoothingsplines}. This method consists in the minimization of the following function:
\begin{equation}
\sum_{i=1}^n \left(y_i - \hat{\mu}(x_i) \right) ^2 + \lambda \int_{x_1}^{x_n} \left(\hat{\mu}^{''}(x)\right)^2 dx,
\end{equation}
with $(x_i,y_i)$ the set of $n$ data, $\hat{\mu}$ the non-parametric fit function (a spline), and $\lambda$ the smoothing parameter. This method corresponds to the common $\rchi$ $^2$ minimization with a second term used to limit quick variations in the fit function. The smoothing parameter was for each cases optimized by hand to obtain a good compromise between the smoothness and the proximity to the data in order to fit resonances but to avoid fitting the noise.

\begin{figure}[ht]
\begin{center}
\includegraphics[width=0.5\columnwidth]{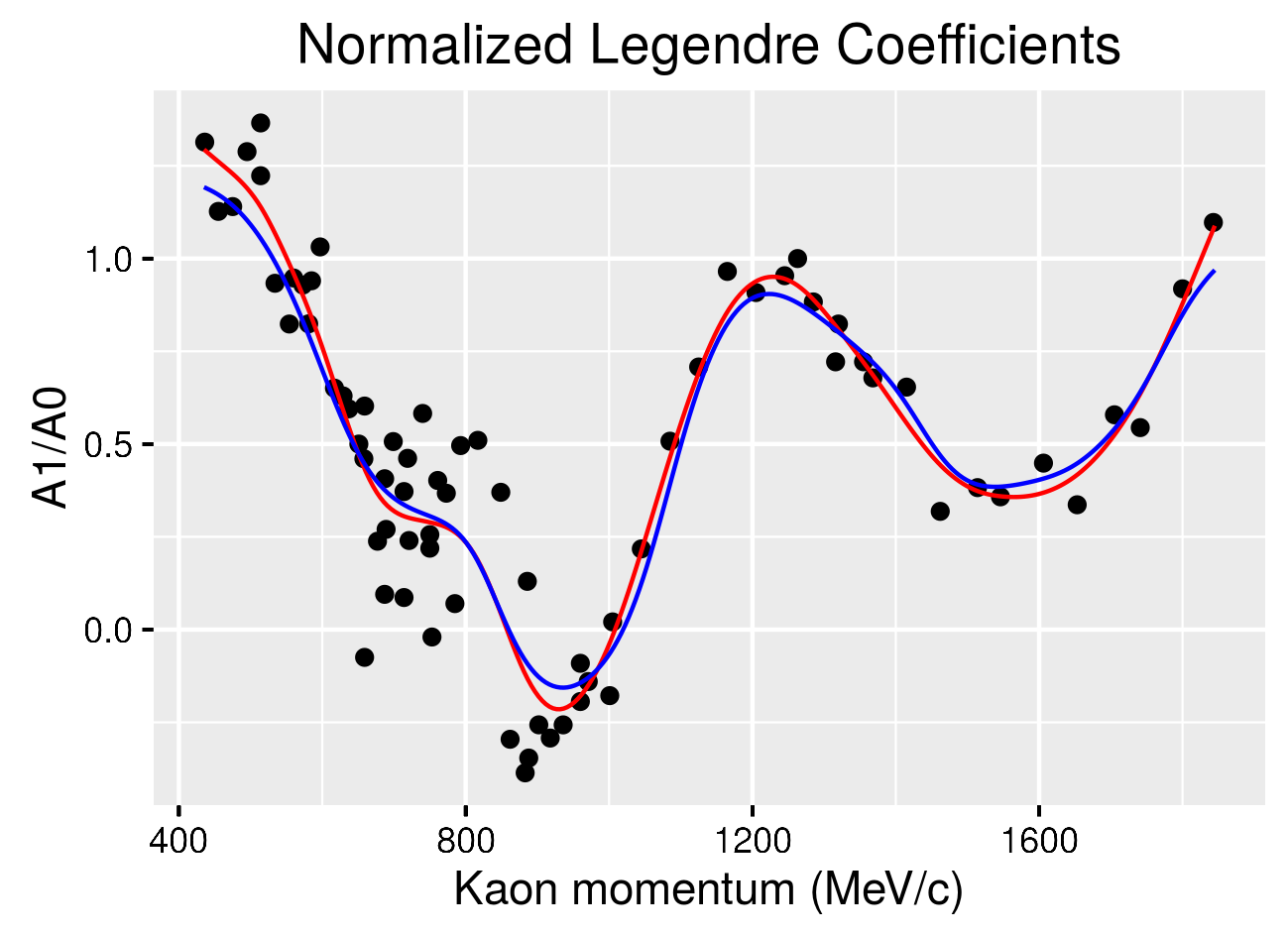}
\end{center}
\caption{Example of $A_1(\sqrt{s})/A_0(\sqrt{s})$ fit in the case $K^- p \rightarrow \Lambda \pi^0$ using Nadaraya-Watson kernel regression (blue), and smoothing spline regression (red).}
\label{regression}
\end{figure}

As already mentioned, there is no fit function for the two non-parametric methods. The result is a tabulation of Legendre coefficients as a function of the momentum with bins as small as needed. An example is shown in \autoref{regression}.

The two methods use completely different ways of fitting but give very similar results, as shown in \autoref{regression}. The choice to use one or the other was made case-by-case. Out of the data range, it was decided to use an isotropic distribution in the energy range below the experimental data and a more and more forward peaked distribution for higher energies

Tables used in INCL are available as electronic supplementary material in ``tabulation.pdf''. Note that the extrapolation of the $A_i(\sqrt s)/A_0(\sqrt s)$ outside the energy range considered here is not reliable and is likely to produce unphysical results.

\section{Comparison with other models}
\label{V}

Here we compare the input parameters determined in this paper, namely cross sections, charge repartition, and phase-space generation, to the same input parameters available in the literature and already used in other models considering strangeness production in the same energy range. These models are: (i) INCL2.0\cite{joseph,deneye}, a version developed to study anti-proton physics and including kaon physics, (ii) the Bertini Cascade model\cite{bertini}, and (iii) the GiBUU model\cite{gibuu}. To do this comparison, different examples will be discussed in order to show the strength and the weakness of each model.

\begin{figure}[b]
\begin{center}
\includegraphics[width=0.5\columnwidth]{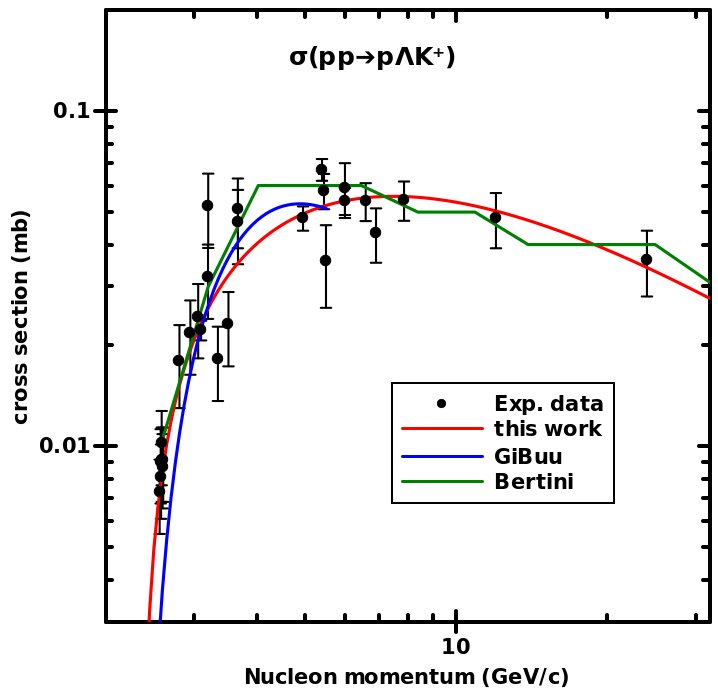}
\end{center}
\captionof{figure}{\label{pptoplkp} The $pp \rightarrow~p \Lambda K^+$ cross section fits from the Bertini cascade model(green line), GiBUU(blue line), and  this work(red line) compared to experimental data(black dots) as a function of the incident  proton momentum. Note that above 5.5 GeV/c GiBUU used Pythia, and so has no proper parametrization.}
\end{figure}

The different models parametrize the reactions using different methods. The Bertini cascade model tabulates the cross sections based on parametrization and calculation at 30 kinetic energies corresponding to as many intervals whose  width is increasing  logarithmically with the incident energy and spanning the $0$ to $32~$GeV domain. In INCL2.0, cross sections were parametrized only for reactions with two particles in the final state. The parametrization is often a fit in one or two parts using a formula like $\sigma = a~p^b$, with $p$ the momentum in the laboratory frame of reference. In the GiBUU model, the energy range is divided in two parts: the \textit{low-energy} part is fitted with parametrizations and the \textit{high-energy} part is treated using PYTHIA \cite{pythia}, which is based on the Lund string model \cite{lund}. The transition between the \textit{low-energy} parametrization and the PYTHIA predictions is a smooth linear transition in an energy transition range. The energy range considered in GiBUU is $\sqrt{s} = 2.2\pm 0.2~$GeV in meson-baryon collisions, which corresponds, in term of momentum, to $2.1\pm 0.5~$GeV/c for pion nucleon collisions and to $1.9\pm 0.2~$GeV/c for kaon nucleon collisions, and $\sqrt{s} = 3.4\pm 0.1~$GeV in baryon-baryon collision, which corresponds to $5.1\pm 0.4~$GeV/c for nucleon-nucleon collisions.

Nucleon-nucleon collisions have a high contribution in the strangeness production. The first open reaction channel with a proton as a projectile is the $pp \rightarrow p \Lambda K^+$ channel, which is important at low energies but which contributes less and less at high energies. As shown in \autoref{pptoplkp}, all models reproduce well the experimental cross sections. However, in the range $3.7-5~$GeV/c, where there are no experimental data, there are significant differences between the different fits. Such differences are very common when experimental data are not available in some energy range and/or are rather inconsistent.

\begin{figure}
\begin{center}
\includegraphics[width=0.5\columnwidth]{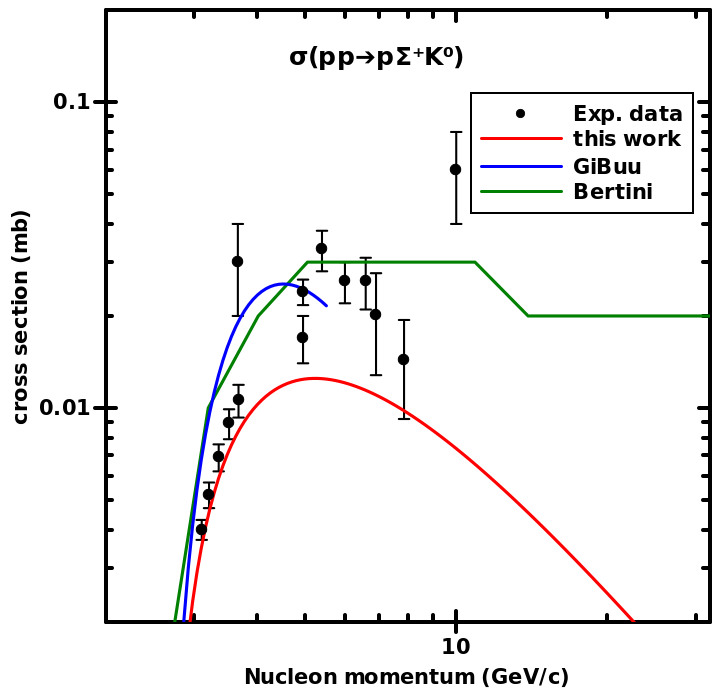}
\end{center}
\captionof{figure}{\label{ppNSK} The $pp \rightarrow p \Sigma^+ K^0$ cross section fits from the Bertini cascade model(green line), GiBUU(blue line), and  this work(red line) compared to experimental data(black dots) as a function of the incident proton momentum. Note that above 5.5 GeV/c GiBUU used Pythia, and so has no proper parametrization.}
\end{figure}

\begin{figure}[t]
\begin{center}
\includegraphics[width=0.45\columnwidth]{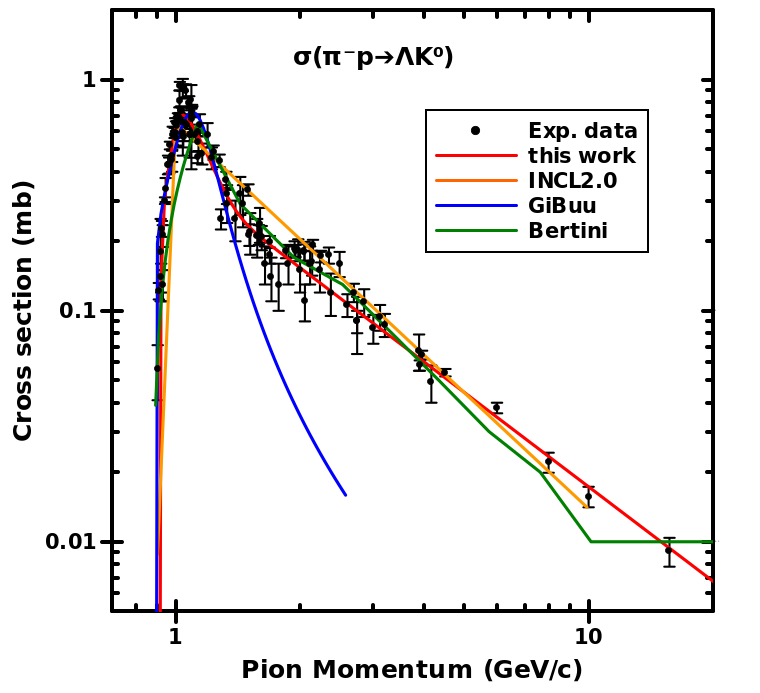}
\includegraphics[width=0.45\columnwidth]{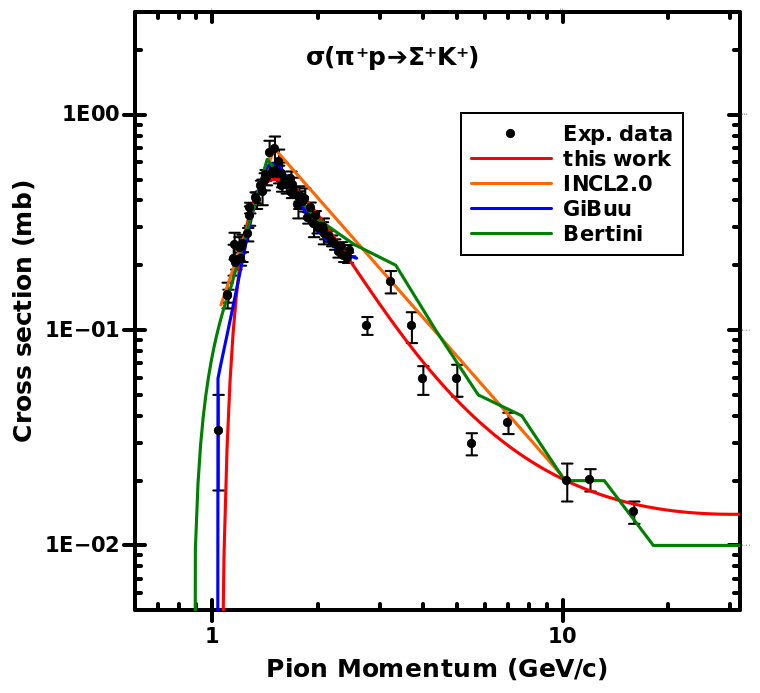}
\end{center}
\caption{The $\pi^- p \rightarrow \Lambda K^0$ and $\pi^+ p \rightarrow \Sigma^+ K^+$ cross sections fits from the Bertini cascade model(green line), GiBUU(blue line), INCL2.0(orange line), and  this work(red line) compared to experimental data(black dots) as a function of the incident pion momentum.}
\label{pipLK}
\end{figure}

A typical problematic channel is $pp \rightarrow~p \Sigma^+ K^0$ with the cross section parametrization shown in \autoref{ppNSK}. The parametrization from our work matches relatively well the experimental data at energies up to $4~$GeV/c but underestimates the high energy part. This is due to the compromise between inclusive calculations from the Fritiof model \cite{fritiof} and exclusive cross section measurements. We have chosen to artificially reduce our fit in order to be consistent with the inclusive cross section data. However, this type of reaction could deserve extra work according to its contribution in INC models. Another crucial point for this type of reaction, which can also be observed in \autoref{ppNSK}, is the inconsistency of the experimental data. For example, the two measurements around $3.7~$GeV/c differ with a factor $3$ and the data point at $10~GeV$ is suspiciously high compared not only to other data from this reaction but also compared to other isospin channels, which seem to show decreasing cross sections with increasing energy. The parametrizations in the other models differ strongly from our work. The Bertini cascade model and the GiBUU model, which uses the formula from \cite{tsushima} divided by a factor $1.5$, are other compromises with the experimental data.

\begin{figure}[t]
\begin{center}
\includegraphics[width=0.45\columnwidth]{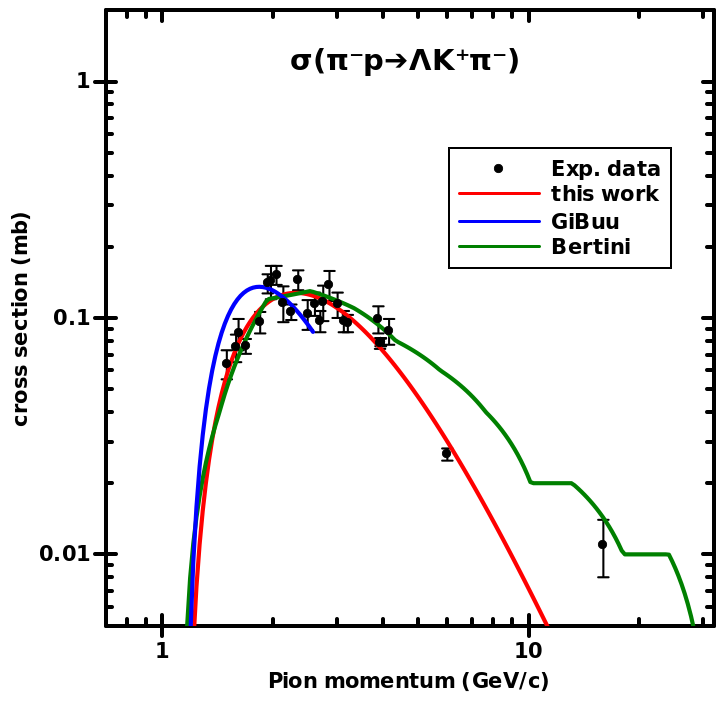}
\end{center}
\captionof{figure}{\label{pipYKpi}The $\pi^- p \rightarrow \Lambda K^+ \pi^-$  cross section fits from the Bertini cascade model(green line), GiBUU(blue line), and  this work(red line) compared to experimental data(black dots) as a function of the incident pion momentum.}
\end{figure}

The \autoref{pipLK} highlights a problem with the INCL2.0 parametrizations. The result of the parametrization describes correctly the magnitude of cross sections but does not give good fits of the energy dependence of the cross sections. As seen in \autoref{pipLK}, the cross section is slightly overestimated in the energy range $1.5-2~$GeV/c for the $\pi^- p \rightarrow \Lambda K^0$ reaction and in the energy range $1.5-10~$GeV/c for the $\pi^+ p \rightarrow \Sigma^+ K^+$ reaction. In the Bertini cascade model tabulations, because of the few energy intervals, quick variations in cross sections as a function of energy can be missed. For example, for the reaction $\pi^- p \rightarrow \Lambda K^0$ shown in \autoref{pipLK}, the Bertini cascade model reproduces well the experimental data near the threshold and at high energies but, the first interval being too wide, some part of the cross section is underestimated. The $\pi^- p \rightarrow \Lambda K^0$ cross section from the GiBUU model is close to the experimental data up to $1.4~$GeV/c but, surprisingly, there are relatively large deviations from the experimental data at higher momenta. However, this deviation is in the energy range of the transition between the parametrization and the PYTHIA model (see above). Note also that the parametrization for the reaction $\pi^+ p \rightarrow \Sigma^+ K^+$ from this work is slightly shifted to higher energies (about $10~$MeV - so seen only at low energies) because the isospin invariance considers an equal mass for all particles belonging to a same multiplet. Here, the mass for a multiplet was considered as the heaviest mass of this multiplet and therefore, can produce this artefact.

\begin{figure}[t]
\begin{center}
\includegraphics[width=0.45\columnwidth]{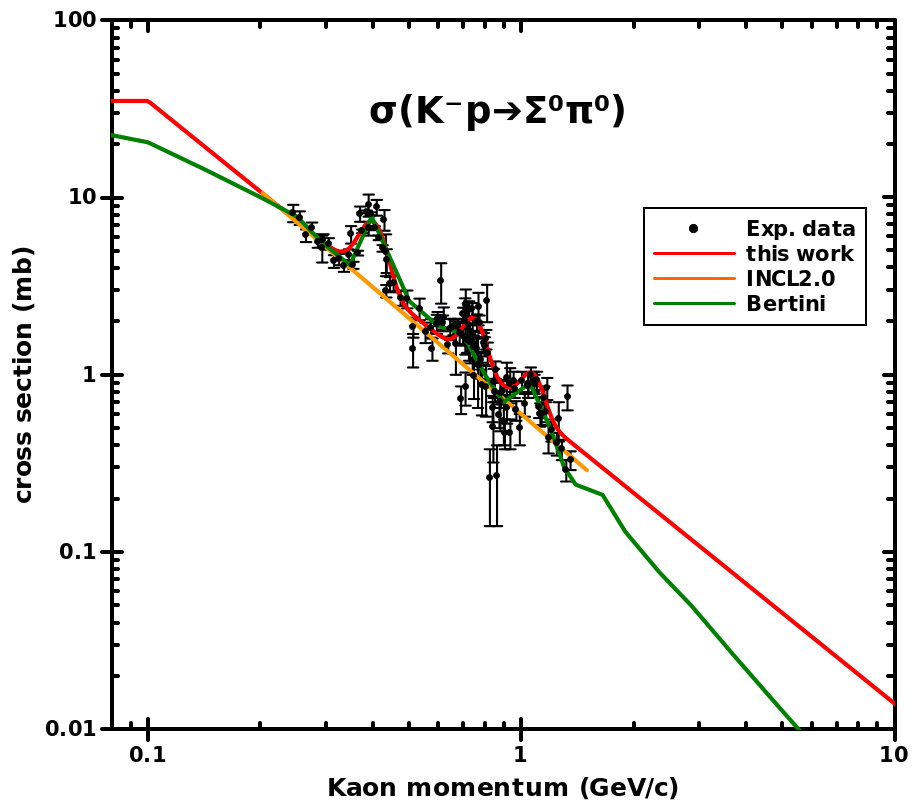}
\includegraphics[width=0.45\columnwidth]{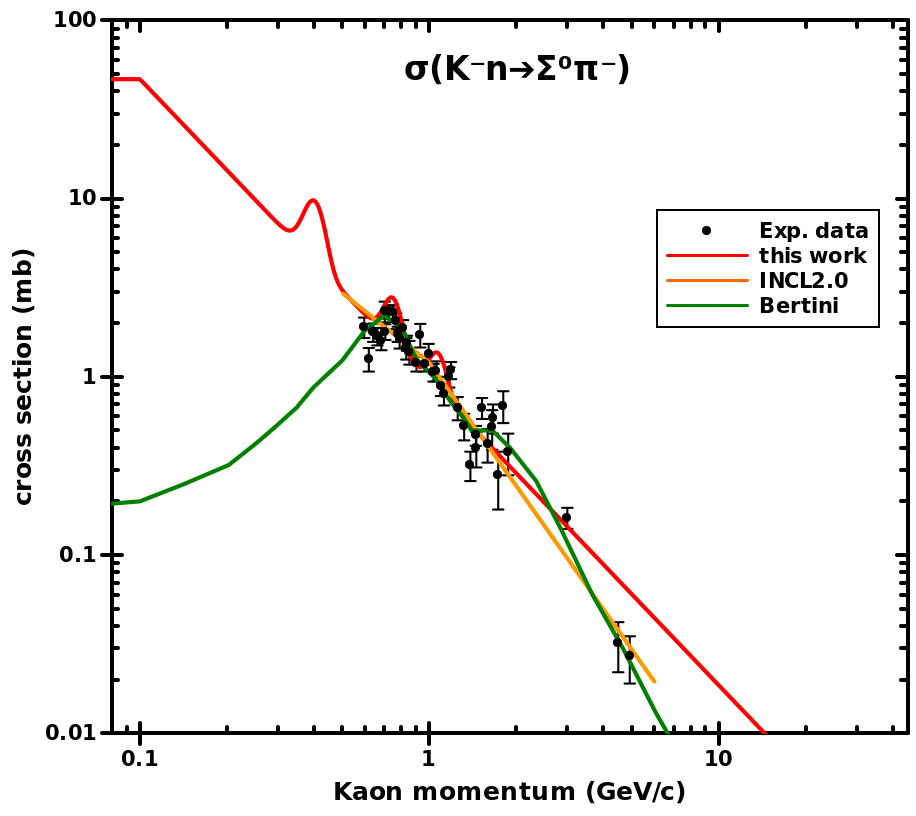}
\end{center}
\captionof{figure}{\label{kmp_spi}The $K^- p(n) \rightarrow \Sigma^{0} \pi^{0(-)}$ cross section fits from the Bertini cascade model(green line), INCL2.0 (orange line), and  this work(red line) compared to experimental data(black dots) as a function of the incident kaon momentum.}
\end{figure}

The \autoref{pipYKpi} illustrates another important result: the predictions at high energies from the Bertini cascade model are significantly different from our results. However, since there are only very few experimental data in this energy range, we cannot state which model is more reliable. This phenomenon is also visible in \autoref{kmp_spi}, though with more physical relevance. Deviations between experimental data and predictions are not very problematic when cross sections are relatively low because other reactions dominate. However, deviations of two orders of magnitude as seen for the reaction $K^- n\rightarrow~\Sigma^{0} \pi^-$ (\autoref{kmp_spi}) are much more significant. Again, looking only at the experimental data, it is not obvious which of the parametrizations are correct. Fortunately, for this special case the deviations have a low impact on the entire cascade because antiKaons, except if they are projectiles, play a minor role (very low production
yield).

\begin{figure}[t]
\begin{center}
\includegraphics[width=0.45\columnwidth]{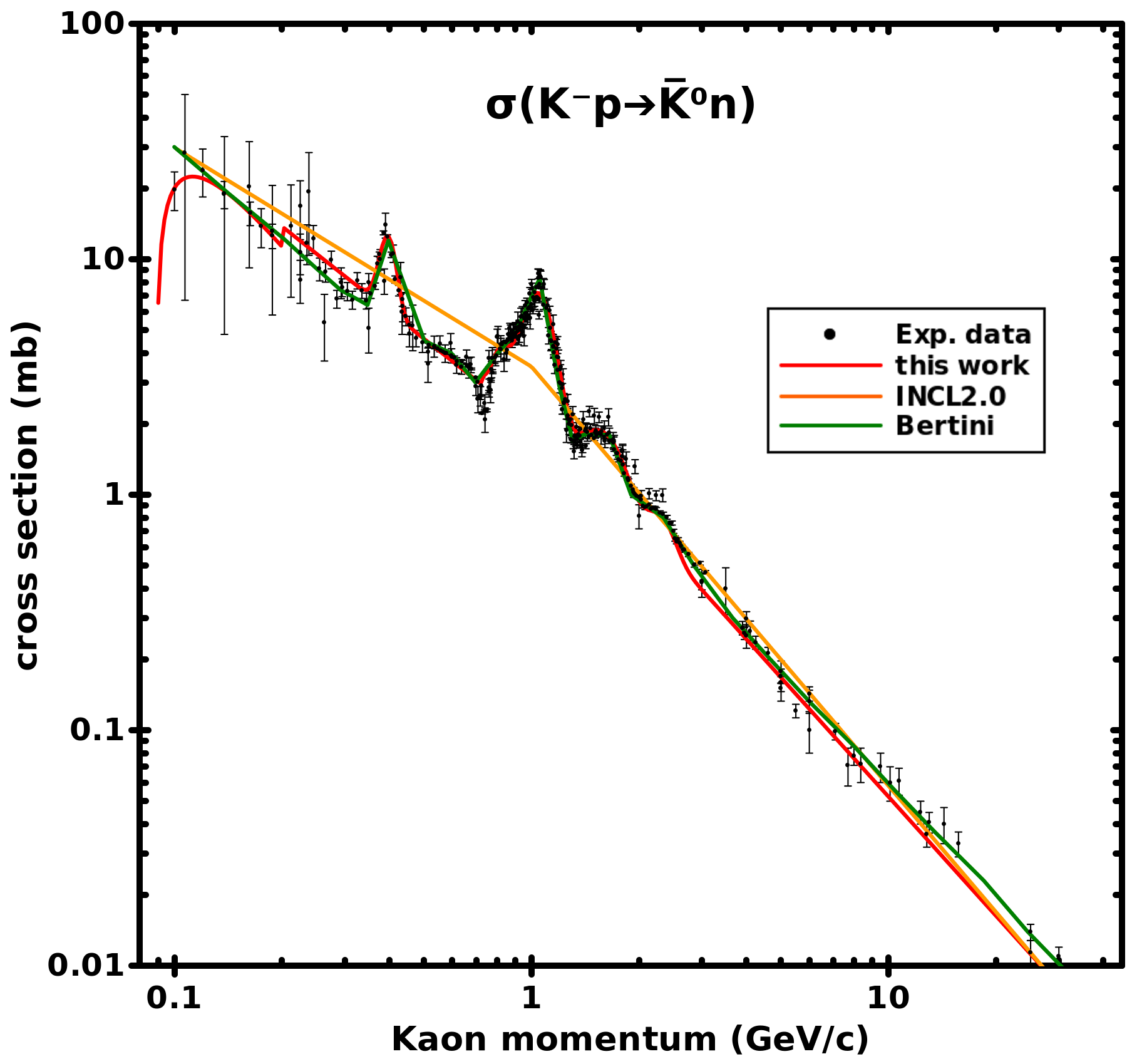}
\end{center}
\captionof{figure}{\label{kmp_quasi}The $K^- p$ quasi-elastic cross section fits from the Bertini cascade model(green line), INCL2.0(orange line), and  this work(red line) compared to experimental data(black dots) as a function of the incident kaon momentum.}
\end{figure}

Resonances are not treated directly in our work. However, they appear as Gaussians in the cross section parametrization.  If the hadron exchange model is used to determine a missing channel, those resonances appear also in the missing channel cross section even if they cannot be the intermediate state because of quantum number considerations. As an example, the resonances fitted for the reaction $K^- p \rightarrow \Sigma^{0} \pi^0$ (\autoref{kmp_spi}) appear also in the $K^- n\rightarrow~\Sigma^{0} \pi^-$ cross section fit, even if the third component of the isospin differs ($0$ for the former and $-1$ for the latter). 
Note that the GiBUU parametrization is not shown in \autoref{kmp_spi} because the reaction is treated in a different way using resonant and non-resonant cross sections. Therefore, no simple formula can be given. The \autoref{kmp_spi} and \autoref{kmp_quasi} also show another problem with the earlier INCL2.0 parametrizations: resonances are not reproduced. In contrast and as an improvement, the parametrizations proposed in this work and in the Bertini cascade model have no difficulties reproducing resonant cross sections.

\begin{figure}
\begin{center}
\includegraphics[width=0.5\columnwidth]{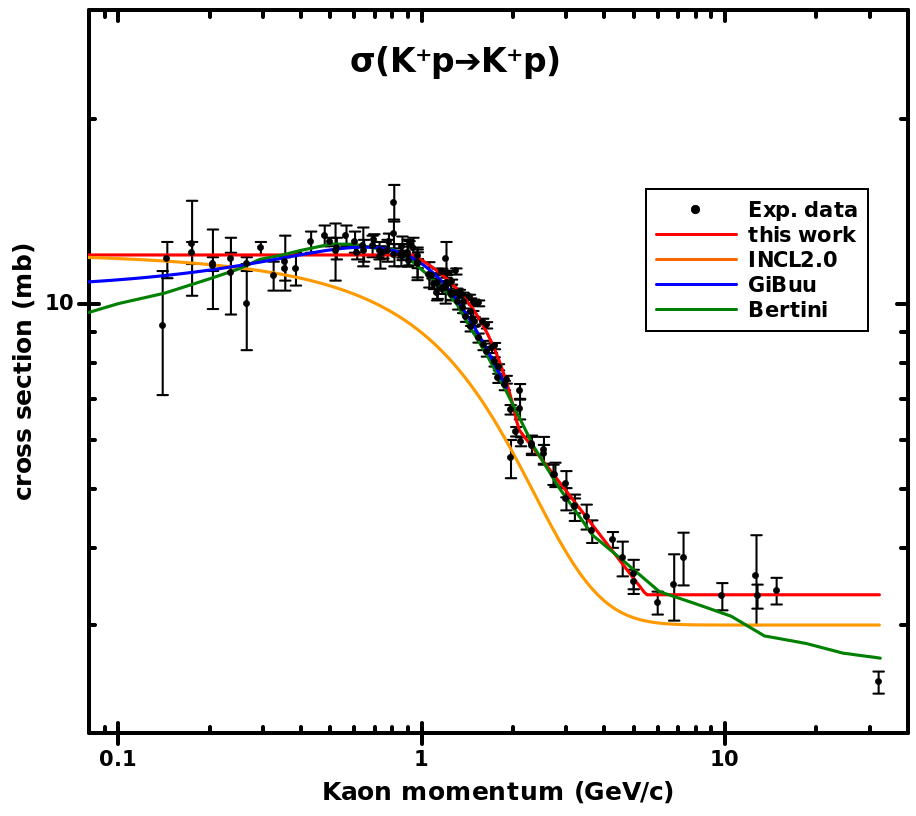}
\end{center}
\captionof{figure}{\label{kpp_ela}The $K^+ p$ elastic cross section fits from the Bertini cascade model(green line), INCL2.0(orange line), GiBUU(blue line), and this work(red line) compared to experimental data(black dots) as a function of the incident kaon momentum.}
\end{figure}

Unlike antiKaon-nucleon collision cross sections discussed above, the $K^+ p$ elastic cross section is important for spallation process with either nucleons or pions as projectiles. This is due to the low production rate of antiKaons compared to Kaons. The \autoref{kpp_ela} shows that the cross section is well reproduced using the results from this work and in the Bertini cascade model. Also the GiBUU model gives a good description of the experimental data. Differences between the three different approaches are observable at low energies where the differences are not very relevant because of the lack of competing processes in this energy range. In contrast, the INCL2.0 model underestimates the cross sections over the entire energy range.

In general, the parametrizations of the three different models fit the experimental data (if available) rather well. However, if experimental data are missing in an energy range, fits can be very different.

The two last subjects developed in our work are charge repartition and phase space generation. Since information about phase space generation in other models is too scarce, a comparison between the different models is not possible. Considering charge repartition, different methods are used by the different models. The Bertini cascade model uses a simplified version \cite{bertini} of the Iljinov et \textit{al.} approach \cite{iljinov}. For the GiBUU model, the charge repartition is determined using isospin rules and, in the case $\pi N \rightarrow NK\overline{K}$, using the hadron exchange model with $K^*$ and $\pi$ exchange diagrams. In INCL2.0, the charge repartition was determined using isospin invariance rules by neglecting interferences.

\section{Conclusion}
\label{VII}

A comprehensive and consistent description of all relevant elementary reactions involving strangeness production, scattering, and absorption when a light particle hit a nucleus was performed. Here we focused on energies below 15 GeV. The considered reactions are compiled in tables 1 and 2. This work was motivated by the implementation of strange particle physics into the intranuclear cascade model INCL with two major goals: refinement of the high-energy modelling (beyond 2-3 GeV) and possibility to contribute to hypernucleus studies.

This description includes parametrization of reaction cross sections, charge repartition, and phase space generation. These parametrizations are based on experimental measurements, when available, in order to be as model independent as possible. Unfortunately, for the reaction cross sections less than 20\% of the needed information can be obtained directly in this way. Therefore hypotheses and models  are used to complete the parametrization. Isospin symmetry allows to parametrize a large number of cross sections by linking known and unknown cross sections. This is applied in two different ways, either by taking into account only the initial and final states (called Bystricky procedure) or by considering the isospin symmetry at each vertex of the Feynman diagrams used in a hadron exchange model. Nevertheless, still roughly one third of the cross sections needed additional information for a full characterization. Then, in few cases where experimental data were rare, it was necessary to use similarities, \textit{e.g.}, in the cross section ratios when one pion is added. Finally two types of reactions were fully based on modelling, \textit{i.e.} without possible confrontation with experimental data : reactions with numerous particles in the final state (with increasing energy) and delta-induced reactions.

For quality control, we compared our cross sections to experimental data and parametrizations used in other models. They reproduce quite well the measurements, but assessing the quality of our cross sections for reactions and in energy ranges where no experimental data exist is still a problem. It is worth to mention that parametrizations differ often where no data point is measured. A typical case is  the $\Delta$-induced reactions that should play an interesting role. No measurements exist and our parametrization relies on a theoretical model stating that those channels contribute in a significant way in kaon and hyperon production [17].

This set of newly parametrized cross sections, dealing with strangeness, will be implemented in the INCL code and formulae are given in appendix B. Calculations of Kaon and hyperon production, as well as of hypernucleus production, from interactions of a light particle with a nucleus will be soon performed and compared to experimental data. According to the available measurements, not only the reliability of the parametrizations obtained in this work  will be estimated, but also the role and the weight of the different elementary reactions analysed.Those comparisons could add new constraints on these latter. 

We hope that this compilation of formulae will be useful, not only for the users of transport codes, but also to model developers and physicists, who are interested in hypernuclear physics.

\section{Acknowledgements}

The authors would like to thank Janus Weil and the GiBUU collaboration, Denis Wright, Nikolai Mokhov and Gudima Konstantin for the different models calculations. We also thank Georg Schnabel and Jose-Luis Rodriguez-Sanchez for useful and productive discussions.

\appendix
\onecolumn
\section{Relations extracted from the Hadron exchange model and from the Bystricky procedure.}
\label{channel}
\numberwithin{equation}{section}

\hspace{5mm}This appendix summarizes the relations obtained from the hadron exchange model (normal style) and the relation obtain from the Bystricky procedure (in bold) (see \autoref{bystricky} and \autoref{HEM}).

\hspace{5mm}In what follows, $N$ represents a nucleon, $\Delta$ a Delta particle, $B$ a nucleon or a Delta particle, $Y$ a hyperon, $\pi$ a pion, $K$ a Kaon (excluding $\overline{K}^0$ and $K^-$), and $\overline{K}$ an antiKaon.

\hspace{5mm}The reliability of equation displayed here are discussed in the paper. In resume, bold equations (coming from the Bystricky procedure) are highly reliable equations. Normal style equations (coming from the  hadron exchange model) often used debatable hypotheses, which could produce surprising results but always consistent with equations in bold.

\begin{center}
\fbox{Reaction type: $ N K (\overline{K}, \Lambda) \rightarrow N K (\overline{K}, \Lambda)$}
\end{center}

The reactions $ N K~\rightarrow~N K$, $ N \overline{K}~\rightarrow~N \overline{K}$, and $ N \Lambda~\rightarrow~N \Lambda$ do not have symmetries, except the trivial ones. They also have threshold effects, therefore the hadron exchange model is not relevant for these reactions.

\begin{center}
\fbox{Reaction type: $ NN \rightarrow NYK$}
\end{center}

\begin{align*}
 \sigma(pp \rightarrow p\Lambda K^+) & = \sigma(nn \rightarrow n\Lambda K^0) \\
 \sigma(pn \rightarrow p\Lambda K^0) & = \sigma(pn \rightarrow n\Lambda K^+)
\end{align*}

\begin{center}
\noindent\rule{8cm}{0.4pt}
\end{center}

\begin{align*}
 4 \sigma(pp \rightarrow p \Sigma^+ K^0) = 4 \sigma(nn \rightarrow n \Sigma^- K^+) & = 8 \sigma(pp \rightarrow p \Sigma^0 K^+) = 8 \sigma(nn \rightarrow n \Sigma^0 K^0) \\
= \sigma(pp \rightarrow n \Sigma^+ K^+) = \sigma(nn \rightarrow p \Sigma^- K^0) & = \frac{8}{5} \sigma(pn \rightarrow p \Sigma^0 K^0) = \frac{8}{5}\sigma(pn \rightarrow n \Sigma^0 K^+) \\
= 4 \sigma(pn \rightarrow p \Sigma^- K^+) & = 4 \sigma(pn \rightarrow n \Sigma^+ K^0)
\end{align*}
\boldmath
\begin{align*}
\sigma(pn \rightarrow p \Sigma^- K^+) + \sigma(pp & \rightarrow n \Sigma^+ K^+) + \sigma(pp \rightarrow p \Sigma^+ K^0) \\
= 2 \sigma(pn \rightarrow p \Sigma^0 K^0) & + 2 \sigma(pp \rightarrow p \Sigma^0 K^+)
\end{align*}
\unboldmath

\begin{center}
\fbox{Reaction type: $ NN \rightarrow NYK\pi$}
\end{center}

Calculation are based on $ NN \rightarrow \Delta YK \rightarrow NYK\pi$.

\begin{align*}
 \frac{4}{9}\sigma(pp \rightarrow p\Lambda K^0 \pi^+) = \frac{4}{9}\sigma(nn \rightarrow n\Lambda K^+ \pi^-) & = 2\sigma(pp \rightarrow p\Lambda K^+ \pi^0) = 2\sigma(nn \rightarrow n\Lambda K^0 \pi^0) \\
= 4\sigma(pp \rightarrow n\Lambda K^+ \pi^+) = 4\sigma(nn \rightarrow p\Lambda K^0 \pi^-) & = 2\sigma(pn \rightarrow p\Lambda K^+ \pi^-) = 2\sigma(pn \rightarrow n\Lambda K^0 \pi^+) \\
= \sigma(pn \rightarrow p\Lambda K^0 \pi^0) & = \sigma(pn \rightarrow n\Lambda K^+ \pi^0)
\end{align*}
\boldmath
\begin{align*}
 \sigma(pn \rightarrow p\Lambda K^+ \pi^-) + \sigma(pp & \rightarrow n\Lambda K^+ \pi^+) + \sigma(pp \rightarrow p\Lambda K^0 \pi^+) \\
 = 2 \sigma(pn \rightarrow p\Lambda K^0 \pi^0) & + 2 \sigma(pp \rightarrow p\Lambda K^+ \pi^0)
\end{align*}\unboldmath

\begin{center}
\noindent\rule{8cm}{0.4pt}
\end{center}

\boldmath
\begin{align*}
 \sigma(pn \rightarrow p \Sigma^- K^+ \pi^0) + \sigma(pp \rightarrow n \Sigma^+ K^+ \pi^0) & + \sigma(pp \rightarrow p \Sigma^+ K^0 \pi^0) \\
 = \sigma(pn \rightarrow p \Sigma^0 K^+ \pi^-) + \sigma(pp \rightarrow n \Sigma^0 K^+ \pi^+) & + \sigma(pp \rightarrow p \Sigma^0 K^0 \pi^+)
\end{align*}
\begin{align*}
 \sigma(pn \rightarrow p \Sigma^- K^0 \pi^+) + \sigma(pn & \rightarrow p \Sigma^+ K^0 \pi^-)  + \sigma(pp \rightarrow n \Sigma^+ K^0 \pi^+) \\
 + \sigma(pp \rightarrow p \Sigma^- K^+ \pi^+) & + \sigma(pp \rightarrow p \Sigma^+ K^+ \pi^-) \\
 = \sigma(pn \rightarrow p \Sigma^0 K^+ \pi^-) + 2 \sigma(pn & \rightarrow p \Sigma^0 K^0 \pi^0) + \sigma(pp \rightarrow n \Sigma^0 K^+ \pi^+) \\
 + 2 \sigma(pp \rightarrow p \Sigma^0 K^+ \pi^0) & + \sigma(pp \rightarrow p \Sigma^0 K^0 \pi^+)
\end{align*}\unboldmath
\begin{align*}
 \sigma(pp \rightarrow p \Sigma^+ K^0 \pi^0) = \sigma(nn \rightarrow n \Sigma^- K^+ \pi^0) & = 2\sigma(pp \rightarrow n \Sigma^+ K^0 \pi^+) = 2\sigma(nn \rightarrow p \Sigma^- K^+ \pi^-) \\
 = \sigma(pp \rightarrow n \Sigma^+ K^+ \pi^0) = \sigma(nn \rightarrow p \Sigma^- K^0 \pi^0) & = 2\sigma(pp \rightarrow p \Sigma^+ K^+ \pi^-) = 2\sigma(nn \rightarrow n \Sigma^- K^0 \pi^+) \\
 = \sigma(pp \rightarrow p \Sigma^0 K^+ \pi^0) = \sigma(nn \rightarrow n \Sigma^0 K^0 \pi^0) & = 2\sigma(pp \rightarrow n \Sigma^0 K^+ \pi^+) = 2\sigma(nn \rightarrow p \Sigma^0 K^0 \pi^-) \\ 
 = \frac{4}{9}\sigma(pp \rightarrow p \Sigma^0 K^0 \pi^+) = \frac{4}{9}\sigma(nn \rightarrow n \Sigma^0 K^+ \pi^-) & = \frac{4}{9}\sigma(pp \rightarrow p \Sigma^- K^+ \pi^+) = \frac{4}{9}\sigma(nn \rightarrow n \Sigma^+ K^0 \pi^-) \\
 = \frac{4}{9}\sigma(pn \rightarrow p \Sigma^- K^0 \pi^+) = \frac{4}{9}\sigma(pn \rightarrow n \Sigma^+ K^+ \pi^-) & = 2\sigma(pn \rightarrow p \Sigma^0 K^0 \pi^0) = 2\sigma(pn \rightarrow n \Sigma^0 K^+ \pi^0) \\
 = 4\sigma(pn \rightarrow p \Sigma^0 K^+ \pi^-) = 4\sigma(pn \rightarrow n \Sigma^0 K^0 \pi^+) & = \sigma(pn \rightarrow p \Sigma^- K^+ \pi^0) = \sigma(pn \rightarrow n \Sigma^+ K^0 \pi^0) \\
 = 2\sigma(pn \rightarrow p \Sigma^+ K^0 \pi^-) & = 2\sigma(pn \rightarrow n \Sigma^- K^+ \pi^+) 
\end{align*}

\begin{center}
\fbox{Reaction type: $ NN \rightarrow NN K\overline{K}$}
\end{center}

\begin{align*}
4 \sigma(pp \rightarrow pp K^+ K^-) = 4 \sigma(nn \rightarrow nn K^0 \overline{K}^0) & = 4 \sigma(pp \rightarrow pp K^0 \overline{K}^0) = 4 \sigma(nn \rightarrow nn K^+ K^-) \\
= \sigma(pp \rightarrow pn K^+ \overline{K}^0) = \sigma(nn \rightarrow pn K^0 K^-) & = \sigma(pn \rightarrow pp K^0 K^-) = \sigma(pn \rightarrow nn K^+ \overline{K}^0) \\
= 4/9 \ \sigma(pn \rightarrow pn K^+ K^-) & = 4/9 \ \sigma(pn \rightarrow pn K^0 \overline{K}^0)
\end{align*}

\textbf{No solution with the Bystricky procedure}

\begin{center}
\fbox{Reaction type: $ NK \rightarrow NK \pi$}
\end{center}

\begin{align*}
0.83 \sigma(p K^+ \rightarrow p K^+ \pi^0) = 0.83 \sigma(n K^0 \rightarrow n K^0 \pi^0) & = \frac{1}{3} \sigma(p K^+ \rightarrow p K^0 \pi^+) = \frac{1}{3} \sigma(n K^0 \rightarrow n K^+ \pi^-) \\
= 1.25 \sigma(p K^+ \rightarrow n K^+ \pi^+) = 1.25 \sigma(n K^0 \rightarrow p K^0 \pi^-) & = \sigma(p K^0 \rightarrow p K^+ \pi^-) = \sigma(n K^+ \rightarrow n K^0 \pi^+) \\
= 1.18 \sigma(p K^0 \rightarrow p K^0 \pi^0) = 1.18 \sigma(n K^+ \rightarrow n K^+ \pi^0) & = 0.68 \sigma(p K^0 \rightarrow n K^+ \pi^0) = 0.68 \sigma(n K^+ \rightarrow p K^0 \pi^0) \\
= 0.45 \sigma(p K^0 \rightarrow n K^0 \pi^+) & = 0.45 \sigma(n K^+ \rightarrow p K^+ \pi^-)
\end{align*}
\boldmath
\begin{align*}
 \sigma(p K^0 \rightarrow n K^0 \pi^+) + \sigma(p K^0 \rightarrow p K^+ \pi^-) & + \sigma(p K^+ \rightarrow n K^+ \pi^+) + \sigma(p K^+ \rightarrow p K^0 \pi^+) \\
 = 2 \sigma(p K^0 \rightarrow n K^+ \pi^0)  + 2 \sigma(p K^0 & \rightarrow p K^0 \pi^0) + 2 \sigma(p K^+ \rightarrow p K^+ \pi^0)
\end{align*}\unboldmath

\begin{center}
\fbox{Reaction type: $ NK \rightarrow NK \pi \pi$}
\end{center}

\begin{align*}
\sigma(p K^+ \rightarrow p K^+ \pi^+ \pi^-) = \sigma(n K^0 \rightarrow n K^0 \pi^+ \pi^-) & = 8 \sigma(p K^+ \rightarrow p K^+ \pi^0 \pi^0) = 8 \sigma(n K^0 \rightarrow n K^0 \pi^0 \pi^0) \\
= \sigma(p K^+ \rightarrow p K^0 \pi^+ \pi^0) = \sigma(n K^0 \rightarrow n K^+ \pi^0 \pi^-) & = 2 \sigma(p K^+ \rightarrow n K^+ \pi^+ \pi^0) = 2 \sigma(n K^0 \rightarrow p K^0 \pi^0 \pi^-) \\
= 4 \sigma(p K^+ \rightarrow n K^0 \pi^+ \pi^+) = 4 \sigma(n K^0 \rightarrow p K^+ \pi^- \pi^-) & = \sigma(p K^0 \rightarrow p K^+ \pi^0 \pi^-) = \sigma(n K^+ \rightarrow n K^0 \pi^+ \pi^0) \\
= \sigma(p K^0 \rightarrow p K^0 \pi^+ \pi^-) = \sigma(n K^+ \rightarrow n K^+ \pi^+ \pi^-) & = 8 \sigma(p K^0 \rightarrow p K^0 \pi^0 \pi^0) = 8 \sigma(n K^+ \rightarrow n K^+ \pi^0 \pi^0) \\
= 4 \sigma(p K^0 \rightarrow n K^+ \pi^+ \pi^-) = 4 \sigma(n K^+ \rightarrow p K^0 \pi^+ \pi^-) & = 4 \sigma(p K^0 \rightarrow n K^+ \pi^0 \pi^0) = 4 \sigma(n K^+ \rightarrow p K^0 \pi^0 \pi^0) \\
= 2 \sigma(p K^0 \rightarrow n K^0 \pi^+ \pi^0) & = 2 \sigma(n K^+ \rightarrow p K^+ \pi^0 \pi^-)
\end{align*}
\boldmath
\begin{align*}
 \sigma(p K^0 \rightarrow n K^0 \pi^+ \pi^0) + 4 \sigma(p K^0 \rightarrow n K^+ \pi^0 \pi^0) & + 4 \sigma(p K^0 \rightarrow p K^0 \pi^0 \pi^0) + \sigma(p K^0 \rightarrow p K^+ \pi^0 \pi^-) \\
 + \sigma(p K^+ \rightarrow n K^+ \pi^+ \pi^0) + \sigma(p K^+ \rightarrow p K^0 \pi^+ \pi^0) & + 4 \sigma(p K^+ \rightarrow p K^+ \pi^0 \pi^0) = 2 \sigma(p K^0 \rightarrow n K^+ \pi^+ \pi^-) \\
 + 2 \sigma(p K^0 \rightarrow p K^0 \pi^+ \pi^-) + 2 \sigma(p K^+ & \rightarrow n K^0 \pi^+ \pi^+) + 2 \sigma(p K^+ \rightarrow p K^+ \pi^+ \pi^-)
\end{align*}\unboldmath

\newpage

\begin{center}
\fbox{Reaction type: $ N\overline{K} \rightarrow N\overline{K} \pi$}
\end{center}

\begin{align*}
 12 \sigma(p\overline{K}^0 \rightarrow p\overline{K}^0 \pi^0) = 12 \sigma(nK^- \rightarrow nK^- \pi^0) & = 6\sigma(p\overline{K}^0 \rightarrow pK^- \pi^+) = 6\sigma(nK^- \rightarrow n\overline{K}^0 \pi^-) \\
= 12 \sigma(p\overline{K}^0 \rightarrow n\overline{K}^0 \pi^+) = 12 \sigma(nK^- \rightarrow pK^- \pi^-) & = 9 \sigma(pK^- \rightarrow p\overline{K}^0 \pi^-) = 9 \sigma(n\overline{K}^0 \rightarrow nK^- \pi^+) \\
= 12 \sigma(pK^- \rightarrow pK^- \pi^0) = 12 \sigma(n\overline{K}^0 \rightarrow n\overline{K}^0 \pi^0) & = 3 \sigma(pK^- \rightarrow n\overline{K}^0 \pi^0) = 3 \sigma(n\overline{K}^0 \rightarrow pK^- \pi^0) \\
= 8 \sigma(pK^- \rightarrow nK^- \pi^+) & = 8 \sigma(n\overline{K}^0 \rightarrow p\overline{K}^0 \pi^-)
\end{align*}
\boldmath
\begin{align*}
 \sigma(pK^- \rightarrow nK^- \pi^+) + \sigma(pK^- \rightarrow p\overline{K}^0 \pi^-) & + \sigma(p\overline{K}^0 \rightarrow n\overline{K}^0 \pi^+) + \sigma(p\overline{K}^0 \rightarrow pK^- \pi^+) \\
 = 2 \sigma(pK^- \rightarrow n\overline{K}^0 \pi^0) + 2 \sigma(pK^- & \rightarrow pK^- \pi^0) + 2 \sigma(p\overline{K}^0 \rightarrow p\overline{K}^0 \pi^0)
\end{align*}\unboldmath

\begin{center}
\fbox{Reaction type: $ N \overline{K} \rightarrow N \overline{K} \pi \pi$}
\end{center}
\begin{align*}
 \sigma(p\overline{K}^0 \rightarrow p\overline{K}^0 \pi^+ \pi^-) = \sigma(nK^- \rightarrow nK^- \pi^+ \pi^-) & = 4 \sigma(p\overline{K}^0 \rightarrow p\overline{K}^0 \pi^0 \pi^0) = 4 \sigma(nK^- \rightarrow nK^- \pi^0 \pi^0) \\
= \sigma(p\overline{K}^0 \rightarrow pK^- \pi^+ \pi^0) = \sigma(nK^- \rightarrow n\overline{K}^0 \pi^0 \pi^-) & = \sigma(p\overline{K}^0 \rightarrow n\overline{K}^0 \pi^+ \pi^0) = \sigma(nK^- \rightarrow pK^- \pi^0 \pi^-) \\
= \sigma(p\overline{K}^0 \rightarrow nK^- \pi^+ \pi^+) = \sigma(nK^- \rightarrow p\overline{K}^0 \pi^- \pi^-) & = \sigma(pK^- \rightarrow p\overline{K}^0 \pi^0 \pi^-) = \sigma(n\overline{K}^0 \rightarrow nK^- \pi^+ \pi^0) \\
= \sigma(pK^- \rightarrow pK^- \pi^+ \pi^-) = \sigma(n\overline{K}^0 \rightarrow n\overline{K}^0 \pi^+ \pi^-) & = 4 \sigma(pK^- \rightarrow pK^- \pi^0 \pi^0) = 4 \sigma(n\overline{K}^0 \rightarrow n\overline{K}^0 \pi^0 \pi^0) \\
= \sigma(pK^- \rightarrow n\overline{K}^0 \pi^+ \pi^-) = \sigma(n\overline{K}^0 \rightarrow pK^- \pi^+ \pi^-) & = 2 \sigma(pK^- \rightarrow n\overline{K}^0 \pi^0 \pi^0) = 2 \sigma(n\overline{K}^0 \rightarrow pK^- \pi^0 \pi^0) \\
= \sigma(pK^- \rightarrow nK^- \pi^+ \pi^0) & = \sigma(n\overline{K}^0 \rightarrow p\overline{K}^0 \pi^0 \pi^-)
\end{align*}
\boldmath
\begin{align*}
\sigma(pK^- \rightarrow nK^- \pi^+ \pi^0) + 4 \sigma(pK^- & \rightarrow n\overline{K}^0 \pi^0 \pi^0) + 4 \sigma(pK^- \rightarrow pK^- \pi^0 \pi^0) \\
+ \sigma(pK^- \rightarrow p\overline{K}^0 \pi^0 \pi^-) & + \sigma(p\overline{K}^0 \rightarrow n\overline{K}^0 \pi^+ \pi^0) \\
+ \sigma(p\overline{K}^0 \rightarrow pK^- \pi^+ \pi^0) & + 4 \sigma(p\overline{K}^0 \rightarrow p\overline{K}^0 \pi^0 \pi^0)\\
= 2 \sigma(pK^- \rightarrow n\overline{K}^0 \pi^+ \pi^-) & + 2 \sigma(pK^- \rightarrow pK^- \pi^+ \pi^-) \\
+ 2 \sigma(p\overline{K}^0 \rightarrow nK^- \pi^+ \pi^+) & + 2 \sigma(p\overline{K}^0 \rightarrow p\overline{K}^0 \pi^+ \pi^-)
\end{align*}\unboldmath

\begin{center}
\fbox{Reaction type: $ N \overline{K} \rightarrow Y \pi$}
\end{center}

\begin{align*}
 \sigma(p\overline{K}^0 \rightarrow \Lambda \pi^+) = \sigma(nK^- \rightarrow \Lambda \pi^-) = 2 \sigma(pK^- \rightarrow \Lambda \pi^0) = 2 \sigma(n\overline{K}^0 \rightarrow \Lambda \pi^0)
\end{align*}
\boldmath
\begin{equation*}
\sigma(p\overline{K}^0 \rightarrow \Lambda \pi^+) = 2 \sigma(pK^- \rightarrow \Lambda \pi^0)
\end{equation*}\unboldmath

\begin{center}
\noindent\rule{8cm}{0.4pt}
\end{center}

\begin{align*}
 \sigma(p\overline{K}^0 \rightarrow \Sigma^+ \pi^0) = \sigma(nK^- \rightarrow \Sigma^- \pi^0) & = \sigma(p\overline{K}^0 \rightarrow \Sigma^0 \pi^+) = \sigma(nK^- \rightarrow \Sigma^0 \pi^-) \\
= \frac{3}{4} \sigma(pK^- \rightarrow \Sigma^+ \pi^-) = \frac{3}{4} \sigma(n\overline{K}^0 \rightarrow \Sigma^- \pi^+) & = \frac{3}{2} \sigma(pK^- \rightarrow \Sigma^0 \pi^0) = \frac{3}{2} \sigma(n\overline{K}^0 \rightarrow \Sigma^0 \pi^0) \\
= \sigma(pK^- \rightarrow \Sigma^- \pi^+) & = \sigma(n\overline{K}^0 \rightarrow \Sigma^+ \pi^-)
\end{align*}
\boldmath
\begin{align*}
\sigma(p\overline{K}^0 \rightarrow \Sigma^+ \pi^0) & =\sigma(p\overline{K}^0 \rightarrow \Sigma^0 \pi^+) \\
\\
\sigma(pK^- \rightarrow \Sigma^- \pi^+) + \sigma(pK^- \rightarrow \Sigma^+ \pi^-) & = 2 \sigma(pK^- \rightarrow \Sigma^0 \pi^0) + \sigma(p\overline{K}^0 \rightarrow \Sigma^+ \pi^0)
\end{align*}\unboldmath

\begin{center}
\fbox{Reaction type: $ N \overline{K} \rightarrow Y \pi \pi$}
\end{center}

\begin{align*}
 \sigma(p\overline{K}^0 \rightarrow \Lambda \pi^+ \pi^0) = \sigma(nK^- \rightarrow \Lambda \pi^0 \pi^-) & = \sigma(pK^- \rightarrow \Lambda \pi^+ \pi^-) = \sigma(n\overline{K}^0 \rightarrow \Lambda \pi^+ \pi^-) \\
= 4 \sigma(pK^- \rightarrow \Lambda \pi^0 \pi^0) & = 4 \sigma(n\overline{K}^0 \rightarrow \Lambda \pi^0 \pi^0)
\end{align*}
\boldmath
\begin{align*}
4 \sigma(pK^- \rightarrow \Lambda \pi^0 \pi^0) + \sigma(p\overline{K}^0 \rightarrow \Lambda \pi^+ \pi^0)= 2 \sigma(pK^- \rightarrow \Lambda \pi^+ \pi^-)
\end{align*}\unboldmath

\begin{center}
\noindent\rule{8cm}{0.4pt}
\end{center}

\begin{align*}
\frac{3}{2} \sigma(p\overline{K}^0 \rightarrow \Sigma^+ \pi^+ \pi^-) = \frac{3}{2} \sigma(nK^- \rightarrow \Sigma^- \pi^+ \pi^-) & = 4 \sigma(p\overline{K}^0 \rightarrow \Sigma^+ \pi^0 \pi^0) = 4 \sigma(nK^- \rightarrow \Sigma^- \pi^0 \pi^0) \\
= \frac{6}{5} \sigma(p\overline{K}^0 \rightarrow \Sigma^0 \pi^+ \pi^0) = \frac{6}{5} \sigma(nK^- \rightarrow \Sigma^0 \pi^0 \pi^-) & = \frac{3}{2} \sigma(p\overline{K}^0 \rightarrow \Sigma^- \pi^+ \pi^+) = \frac{3}{2} \sigma(nK^- \rightarrow \Sigma^+ \pi^- \pi^-) \\
= \sigma(pK^- \rightarrow \Sigma^+ \pi^0 \pi^-) = \sigma(n\overline{K}^0 \rightarrow \Sigma^- \pi^+ \pi^0) & = \frac{3}{2} \sigma(pK^- \rightarrow \Sigma^0 \pi^+ \pi^-) = \frac{3}{2} \sigma(n\overline{K}^0 \rightarrow \Sigma^0 \pi^+ \pi^-) \\
= 8 \sigma(pK^- \rightarrow \Sigma^0 \pi^0 \pi^0) = 8 \sigma(n\overline{K}^0 \rightarrow \Sigma^0 \pi^0 \pi^0) & = \frac{3}{2} \sigma(pK^- \rightarrow \Sigma^- \pi^+ \pi^0) = \frac{3}{2} \sigma(n\overline{K}^0 \rightarrow \Sigma^+ \pi^0 \pi^-)
\end{align*}
\boldmath
\begin{align*}
 \sigma(pK^- \rightarrow \Sigma^- \pi^+ \pi^0) + \sigma(pK^- & \rightarrow \Sigma^+ \pi^0 \pi^-)
 + 2 \sigma(p\overline{K}^0 \rightarrow \Sigma^+ \pi^0 \pi^0)\\
 = 2 \sigma(pK^- \rightarrow \Sigma^0 \pi^+ \pi^-) & + \sigma(p\overline{K}^0 \rightarrow \Sigma^0 \pi^+ \pi^0) \\
 \\
 \sigma(p\overline{K}^0 \rightarrow \Sigma^- \pi^+ \pi^+) & + \sigma(p\overline{K}^0 \rightarrow \Sigma^+ \pi^+ \pi^-) \\
 = 2 \sigma(pK^- \rightarrow \Sigma^0 \pi^0 \pi^0) + \sigma(p\overline{K}^0 & \rightarrow \Sigma^0 \pi^+ \pi^0) + \sigma(p\overline{K}^0 \rightarrow \Sigma^+ \pi^0 \pi^0)
\end{align*}\unboldmath

\begin{center}
\fbox{Reaction type: $ NY \rightarrow N'Y'$}
\end{center}

\begin{align*}
2 \sigma(p\Lambda \rightarrow p \Sigma^0) = 2 \sigma(n\Lambda \rightarrow n \Sigma^0) = \sigma(p\Lambda \rightarrow n \Sigma^+) = \sigma(n\Lambda \rightarrow p \Sigma^-)
\end{align*}
\boldmath
\begin{equation*}
\sigma(p\Lambda \rightarrow n \Sigma^+) = 2 \sigma(p\Lambda \rightarrow p \Sigma^0)
\end{equation*}\unboldmath

\begin{center}
\noindent\rule{8cm}{0.4pt}
\end{center}
\begin{align*}
2 \sigma(p \Sigma^0 \rightarrow p \Lambda) = 2 \sigma(n \Sigma^0 \rightarrow n \Lambda) = \sigma(p \Sigma^- \rightarrow n \Lambda) = \sigma(n \Sigma^+ \rightarrow p \Lambda)
\end{align*}
\boldmath
\begin{equation*}
\sigma(p \Sigma^- \rightarrow n \Lambda) = 2 \sigma(p \Sigma^0 \rightarrow p \Lambda)
\end{equation*}\unboldmath

\begin{center}
\noindent\rule{8cm}{0.4pt}
\end{center}
\begin{align*}
\sigma(p \Sigma^- \rightarrow p \Sigma^-) = \sigma(n \Sigma^+ \rightarrow n \Sigma^+) & = \sigma(p \Sigma^+ \rightarrow p \Sigma^+) = \sigma(n \Sigma^- \rightarrow n \Sigma^-)\\
\sigma(p \Sigma^0 \rightarrow n \Sigma^+) = \sigma(n \Sigma^0 \rightarrow p \Sigma^-) & = \sigma(p \Sigma^0 \rightarrow p \Sigma^0) = \sigma(n \Sigma^0 \rightarrow n \Sigma^0)
\end{align*}

\newpage

\begin{center}
\fbox{Reaction type: $ \Delta N \rightarrow NNK\overline{K}$}
\end{center}

\begin{align*}
\sigma( \Delta^{++}  p \rightarrow p  p  K^+  \overline{K}^0 ) = \sigma( \Delta^-  n \rightarrow n  n  K^0  K^- ) & = 2 \sigma( \Delta^{++}  n \rightarrow p  p  K^+  K^- ) = 2 \sigma( \Delta^-  p \rightarrow n  n  K^0  \overline{K}^0 ) \\
= 2 \sigma( \Delta^{++}  n \rightarrow p  n  K^+  \overline{K}^0 ) = 2 \sigma( \Delta^-  p \rightarrow n  p  K^0  K^- )  & = 2 \sigma( \Delta^{++}  n \rightarrow p  p  K^0  \overline{K}^0 ) = 2 \sigma( \Delta^-  p \rightarrow n  n  K^+  K^- ) \\
= 2 \sigma( \Delta^+  p \rightarrow p  p  K^+  K^- ) = 2 \sigma( \Delta^0  n \rightarrow n  n  K^0  \overline{K}^0 )  & = 6 \sigma( \Delta^+  p \rightarrow p  p  K^0  \overline{K}^0 ) = 6 \sigma( \Delta^0  n \rightarrow n  n  K^+  K^- ) \\
= 2 \sigma( \Delta^+  p \rightarrow p  n  K^+  \overline{K}^0 ) = 2 \sigma( \Delta^0  n \rightarrow n  p  K^0  K^- )  & = 3 \sigma( \Delta^+  n \rightarrow p  p  K^0  K^- ) = 3 \sigma( \Delta^0  p \rightarrow n  n  K^+  \overline{K}^0 ) \\
= 6 \sigma( \Delta^+  n \rightarrow p  n  K^+  K^- ) = 6 \sigma( \Delta^0  p \rightarrow n  p  K^0  \overline{K}^0 )  & = 3 \sigma( \Delta^+  n \rightarrow p  n  K^0  \overline{K}^0 ) = 3 \sigma( \Delta^0  p \rightarrow n  p  K^+  K^- ) \\
= 2 \sigma( \Delta^+  n \rightarrow n  n  K^+  \overline{K}^0 ) &= 2 \sigma( \Delta^0  p \rightarrow p  p  K^0  K^- )
\end{align*}
\boldmath
\begin{align*}
3 \sigma( \Delta^+ n \rightarrow nn K^+ \overline{K}^0 ) &= 2 \sigma( \Delta^{++} n \rightarrow pn K^+ \overline{K}^0 )  \\
\\
3 \sigma( \Delta^+ n \rightarrow pn K^0 \overline{K}^0 ) &+ \sigma( \Delta^{++} n \rightarrow pn K^+ \overline{K}^0 ) \\
= 3 \sigma( \Delta^+ p \rightarrow pp K^+ K^- ) &+ \sigma( \Delta^{++} n \rightarrow pp K^0 \overline{K}^0 ) \\
\\
3 \sigma( \Delta^+ n \rightarrow pn K^+ K^- ) &+ \sigma( \Delta^{++} n \rightarrow pn K^+ \overline{K}^0 ) \\
= 3 \sigma( \Delta^+ p \rightarrow pp K^0 \overline{K}^0 ) &+ \sigma( \Delta^{++} n \rightarrow pp K^+ K^- )  \\
\end{align*}
\begin{align*}
3 \sigma( \Delta^+ n \rightarrow pp K^0 K^- ) + 3 \sigma( \Delta^+ p & \rightarrow pp K^0 \overline{K}^0 ) + 3 \sigma( \Delta^+ p \rightarrow pp K^+ K^- ) \\
= 2 \sigma( \Delta^{++} n \rightarrow pn K^+ \overline{K}^0 )  & + \sigma( \Delta^{++} n \rightarrow pp K^0 \overline{K}^0 ) \\
+ \sigma( \Delta^{++} n \rightarrow pp K^+ K^- ) & + \sigma( \Delta^{++} p \rightarrow pp K^+ \overline{K}^0 )  \\
\\
3 \sigma( \Delta^+ p \rightarrow pn K^+ \overline{K}^0 ) + 3 \sigma( \Delta^+ p & \rightarrow pp K^0 \overline{K}^0 ) + 3 \sigma( \Delta^+ p \rightarrow pp K^+ K^- ) \\
= \sigma( \Delta^{++} n \rightarrow pn K^+ \overline{K}^0 ) & + \sigma( \Delta^{++} n \rightarrow pp K^0 \overline{K}^0 )\\
+ \sigma( \Delta^{++} n \rightarrow pp K^+ K^- ) & + 2 \sigma( \Delta^{++} p \rightarrow pp K^+ \overline{K}^0 ) 
\end{align*}\unboldmath

\begin{center}
\fbox{Reaction type: $ \Delta N \rightarrow BYK$}
\end{center}

\begin{align*}
\sigma(\Delta^{++} n \rightarrow p \Lambda K^+) = \sigma(\Delta^- p \rightarrow n \Lambda K^0) & =3 \sigma(\Delta^{+} p \rightarrow p \Lambda K^+) = 3\sigma(\Delta^0 n \rightarrow n \Lambda K^0) \\
=3 \sigma(\Delta^{+} n \rightarrow p \Lambda K^0) = 3\sigma(\Delta^0 p \rightarrow n \Lambda K^+) & =3 \sigma(\Delta^{+} n \rightarrow n \Lambda K^+) = 3\sigma(\Delta^0 p \rightarrow p \Lambda K^0)
\end{align*}
\boldmath
\begin{align*}
\sigma(\Delta^{++} n \rightarrow p \Lambda K^+) =3 \sigma(\Delta^{+} p \rightarrow p \Lambda K^+)=3 \sigma(\Delta^{+} p \rightarrow p \Lambda K^0) =3 \sigma(\Delta^{+} n \rightarrow p \Lambda K^+)
\end{align*}\unboldmath

\begin{center}
\noindent\rule{8cm}{0.4pt}
\end{center}
\begin{align*}
\sigma(\Delta^{++}  p \rightarrow p \Sigma^+ K^+) = \sigma(\Delta^- n \rightarrow n \Sigma^- K^0) & = 2\sigma(\Delta^{++}  n \rightarrow p \Sigma^+ K^0) = 2 \sigma(\Delta^- p \rightarrow n \Sigma^- K^+) \\
= 2\sigma(\Delta^{++}  n \rightarrow p \Sigma^0 K^+) = 2 \sigma(\Delta^- p \rightarrow n \Sigma^0 K^0) & = 2\sigma(\Delta^{++}  n \rightarrow n \Sigma^+ K^+) = 2 \sigma(\Delta^- p \rightarrow p \Sigma^- K^0) \\
= 3\sigma(\Delta^+  p \rightarrow p \Sigma^+ K^0) = 3 \sigma(\Delta^0 n \rightarrow n \Sigma^- K^+) & = 3\sigma(\Delta^+  p \rightarrow p \Sigma^0 K^+) = 3 \sigma(\Delta^0 n \rightarrow n \Sigma^0 K^0) \\
= 2\sigma(\Delta^+  p \rightarrow n \Sigma^+ K^+) = 2 \sigma(\Delta^0 n \rightarrow p \Sigma^- K^0) & = 2\sigma(\Delta^+  n \rightarrow p \Sigma^0 K^0) = 2 \sigma(\Delta^0 p \rightarrow n \Sigma^0 K^+) \\
= 3\sigma(\Delta^+  n \rightarrow p \Sigma^- K^+) = 3 \sigma(\Delta^0 p \rightarrow n \Sigma^+ K^0) & = 3\sigma(\Delta^+  n \rightarrow n \Sigma^+ K^0) = 3 \sigma(\Delta^0 p \rightarrow p \Sigma^- K^+) \\
= 3\sigma(\Delta^+  n \rightarrow n \Sigma^0 K^+) &= 3 \sigma(\Delta^0 p \rightarrow p \Sigma^0 K^0)
\end{align*}

\boldmath
\begin{align*}
3 \sigma(\Delta^+n \rightarrow  n \Sigma^0 K^+) + \sigma(\Delta^{++}n \rightarrow  p \Sigma^0 K^+) & = 3 \sigma(\Delta^+p \rightarrow  p \Sigma^+ K^0) + \sigma(\Delta^{++}n \rightarrow  n \Sigma^+ K^+) \\
\\
3 \sigma(\Delta^+n \rightarrow  n \Sigma^+ K^0) + \sigma(\Delta^{++}p \rightarrow  p \Sigma^+ K^+) & = 3 \sigma(\Delta^+p \rightarrow  p \Sigma^0 K^+) + \sigma(\Delta^{++}n \rightarrow  p \Sigma^0 K^+)
\end{align*}
\begin{align*}
3 \sigma(\Delta^+n \rightarrow  p \Sigma^- K^+) &= 2 \sigma(\Delta^{++}n \rightarrow  p \Sigma^0 K^+) \\
\\
3 \sigma(\Delta^+n \rightarrow  p \Sigma^0 K^0) + 3 \sigma(\Delta^+p & \rightarrow  p \Sigma^0 K^+) + 3 \sigma(\Delta^+p \rightarrow  p \Sigma^+ K^0) \\
= \sigma(\Delta^{++}n \rightarrow  n \Sigma^+ K^+) + 2 \sigma(\Delta^{++}n & \rightarrow  p \Sigma^+ K^0) + 2 \sigma(\Delta^{++}p \rightarrow  p \Sigma^+ K^+) \\
\\
3 \sigma(\Delta^+p \rightarrow  n \Sigma^+ K^+) + 3 \sigma(\Delta^+p & \rightarrow  p \Sigma^0 K^+) + 3 \sigma(\Delta^+p \rightarrow  p \Sigma^+ K^0) \\
= \sigma(\Delta^{++}n \rightarrow  n \Sigma^+ K^+) & + \sigma(\Delta^{++}n \rightarrow  p \Sigma^0 K^+) \\
+ \sigma(\Delta^{++}n \rightarrow  p \Sigma^+ K^0) & + 2 \sigma(\Delta^{++}p \rightarrow p \Sigma^+ K^+)
\end{align*}\unboldmath

\begin{center}
\noindent\rule{8cm}{0.4pt}
\end{center}

\begin{align*}
3 \sigma( \Delta^{++}  p  \rightarrow \Lambda K^+  \Delta^{++} ) = 3 \sigma( \Delta^-   n \rightarrow \Lambda K^0  \Delta^- ) & = 4\sigma( \Delta^{++}   n \rightarrow \Lambda K^+  \Delta^+ ) = 4\sigma( \Delta^-  p  \rightarrow \Lambda K^0  \Delta^0 ) \\
= 3 \sigma( \Delta^{++}   n \rightarrow \Lambda K^0  \Delta^{++} ) = 3\sigma( \Delta^-  p  \rightarrow \Lambda K^+  \Delta^- ) & = 4\sigma( \Delta^+  p  \rightarrow \Lambda K^+  \Delta^+ ) = 4\sigma( \Delta^0   n \rightarrow \Lambda K^0  \Delta^0 ) \\
= 6\sigma( \Delta^+  p  \rightarrow \Lambda K^0  \Delta^{++} ) = 6\sigma( \Delta^0   n \rightarrow \Lambda K^+  \Delta^- ) & = 3 \sigma( \Delta^+   n \rightarrow \Lambda K^+  \Delta^0 ) = 3\sigma( \Delta^0  p  \rightarrow \Lambda K^0  \Delta^+ ) \\
= 6 \sigma( \Delta^+   n \rightarrow \Lambda K^0  \Delta^+ ) &= 6\sigma( \Delta^0  p  \rightarrow \Lambda K^+  \Delta^0 )
\end{align*}     
\boldmath
\begin{align*}
3 \sigma( \Delta^+ n \rightarrow \Lambda K^0  \Delta^+ ) + 2 \sigma( \Delta^{++} n \rightarrow \Lambda K^+  \Delta^+ ) & = 2 \sigma( \Delta^{++} n \rightarrow \Lambda K^0  \Delta^{++} ) +  \sigma( \Delta^{++} p \rightarrow \Lambda K^+  \Delta^{++} ) \\
\\
3 \sigma( \Delta^+ n \rightarrow \Lambda K^+  \Delta^0 ) &= 4 \sigma( \Delta^{++} n \rightarrow \Lambda K^+  \Delta^+ )\\
\\
3 \sigma( \Delta^+ p \rightarrow \Lambda K^+  \Delta^+ ) + 2 \sigma( \Delta^{++} n \rightarrow \Lambda K^+  \Delta^+ ) & =  \sigma( \Delta^{++} n \rightarrow \Lambda K^0  \Delta^{++} ) + 2 \sigma( \Delta^{++} p \rightarrow \Lambda K^+  \Delta^{++} )      
\end{align*}\unboldmath

\begin{center}
\noindent\rule{8cm}{0.4pt}
\end{center}

\begin{align*}
6 \sigma( \Delta^{++}  p  \rightarrow \Sigma^0  K^+  \Delta^{++} ) = 6 \sigma( \Delta^-   n \rightarrow \Sigma^0  K^0  \Delta^- ) &= 6 \sigma( \Delta^{++}   n \rightarrow \Sigma^0  K^+  \Delta^+ ) = 6 \sigma( \Delta^-  p  \rightarrow \Sigma^0  K^0  \Delta^0 ) \\
= 12 \sigma( \Delta^{++}  p  \rightarrow \Sigma^+  K^0  \Delta^{++} ) = 12 \sigma( \Delta^-   n \rightarrow \Sigma^-  K^+  \Delta^- ) &= 12 \sigma( \Delta^{++}   n \rightarrow \Sigma^0  K^0  \Delta^{++} ) = 12 \sigma( \Delta^-  p  \rightarrow \Sigma^0  K^+  \Delta^- ) \\
= 2 \sigma( \Delta^{++}   n \rightarrow \Sigma^+  K^+  \Delta^0 ) = 2 \sigma( \Delta^-  p  \rightarrow \Sigma^-  K^0  \Delta^+ ) &= 2 \sigma( \Delta^{++}   n \rightarrow \Sigma^+  K^0  \Delta^+ ) = 2 \sigma( \Delta^-  p  \rightarrow \Sigma^-  K^+  \Delta^0 ) \\
= 3 \sigma( \Delta^{++}   n \rightarrow \Sigma^-  K^+  \Delta^{++} ) = 3 \sigma( \Delta^-  p  \rightarrow \Sigma^+  K^0  \Delta^- ) &= 3 \sigma( \Delta^+  p  \rightarrow \Sigma^0  K^0  \Delta^{++} ) = 3 \sigma( \Delta^0   n \rightarrow \Sigma^0  K^+  \Delta^- ) \\
= 6 \sigma( \Delta^+  p  \rightarrow \Sigma^+  K^+  \Delta^0 ) = 6 \sigma( \Delta^0   n \rightarrow \Sigma^-  K^0  \Delta^+ ) &= 6 \sigma( \Delta^+  p  \rightarrow \Sigma^-  K^+  \Delta^{++} ) = 6 \sigma( \Delta^0   n \rightarrow \Sigma^+  K^0  \Delta^- ) \\
= 12 \sigma( \Delta^+  p  \rightarrow \Sigma^0  K^+  \Delta^+ ) = 12 \sigma( \Delta^0   n \rightarrow \Sigma^0  K^0  \Delta^0 ) &= 12 \sigma( \Delta^+   n \rightarrow \Sigma^0  K^0  \Delta^+ ) = 12 \sigma( \Delta^0  p  \rightarrow \Sigma^0  K^+  \Delta^0 ) \\
= 6 \sigma( \Delta^+  p  \rightarrow \Sigma^+  K^0  \Delta^+ ) = 6 \sigma( \Delta^0   n \rightarrow \Sigma^-  K^+  \Delta^0 ) &= 6 \sigma( \Delta^+   n \rightarrow \Sigma^+  K^+  \Delta^- ) = 6 \sigma( \Delta^0  p  \rightarrow \Sigma^-  K^0  \Delta^{++} ) \\
= 3 \sigma( \Delta^+   n \rightarrow \Sigma^0  K^+  \Delta^0 ) = 3 \sigma( \Delta^0  p  \rightarrow \Sigma^0  K^0  \Delta^+ ) &= 6 \sigma( \Delta^+   n \rightarrow \Sigma^-  K^+  \Delta^+ ) = 6 \sigma( \Delta^0  p  \rightarrow \Sigma^+  K^0  \Delta^0 ) \\
= 6 \sigma( \Delta^+   n \rightarrow \Sigma^+  K^0  \Delta^0 ) = 6 \sigma( \Delta^0  p  \rightarrow \Sigma^-  K^+  \Delta^+ ) &= 6 \sigma( \Delta^+   n \rightarrow \Sigma^-  K^0  \Delta^{++} ) = 6 \sigma( \Delta^0  p  \rightarrow \Sigma^+  K^+  \Delta^- )
\end{align*}
\boldmath
\begin{align*}
2 \sigma(\Delta^+ n \rightarrow \Sigma^- K^0 \Delta^{++} ) & + 2 \sigma(\Delta^{++}p \rightarrow \Sigma^+ K^+ \Delta^+ ) \\
= 3 \sigma(\Delta^+ p \rightarrow \Sigma^+ K^+ \Delta^0) & +  \sigma(\Delta^{++}n \rightarrow \Sigma^+ K^+ \Delta^0)
\end{align*}
\begin{align*}
12 \sigma(\Delta^+ n \rightarrow \Sigma^0 K^0 \Delta^+ ) + 15 \sigma(\Delta^+ p & \rightarrow \Sigma^+ K^+ \Delta^0) + 2 \sigma(\Delta^{++}n \rightarrow \Sigma^+ K^0 \Delta^+ ) \\
+ 2 \sigma(\Delta^{++}n \rightarrow \Sigma^- K^+ \Delta^{++} ) + 9 \sigma(\Delta^{++}n & \rightarrow \Sigma^+ K^+ \Delta^0) + 2 \sigma(\Delta^{++}p \rightarrow \Sigma^+ K^0 \Delta^{++} ) \\
+ 4 \sigma(\Delta^{++}p \rightarrow \Sigma^0 K^+ \Delta^{++} ) & = 18 \sigma(\Delta^+ p \rightarrow \Sigma^0 K^+ \Delta^+ ) \\
+ 6 \sigma(\Delta^{++}n \rightarrow \Sigma^0 K^0 \Delta^{++} ) + 8 \sigma(\Delta^{++}n & \rightarrow \Sigma^0 K^+ \Delta^+ ) + 18 \sigma(\Delta^{++}p \rightarrow \Sigma^+ K^+ \Delta^+ )
\end{align*}
\begin{align*}
6 \sigma(\Delta^+ n \rightarrow \Sigma^+ K^0 \Delta^0 ) + 9 \sigma(\Delta^+ p & \rightarrow \Sigma^+ K^+ \Delta^0 ) + 2 \sigma(\Delta^{++}n \rightarrow \Sigma^- K^+ \Delta^{++} ) \\
+ 9 \sigma(\Delta^{++}n \rightarrow \Sigma^+ K^+ \Delta^0 ) & + 2 \sigma(\Delta^{++}p \rightarrow \Sigma^+ K^0 \Delta^{++} ) \\
= 6 \sigma(\Delta^+ p \rightarrow \Sigma^0 K^+ \Delta^+ ) + 2 \sigma(\Delta^{++}n & \rightarrow \Sigma^0 K^0 \Delta^{++} )  + 6 \sigma(\Delta^{++}n \rightarrow \Sigma^+ K^0 \Delta^+ ) \\
+ 8 \sigma(\Delta^{++}n \rightarrow \Sigma^0 K^+ \Delta^+ ) & + 10 \sigma(\Delta^{++}p \rightarrow \Sigma^+ K^+ \Delta^+ )
\end{align*}
\begin{align*}
12 \sigma(\Delta^+ n \rightarrow \Sigma^- K^+ \Delta^+ ) + 18 \sigma(\Delta^+ p & \rightarrow \Sigma^0 K^+ \Delta^+ ) + 6 \sigma(\Delta^{++}n \rightarrow \Sigma^+ K^0 \Delta^+ ) \\
+ 16 \sigma(\Delta^{++}n \rightarrow \Sigma^0 K^+ \Delta^+ ) &+ 18 \sigma(\Delta^{++}p \rightarrow \Sigma^+ K^+ \Delta^+ )\\
= 9 \sigma(\Delta^+ p \rightarrow \Sigma^+ K^+ \Delta^0 ) + 2 \sigma(\Delta^{++}n & \rightarrow \Sigma^0 K^0 \Delta^{++} ) + 10 \sigma(\Delta^{++}n \rightarrow \Sigma^- K^+ \Delta^{++} ) \\
+ 15 \sigma(\Delta^{++}n \rightarrow \Sigma^+ K^+ \Delta^0 ) + 6 \sigma(\Delta^{++}p & \rightarrow \Sigma^+ K^0 \Delta^{++} ) + 8 \sigma(\Delta^{++}p \rightarrow \Sigma^0 K^+ \Delta^{++} )
\end{align*}
\begin{align*}
6 \sigma(\Delta^+ n \rightarrow \Sigma^0 K^+ \Delta^0 ) + 6 \sigma(\Delta^+ p & \rightarrow \Sigma^0 K^+ \Delta^+ ) + 2 \sigma(\Delta^{++}n \rightarrow \Sigma^0 K^0 \Delta^{++} ) \\
+ 2 \sigma(\Delta^{++}p \rightarrow \Sigma^+ K^+ \Delta^+ ) = 3 \sigma(\Delta^+ p & \rightarrow \Sigma^+ K^+ \Delta^0 ) + 2 \sigma(\Delta^{++}n \rightarrow \Sigma^+ K^0 \Delta^+ ) \\
+ 2 \sigma(\Delta^{++}n  \rightarrow \Sigma^- K^+ \Delta^{++} ) + \sigma(\Delta^{++}n & \rightarrow \Sigma^+ K^+ \Delta^0 ) + 2 \sigma(\Delta^{++}p \rightarrow \Sigma^+ K^0 \Delta^{++} )
\end{align*}
\begin{align*}
4 \sigma(\Delta^+ p \rightarrow \Sigma^0 K^0 \Delta^{++} ) + 6 \sigma(\Delta^+ p & \rightarrow \Sigma^0 K^+ \Delta^+ ) + 2 \sigma(\Delta^{++}n \rightarrow \Sigma^0 K^0 \Delta^{++} ) \\
+ 4 \sigma(\Delta^{++}n \rightarrow \Sigma^0 K^+ \Delta^+ ) & + 2 \sigma(\Delta^{++}p \rightarrow \Sigma^+ K^+ \Delta^+ ) \\
= 3 \sigma(\Delta^+ p \rightarrow \Sigma^+ K^+ \Delta^0) + 2 \sigma(\Delta^{++}n & \rightarrow \Sigma^+ K^0 \Delta^+ ) + 2 \sigma(\Delta^{++}n \rightarrow \Sigma^- K^+ \Delta^{++} ) \\
+ 5 \sigma(\Delta^{++}n \rightarrow \Sigma^+ K^+ \Delta^0 ) & + 2 \sigma(\Delta^{++}p \rightarrow \Sigma^+ K^0 \Delta^{++} ) 
\end{align*}
\begin{align*}
6 \sigma(\Delta^+ p \rightarrow \Sigma^+ K^0 \Delta^+ ) + 6 \sigma(\Delta^+ p & \rightarrow \Sigma^0 K^+ \Delta^+ ) + 4 \sigma(\Delta^{++}n \rightarrow \Sigma^+ K^0 \Delta^+ ) \\
+ 4 \sigma(\Delta^{++}n \rightarrow \Sigma^0 K^+ \Delta^+ ) & + 8 \sigma(\Delta^{++}p \rightarrow \Sigma^+ K^+ \Delta^+ ) \\
= 3 \sigma(\Delta^+ p \rightarrow \Sigma^+ K^+ \Delta^0) + 2 \sigma(\Delta^{++}n & \rightarrow \Sigma^0 K^0 \Delta^{++} ) + 2 \sigma(\Delta^{++}n \rightarrow \Sigma^- K^+ \Delta^{++} ) \\
+ 5 \sigma(\Delta^{++}n \rightarrow \Sigma^+ K^+ \Delta^0 ) + 4 \sigma(\Delta^{++}p & \rightarrow \Sigma^+ K^0 \Delta^{++} ) + 4 \sigma(\Delta^{++}p \rightarrow \Sigma^0 K^+ \Delta^{++} )
\end{align*}
\begin{align*}
4 \sigma(\Delta^+ p \rightarrow \Sigma^- K^+ \Delta^{++} ) + 9 \sigma(\Delta^+ p & \rightarrow \Sigma^+ K^+ \Delta^0) + 2 \sigma(\Delta^{++}n \rightarrow \Sigma^- K^+ \Delta^{++} ) \\
+ 7 \sigma(\Delta^{++}n \rightarrow \Sigma^+ K^+ \Delta^0) & + 2 \sigma(\Delta^{++}p \rightarrow \Sigma^+ K^0 \Delta^{++} ) \\
= 6 \sigma(\Delta^+ p \rightarrow \Sigma^0 K^+ \Delta^+ ) + 2 \sigma(\Delta^{++}n & \rightarrow \Sigma^0 K^0 \Delta^{++} ) + 2 \sigma(\Delta^{++}n \rightarrow \Sigma^+ K^0 \Delta^+ ) \\
+ 8 \sigma(\Delta^{++}n \rightarrow \Sigma^0 K^+ \Delta^+ ) & + 10 \sigma(\Delta^{++}p \rightarrow \Sigma^+ K^+ \Delta^+ )     
\end{align*}\unboldmath

\begin{center}
\fbox{Reaction type: $ \pi N \rightarrow YK$}
\end{center}

\begin{align*}
 2 \sigma(\pi^0 p \rightarrow \Lambda K^+) = 2 \sigma(\pi^0 n \rightarrow \Lambda K^0) = \sigma(\pi^- p \rightarrow \Lambda K^0) = \sigma(\pi^+ n \rightarrow \Lambda K^+)
\end{align*}
\boldmath
\begin{equation*}
 \sigma(\pi^- p \rightarrow \Lambda K^0) = 2 \sigma(\pi^0 p \rightarrow \Lambda K^+)
\end{equation*}\unboldmath

The case of the reaction $\pi N \rightarrow \Sigma K$ is detailed in \autoref{HEM}.

\newpage
\begin{center}
\fbox{Reaction type: $ \pi N \rightarrow Y K\pi$}
\end{center}

\begin{align*}
 \sigma(\pi^+ p \rightarrow \Lambda K^+ \pi^+) = \sigma(\pi^- n \rightarrow \Lambda K^0 \pi^-) & = \sigma(\pi^0 p \rightarrow \Lambda K^0 \pi^+) = \sigma(\pi^- p \rightarrow \Lambda K^0 \pi^0) \\
= \sigma(\pi^+ n \rightarrow \Lambda K^+ \pi^0) = \sigma(\pi^0 n \rightarrow \Lambda K^+ \pi^-) & = 2\sigma(\pi^0 p \rightarrow \Lambda K^+ \pi^0) = 2 \sigma(\pi^0 n \rightarrow \Lambda K^0 \pi^0) \\
= \sigma(\pi^- p \rightarrow \Lambda K^+ \pi^-) & = \sigma(\pi^+ n \rightarrow \Lambda K^0 \pi^+)
\end{align*}
\boldmath
\begin{align*}
 \sigma(\pi^0 p \rightarrow \Lambda K^0 \pi^+) & = \sigma(\pi^- p \rightarrow \Lambda K^0 \pi^0)\\
 \\
 \sigma(\pi^- p \rightarrow \Lambda \pi^- K^+) + \sigma(\pi^+ p \rightarrow \Lambda \pi^+ K^+) & = 2 \sigma(\pi^0 p \rightarrow \Lambda \pi^0 K^+) + \sigma(\pi^0 p \rightarrow \Lambda \pi^+ K^0)
\end{align*}\unboldmath

\begin{center}
\noindent\rule{8cm}{0.4pt}
\end{center}

\begin{align*}
\frac{4}{5} \sigma(\pi^+ p \rightarrow \Sigma^+ K^0 \pi^+) = \frac{4}{5} \sigma(\pi^- n \rightarrow \Sigma^- K^+ \pi^-) & = \frac{4}{3} \sigma(\pi^+ p \rightarrow \Sigma^+ K^+ \pi^0) = \frac{4}{3} \sigma(\pi^- n \rightarrow \Sigma^- K^0 \pi^0) \\
= 4 \sigma(\pi^+ p \rightarrow \Sigma^0 K^+ \pi^+) = 4 \sigma(\pi^- n \rightarrow \Sigma^0 K^0 \pi^-) & = 2 \sigma(\pi^0 p \rightarrow \Sigma^+ K^0 \pi^0) = 2 \sigma(\pi^0 n \rightarrow \Sigma^- K^+ \pi^0) \\
= 2 \sigma(\pi^0 p \rightarrow \Sigma^+ K^+ \pi^-) = 2 \sigma(\pi^0 n \rightarrow \Sigma^- K^0 \pi^+) & = \frac{4}{3} \sigma(\pi^0 p \rightarrow \Sigma^0 K^0 \pi^+) = \frac{4}{3} \sigma(\pi^0 n \rightarrow \Sigma^0 K^+ \pi^-) \\
= \frac{8}{3} \sigma(\pi^0 p \rightarrow \Sigma^0 K^+ \pi^0) = \frac{8}{3} \sigma(\pi^0 n \rightarrow \Sigma^0 K^0 \pi^0) & = 2 \sigma(\pi^0 p \rightarrow \Sigma^- K^+ \pi^+) = 2 \sigma(\pi^0 n \rightarrow \Sigma^+ K^0 \pi^-) \\
= \frac{8}{3} \sigma(\pi^- p \rightarrow \Sigma^+ K^0 \pi^-) = \frac{8}{3} \sigma(\pi^+ n \rightarrow \Sigma^- K^+ \pi^+) & = \frac{8}{5} \sigma(\pi^- p \rightarrow \Sigma^0 K^0 \pi^0) = \frac{8}{5} \sigma(\pi^+ n \rightarrow \Sigma^0 K^+ \pi^0) \\
= \frac{8}{5} \sigma(\pi^- p \rightarrow \Sigma^0 K^+ \pi^-) = \frac{8}{5} \sigma(\pi^+ n \rightarrow \Sigma^0 K^0 \pi^+) & = \sigma(\pi^- p \rightarrow \Sigma^- K^0 \pi^+) = \sigma(\pi^+ n \rightarrow \Sigma^+ K^+ \pi^-) \\
= \frac{8}{3} \sigma(\pi^- p \rightarrow \Sigma^- K^+ \pi^0) & = \frac{8}{3} \sigma(\pi^+ n \rightarrow \Sigma^+ K^0 \pi^0)
\end{align*}

\boldmath
\begin{align*}
\sigma(\pi^- p \rightarrow \Sigma^- \pi^0 K^+) + \sigma(\pi^- p & \rightarrow \Sigma^0 \pi^0 K^0) 
+ \sigma(\pi^+ p \rightarrow \Sigma^+ \pi^0 K^+) \\
= \sigma(\pi^0 p \rightarrow \Sigma^- \pi^+ K^+) + \sigma(\pi^0 p & \rightarrow \Sigma^0 \pi^+ K^0) + \sigma(\pi^0 p \rightarrow \Sigma^+ \pi^- K^+)
\end{align*}

\begin{align*}
\sigma(\pi^- p \rightarrow \Sigma^- \pi^+ K^0) + \sigma(\pi^- p & \rightarrow \Sigma^+ \pi^- K^0) + \sigma(\pi^+ p \rightarrow \Sigma^+ \pi^+ K^0) \\
= \sigma(\pi^- p \rightarrow \Sigma^0 \pi^0 K^0) + 2 \sigma(\pi^0 p & \rightarrow \Sigma^0 \pi^0 K^+) + \sigma(\pi^0 p \rightarrow \Sigma^0 \pi^+ K^0) + \sigma(\pi^0 p \rightarrow \Sigma^+ \pi^0 K^0)
\end{align*}

\begin{align*}
\sigma(\pi^- p \rightarrow \Sigma^0 \pi^- K^+) + \sigma(\pi^- p & \rightarrow \Sigma^0 \pi^0 K^0) + \sigma(\pi^+ p \rightarrow \Sigma^0 \pi^+ K^+) \\
= \sigma(\pi^0 p \rightarrow \Sigma^- \pi^+ K^+) + \sigma(\pi^0 p & \rightarrow \Sigma^+ \pi^- K^+) + \sigma(\pi^0 p \rightarrow \Sigma^+ \pi^0 K^0)
\end{align*}\unboldmath

\begin{center}
\fbox{Reaction type: $ \pi N \rightarrow N K\overline{K}$}
\end{center}

\begin{align*}
2 \sigma(\pi^+ p \rightarrow p K^+ \overline{K}^0) = 2 \sigma(\pi^- n \rightarrow n K^0 K^-) & = 4 \sigma(\pi^0 p \rightarrow p K^+ K^-) = 4 \sigma(\pi^0 n \rightarrow n K^0 \overline{K}^0) \\
= 4 \sigma(\pi^0 p \rightarrow p K^0 \overline{K}^0) = 4 \sigma(\pi^0 n \rightarrow n K^+ K^-) & = \sigma(\pi^0 p \rightarrow n K^+ \overline{K}^0) = \sigma(\pi^0 n \rightarrow p K^0 K^-) \\
= 2 \sigma(\pi^- p \rightarrow p K^0 K^-) = 2 \sigma(\pi^+ n \rightarrow n K^+ \overline{K}^0) & = \sigma(\pi^- p \rightarrow n K^+ K^-) = \sigma(\pi^+ n \rightarrow p K^0 \overline{K}^0) \\
= \sigma(\pi^- p \rightarrow n K^0 \overline{K}^0) & = \sigma(\pi^+ n \rightarrow p K^+ K^-)
\end{align*}
\boldmath
\begin{align*}
 \sigma(\pi^- p \rightarrow n K^0 \overline{K}^0) + \sigma(\pi^- p \rightarrow n K^+ K^-) & + \sigma(\pi^- p \rightarrow p K^0 K^-) + \sigma(\pi^+ p \rightarrow p K^+ \overline{K}^0) \\
 = 2 \sigma(\pi^0 p \rightarrow n K^+ \overline{K}^0) + 2 \sigma(\pi^0 p & \rightarrow p K^0 \overline{K}^0) + 2 \sigma(\pi^0 p \rightarrow p K^+ K^-)
\end{align*}\unboldmath

\onecolumn
\begin{center}
\fbox{Reaction type: $ \pi N \rightarrow Y K \pi \pi$}
\end{center}

\begin{align*}
 \sigma(\pi^+ p \rightarrow \Lambda K^0 \pi^+ \pi^+) & = \sigma(\pi^- n \rightarrow \Lambda K^+ \pi^- \pi^-) &
= \sigma(\pi^+ p \rightarrow \Lambda K^+ \pi^+ \pi^0) & = \sigma(\pi^- n \rightarrow \Lambda K^0 \pi^0 \pi^-) \\
= 2 \sigma(\pi^0 p \rightarrow \Lambda K^0 \pi^+ \pi^0) & = 2 \sigma(\pi^0 n \rightarrow \Lambda K^+ \pi^0 \pi^-) &
= \sigma(\pi^0 p \rightarrow \Lambda K^+ \pi^+ \pi^-) & = \sigma(\pi^0 n \rightarrow \Lambda K^0 \pi^+ \pi^-) \\
= 4 \sigma(\pi^0 p \rightarrow \Lambda K^+ \pi^0 \pi^0) & = 4 \sigma(\pi^0 n \rightarrow \Lambda K^0 \pi^0 \pi^0) &
= \sigma(\pi^- p \rightarrow \Lambda K^0 \pi^+ \pi^-) & = \sigma(\pi^+ n \rightarrow \Lambda K^+ \pi^+ \pi^-) \\
= 2 \sigma(\pi^- p \rightarrow \Lambda K^0 \pi^0 \pi^0) & = 2 \sigma(\pi^+ n \rightarrow \Lambda K^+ \pi^0 \pi^0) &
= \sigma(\pi^- p \rightarrow \Lambda K^+ \pi^0 \pi^-) & = \sigma(\pi^+ n \rightarrow \Lambda K^0 \pi^+ \pi^0)
\end{align*}
\boldmath
\begin{align*}
 \sigma(\pi^- p \rightarrow \Lambda K^+ \pi^0 \pi^-) & + 2 \sigma(\pi^- p \rightarrow \Lambda K^0 \pi^+ \pi^-) \\
 + \sigma(\pi^+ p \rightarrow \Lambda K^+ \pi^+ \pi^0) & + 2 \sigma(\pi^+ p \rightarrow \Lambda K^0 \pi^+ \pi^+) \\
 = 4 \sigma(\pi^0 p \rightarrow \Lambda K^+ \pi^0 \pi^0) + 2 \sigma(\pi^0 p & \rightarrow \Lambda K^+ \pi^+ \pi^-) + 3 \sigma(\pi^0 p \rightarrow \Lambda K^0 \pi^+ \pi^0)
\end{align*}
\begin{align*}
\sigma(\pi^- p \rightarrow & \Lambda K^0 \pi^0 \pi^0) + 2 \sigma(\pi^0 p \rightarrow \Lambda K^+ \pi^0 \pi^0) + \sigma(\pi^0 p \rightarrow \Lambda K^0 \pi^+ \pi^0) \\
& = \sigma(\pi^- p \rightarrow \Lambda K^0 \pi^+ \pi^-) + \sigma(\pi^+ p \rightarrow \Lambda K^0 \pi^+ \pi^+)
\end{align*}\unboldmath

\begin{align*}
\sigma(\pi^+ p \rightarrow \Sigma^+ K^+ \pi^+ \pi^-) & = \sigma(\pi^- n \rightarrow \Sigma^- K^0 \pi^+ \pi^-) = 4 \sigma(\pi^+ p \rightarrow \Sigma^+ K^+ \pi^0 \pi^0) \\
= 4 \sigma(\pi^- n \rightarrow \Sigma^- K^0 \pi^0 \pi^0) & = 2 \sigma(\pi^+ p \rightarrow \Sigma^0 K^+ \pi^+ \pi^0) = 2 \sigma(\pi^- n \rightarrow \Sigma^0 K^0 \pi^0 \pi^-) \\
= 4 \sigma(\pi^+ p \rightarrow \Sigma^- K^+ \pi^+ \pi^+) & = 4 \sigma(\pi^- n \rightarrow \Sigma^+ K^0 \pi^- \pi^-) = \sigma(\pi^+ p \rightarrow \Sigma^+ K^0 \pi^+ \pi^0) \\
= \sigma(\pi^- n \rightarrow \Sigma^- K^+ \pi^0 \pi^-) & = 4 \sigma(\pi^+ p \rightarrow \Sigma^0 K^0 \pi^+ \pi^+) = 4 \sigma(\pi^- n \rightarrow \Sigma^0 K^+ \pi^- \pi^-) \\
= 2 \sigma(\pi^0 p \rightarrow \Sigma^+ K^+ \pi^0 \pi^-) & = 2 \sigma(\pi^0 n \rightarrow \Sigma^- K^0 \pi^+ \pi^0) = 2 \sigma(\pi^0 p \rightarrow \Sigma^0 K^+ \pi^+ \pi^-) \\
= 2 \sigma(\pi^0 n \rightarrow \Sigma^0 K^0 \pi^+ \pi^-) & = 4 \sigma(\pi^0 p \rightarrow \Sigma^0 K^+ \pi^0 \pi^0) = 4 \sigma(\pi^0 n \rightarrow \Sigma^0 K^0 \pi^0 \pi^0) \\
= 4 \sigma(\pi^0 p \rightarrow \Sigma^- K^+ \pi^+ \pi^0) & = 4 \sigma(\pi^0 n \rightarrow \Sigma^+ K^0 \pi^0 \pi^-) = \sigma(\pi^0 p \rightarrow \Sigma^+ K^0 \pi^+ \pi^-) \\
= \sigma(\pi^0 n \rightarrow \Sigma^- K^+ \pi^+ \pi^-) & = 4 \sigma(\pi^0 p \rightarrow \Sigma^+ K^0 \pi^0 \pi^0) = 4 \sigma(\pi^0 n \rightarrow \Sigma^- K^+ \pi^0 \pi^0) \\
= 4 \sigma(\pi^0 p \rightarrow \Sigma^0 K^0 \pi^+ \pi^0) & = 4 \sigma(\pi^0 n \rightarrow \Sigma^0 K^+ \pi^0 \pi^-) = 2 \sigma(\pi^0 p \rightarrow \Sigma^- K^0 \pi^+ \pi^+) \\
= 2 \sigma(\pi^0 n \rightarrow \Sigma^+ K^+ \pi^- \pi^-) & = 4 \sigma(\pi^- p \rightarrow \Sigma^+ K^+ \pi^- \pi^-) = 4 \sigma(\pi^+ n \rightarrow \Sigma^- K^0 \pi^+ \pi^+) \\
= 2 \sigma(\pi^- p \rightarrow \Sigma^0 K^+ \pi^0 \pi^-) & = 2 \sigma(\pi^+ n \rightarrow \Sigma^0 K^0 \pi^+ \pi^0) = 4 \sigma(\pi^- p \rightarrow \Sigma^- K^+ \pi^+ \pi^-) \\
= 4 \sigma(\pi^+ n \rightarrow \Sigma^+ K^0 \pi^+ \pi^-) & = 4 \sigma(\pi^- p \rightarrow \Sigma^- K^+ \pi^0 \pi^0) = 4 \sigma(\pi^+ n \rightarrow \Sigma^+ K^0 \pi^0 \pi^0) \\
= 2 \sigma(\pi^- p \rightarrow \Sigma^+ K^0 \pi^0 \pi^-) & = 2 \sigma(\pi^+ n \rightarrow \Sigma^- K^+ \pi^+ \pi^0) = \sigma(\pi^- p \rightarrow \Sigma^0 K^0 \pi^+ \pi^-) \\
= \sigma(\pi^+ n \rightarrow \Sigma^0 K^+ \pi^+ \pi^-) & = 2 \sigma(\pi^- p \rightarrow \Sigma^0 K^0 \pi^0 \pi^0) = 2 \sigma(\pi^+ n \rightarrow \Sigma^0 K^+ \pi^0 \pi^0) \\
= 2 \sigma(\pi^- p \rightarrow \Sigma^- K^0 \pi^+ \pi^0) & = 2 \sigma(\pi^+ n \rightarrow \Sigma^+ K^+ \pi^0 \pi^-)
\end{align*}
\boldmath
\begin{align*}
\sigma(\pi^- p \rightarrow \Sigma^- K^0 \pi^+ \pi^0) &+ \sigma(\pi^- p \rightarrow \Sigma^+ K^0 \pi^0 \pi^-) + \sigma(\pi^- p \rightarrow \Sigma^- K^+ \pi^0 \pi^0) \\
+ \sigma(\pi^- p \rightarrow \Sigma^- K^+ \pi^+ \pi^-) &+ \sigma(\pi^- p \rightarrow \Sigma^+ K^+ \pi^- \pi^-)  + \sigma(\pi^+ p \rightarrow \Sigma^+ K^0 \pi^+ \pi^0) \\
+ \sigma(\pi^+ p \rightarrow \Sigma^- K^+ \pi^+ \pi^+) &+ \sigma(\pi^+ p \rightarrow \Sigma^+ K^+ \pi^0 \pi^0) + \sigma(\pi^+ p \rightarrow \Sigma^+ K^+ \pi^+ \pi^-) \\
= \sigma(\pi^0 p \rightarrow \Sigma^- K^0 \pi^+ \pi^+) &+ 2 \sigma(\pi^0 p \rightarrow \Sigma^0 K^0 \pi^+ \pi^0)  + \sigma(\pi^0 p \rightarrow \Sigma^+ K^0 \pi^0 \pi^0) \\
+ \sigma(\pi^0 p \rightarrow \Sigma^+ K^0 \pi^+ \pi^-) &+ \sigma(\pi^0 p \rightarrow \Sigma^- K^+ \pi^+ \pi^0) + 2 \sigma(\pi^0 p \rightarrow \Sigma^0 K^+ \pi^0 \pi^0) \\
+ 2 \sigma(\pi^0 p \rightarrow & \Sigma^0 K^+ \pi^+ \pi^-) + \sigma(\pi^0 p \rightarrow \Sigma^+ K^+ \pi^0 \pi^-)
\end{align*}
\begin{align*}
\sigma(\pi^- p \rightarrow \Sigma^- K^+ \pi^+ \pi^-) & + \sigma(\pi^- p \rightarrow \Sigma^+ K^+ \pi^- \pi^-) + \sigma(\pi^0 p \rightarrow \Sigma^- K^0 \pi^+ \pi^+) \\
+ \sigma(\pi^0 p \rightarrow \Sigma^+ K^0 \pi^+ \pi^-) & + \sigma(\pi^+ p \rightarrow \Sigma^- K^+ \pi^+ \pi^+)  + \sigma(\pi^+ p \rightarrow \Sigma^+ K^+ \pi^+ \pi^-) \\
= 2 \sigma(\pi^- p \rightarrow \Sigma^0 K^0 \pi^0 \pi^0) & + \sigma(\pi^- p \rightarrow \Sigma^- K^+ \pi^0 \pi^0) + \sigma(\pi^- p \rightarrow \Sigma^0 K^+ \pi^0 \pi^-) \\
+ \sigma(\pi^0 p \rightarrow \Sigma^0 K^0 \pi^+ \pi^0) & + \sigma(\pi^0 p \rightarrow \Sigma^+ K^0 \pi^0 \pi^0)  + 2 \sigma(\pi^0 p \rightarrow \Sigma^0 K^+ \pi^0 \pi^0) \\
+ \sigma(\pi^+ p \rightarrow  & \Sigma^0 K^+ \pi^+ \pi^0) + \sigma(\pi^+ p \rightarrow \Sigma^+ K^+ \pi^0 \pi^0)
\end{align*}
\begin{align*}
\sigma(\pi^- p \rightarrow \Sigma^- K^+ \pi^0 \pi^0) & + \sigma(\pi^0 p \rightarrow \Sigma^- K^0 \pi^+ \pi^+) + \sigma(\pi^0 p \rightarrow \Sigma^0 K^0 \pi^+ \pi^0) \\
+ 3 \sigma(\pi^0 p \rightarrow \Sigma^+ K^0 \pi^0 \pi^0) & +\sigma(\pi^0 p \rightarrow \Sigma^+ K^0 \pi^+ \pi^-)  + 2 \sigma(\pi^0 p \rightarrow \Sigma^- K^+ \pi^+ \pi^0) \\
+ 2 \sigma(\pi^0 p \rightarrow \Sigma^0 K^+ \pi^0 \pi^0) & + 2 \sigma(\pi^0 p \rightarrow \Sigma^+ K^+ \pi^0 \pi^-)  + \sigma(\pi^+ p \rightarrow \Sigma^+ K^+ \pi^0 \pi^0) \\
= 2 \sigma(\pi^- p \rightarrow \Sigma^0 K^0 \pi^+ \pi^-) & + \sigma(\pi^- p \rightarrow \Sigma^- K^+ \pi^+ \pi^-)  + \sigma(\pi^- p \rightarrow \Sigma^0 K^+ \pi^0 \pi^-) \\
+ \sigma(\pi^- p \rightarrow \Sigma^+ K^+ \pi^- \pi^-) & + 2 \sigma(\pi^+ p \rightarrow \Sigma^0 K^0 \pi^+ \pi^+) + \sigma(\pi^+ p \rightarrow \Sigma^- K^+ \pi^+ \pi^+) \\
+ \sigma(\pi^+ p \rightarrow & \Sigma^0 K^+ \pi^+ \pi^0) + \sigma(\pi^+ p \rightarrow \Sigma^+ K^+ \pi^+ \pi^-)
\end{align*}\unboldmath

\onecolumn
\section{Parametrizations of elementary cross sections involving strange particles ($\mathbf{K}$ $\mathbf{\overline{K}}$ $\mathbf{\Lambda}$ $\mathbf{\Sigma}$) for incident energies from threshold up to 15 GeV.}
\label{param}

Here we give for each considered new channel the full parametrization. If several cross sections are linked by a symmetry only one of the cross section parametrization is given. See \autoref{channel} for a complete list of symmetries between the cross sections.

In the following $P_{lab}$ is the momentum in the target nucleon frame of reference. Note that in INCL protons and neutrons are considered to have the same mass inside the nucleus then formulae given below are valid for proton and neutron targets.

Pions also have been given the same mass. Lambdas (anti)Kaons and Sigmas are considered with their real masses. The threshold for every channels of the same reaction is the same (the highest one calculated with the INCL masses) in order to remain consistent with the isospin invariance hypothesis.

Cross sections are always given in mb.

\subsection{Elastic}

Considering that data available for the elastic and quasi-elastic reactions $\Sigma N \rightarrow \Sigma N$ $\Sigma N \rightarrow \Lambda N$ and $\Lambda N \rightarrow \Sigma N$ are very scarce and with big uncertainties the choice to consider them as equivalent to the $\Lambda N \rightarrow \Lambda N$ was made.

\begin{center}
\fbox{\texorpdfstring{$Y N \rightarrow Y N$}{Lg}}
\end{center}
\begin{equation}
\text{\hspace*{-2cm}}\sigma = \left\{
      \begin{aligned}
         &200   & \ P_{lab} < 145 \ MeV/c \\
         &869 \exp(-P_{lab}[MeV/c]/100)  &  \ 145 \ MeV/c  \leq  \ P_{lab} < 425 \ MeV/c \\
         &12.8 \exp(-6.2 \ 10^{-5} \ P_{lab}[MeV/c])   &  \ 425 \ MeV/c  \leq  \ P_{lab} < 30 \ GeV/c \\
      \end{aligned}
    \right.
\end{equation}

\begin{center}
\fbox{\texorpdfstring{$K N \rightarrow K N $}{Lg}}
\end{center}
\begin{equation}
\text{\hspace*{-2cm}}\sigma = \left\{
      \begin{aligned}
        &12   & \ P_{lab} < 935 \ MeV/c \\
        &17.4-3\exp(6.3 \ 10^{-4} \ P_{lab}[MeV/c])  &  \ 935 \ MeV/c  \leq  \ P_{lab} < 2080 \ MeV/c \\
        &832 \ P_{lab}[MeV/c]^{-0.64}   &  \ 2080 \ MeV/c  \leq  \ P_{lab} < 5.5 \ GeV/c \\
        &3.36   &  \ 5.5 \ GeV/c  \leq  \ P_{lab} < 30 \ GeV/c \\
      \end{aligned}
    \right.
\end{equation}

\begin{center}
\fbox{\texorpdfstring{$\overline{K} N \rightarrow \overline{K} N $}{Lg}}
\end{center}
\begin{equation}
\left.
	\begin{aligned}
		\sigma& = 6.132 P_{lab}[GeV/c]^{-0.2437}+12.98 \exp \frac{-(P_{lab}[GeV/c]-0.9902)^2}{0.05558} \\
		&+2.928 \exp\frac{-(P_{lab}[GeV/c]-1.649)^2}{0.772}+564.3\exp \frac{-(P_{lab}[GeV/c]+0.9901)^2}{0.5995}
	\end{aligned}
\right.
\end{equation}
\begin{center}

\end{center}

\subsection{Inelastic}

In this section if non specified momentum is in $GeV/c$.

\begin{center}
\fbox{\texorpdfstring{$NN \rightarrow N \Lambda K $}{Lg}}
\end{center}
\begin{equation}
\sigma(pp \rightarrow p \Lambda K^+) = 1.11875 \frac{(P_{lab}-2.3393)^{1.0951}}{(P_{lab}+2.3393)^{2.0958}}  \ \ 2.3393 \leq \ P_{lab}< 30 \ GeV/c
\end{equation}

\begin{center}
\fbox{\texorpdfstring{$NN \rightarrow N \Sigma K $}{Lg}}
\end{center}

\begin{equation}
\sigma(pp \rightarrow n \Sigma^+ K^+) = 6.38 (P_{lab}-2.593)^{2.1}/P_{lab}^{4.162}  \ \ P_{lab}\geq2.593 \ GeV/c
\end{equation}

\begin{center}
\fbox{\texorpdfstring{$N N \rightarrow N N K \overline{K} $}{Lg}}
\end{center}
\begin{equation}
\sigma(pp \rightarrow p p K^+ K^-) = 3/38 \left(1-\frac{2.872^2}{s[GeV^2]}\right)^{3} \left(\frac{2.872^2}{s[GeV^2]} \right)^{0.8}  \ \sqrt{s} \geq 2.872 \ GeV
\end{equation}

\begin{center}
\fbox{\texorpdfstring{$
          N N \rightarrow N Y K + x\pi | x \geq 3 \text{ and }
          N N \rightarrow N N K \overline{K} + x\pi | x \geq 1
$}{Lg}}

~

The 2 following formulae are coming from the results of the Fritiof model \cite{fritiof}.
\end{center}
\begin{equation}
  \left.
      \begin{aligned}
        \sigma(pp) &= 8.12 (P_{lab}-6)^{2.157}/P_{lab}^{2.333} \\
        \sigma(pn) &= 10.15 (P_{lab}-6)^{2.157}/P_{lab}^{2.333} \\
      \end{aligned}
 \right\} \ P_{lab}\geq 6 \ GeV
\end{equation}

\begin{center}
\fbox{\texorpdfstring{$\Delta N \rightarrow N N K \overline{K} $}{Lg}}
\end{center}
\begin{equation}
\sigma(\Delta^{++} p \rightarrow p p K^+ \overline{K}^0) = 6.6 \left(1-\frac{2.872^2}{s[GeV^2]}\right)^{3} \left(\frac{2.872^2}{s[GeV^2]} \right)^{0.8}  \ \ \sqrt{s}\geq2.872 \ GeV\\
\end{equation}

\begin{center}
\fbox{\texorpdfstring{$\pi N \rightarrow \Lambda K $}{Lg}}
\end{center}
\begin{equation}
  \left.
      \begin{aligned}
        \sigma(\pi^- p  \rightarrow \Lambda K^0) &= 0.3936 \ P_{lab}^{-1.357}\\
&-6.052 \exp(-(P_{Lab}-0.7154)^2/0.02026) \\
&+0.489 \exp(-(P_{Lab}-0.8886)^2/0.08378) \\
&-0.16 \exp(-(P_{Lab}-0.9684)^2/0.001432) \\
      \end{aligned}
    \right.
\ P_{lab} \geq 0.911 \ GeV/c
\end{equation}

\begin{center}
\fbox{\texorpdfstring{$\pi N \rightarrow \Sigma K $}{Lg}}
\end{center}
\begin{equation}
\sigma(\pi^- p  \rightarrow \Sigma^- K^+) = 4.352 \frac{(P_{Lab}-1.0356)^{1.006}}{(P_{lab}+1.0356)^{0.0978} P_{lab}^{5.375}} \qquad \ P_{lab} \geq 1.0356 \ GeV/c
\end{equation}
\begin{equation}
\sigma(\pi^+ p  \rightarrow \Sigma^+ K^+) = 1.897 \ 10^{-3} \frac{(P_{Lab}-1.0428)^{2.869} \left(P_{Lab}+1.0428\right)^{16.68}}{P_{Lab}^{19.1}} \qquad \ P_{lab} \geq 1.0428 \ GeV/c
\end{equation}
\begin{equation}
\sigma(\pi^- p  \rightarrow \Sigma^0 K^0) = 0.3474 (P_{Lab}-1.034)^{0.07678}/ \ P_{lab}^{1.627} \qquad \ P_{lab} \geq 1.034 \ GeV/c
\end{equation}
\begin{equation}
\sigma(\pi^0 p  \rightarrow \Sigma^0 K^+) = 3.624 (P_{Lab}-1.0356)^{1.4}/ \ P_{lab}^{5.14} \qquad \ P_{lab} \geq 1.0356 \ GeV/c
\end{equation}

\begin{center}
\fbox{\texorpdfstring{$\pi N \rightarrow \Lambda K \pi$}{Lg}}
\end{center}
\begin{equation}
\sigma(\pi^+ p  \rightarrow \Lambda K^+ \pi ^+) = 146.2 \frac{(P_{Lab}-1.147)^{1.996}}{(P_{Lab}+1.147)^{5.921}} \qquad P_{Lab}\geq1.147 \ GeV/c 
\end{equation}

\begin{center}
\fbox{\texorpdfstring{$\pi N \rightarrow \Sigma K \pi$}{Lg}}
\end{center}
\begin{equation}
\sigma(\pi^- p  \rightarrow \Sigma^- K^0 \pi ^+) = 8.139 \frac{(P_{Lab}-1.3041)^{2.431}}{(P_{Lab})^{5.298}} \qquad \ P_{Lab}\geq1.3041 \ GeV/c
\end{equation}

\begin{center}
\fbox{\texorpdfstring{$\pi N \rightarrow \Lambda K \pi\pi$}{Lg}}
\end{center}
\begin{equation}
\sigma(\pi^+ p  \rightarrow \Lambda K^+ \pi ^+ \pi^0) = 18.77 \frac{(P_{Lab}-1.4162)^{4.597}}{(P_{Lab})^{6.877}} \qquad \ P_{Lab}\geq1.4162 \ GeV/c
\end{equation}

\begin{center}
\fbox{\texorpdfstring{$\pi N \rightarrow \Sigma K \pi\pi$}{Lg}}
\end{center}
\begin{equation}
\sigma(\pi^+ p  \rightarrow \Sigma^+ K^+ \pi ^+ \pi^-) = 137.6 \frac{(P_{Lab}-1.5851)^{5.856}}{(P_{Lab})^{9.295}} \qquad \ P_{Lab}\geq1.5851 \ GeV/c
\end{equation}

\begin{center}
\fbox{\texorpdfstring{$\pi N \rightarrow N K \overline{K}$}{Lg}}
\end{center}
\begin{equation}
\sigma(\pi^0 p  \rightarrow n K^+ \overline{K}^0) = 2.996 \frac{(P_{Lab}-1.5066)^{1.929}}{(P_{Lab})^{3.582}}  \qquad \ P_{lab} \geq1.5066 \ GeV/c
\end{equation}

\newpage

\begin{center}
\fbox{\texorpdfstring{$
          \pi N \rightarrow Y K + x\pi | x \geq 3 \text{ and }
          \pi N \rightarrow N K \overline{K} + x\pi | x \geq 1
$}{Lg}}

~

The 3 following formulae are coming from the results of the Fritiof model \cite{fritiof}.
\end{center}
\begin{equation}
  \left.
      \begin{aligned}
        \sigma(\pi^+ p) &= 3.851 \frac{(P_{Lab} - 2.2)^2}{P_{Lab}^{1.88286}} \\
        \sigma(\pi^0 p) &= 4.4755 \frac{(P_{Lab} - 2.2)^{1.927}}{P_{Lab}^{1.89343}} \\
        \sigma(\pi^- p) &= 5.1 \frac{(P_{Lab} - 2.2)^{1.854}}{P_{Lab}^{1.904}} \\
      \end{aligned}
 \right\} \ P_{lab} \geq 2.2 \ GeV/c
\end{equation}

\begin{center}
\fbox{\texorpdfstring{$\Lambda N \rightarrow \Sigma N$}{Lg}}
\end{center}
\begin{equation}
\sigma(p \Lambda \rightarrow \Sigma^+ n) = 8.74 \frac{(P_{Lab}-0.664)^{0.438}}{(P_{Lab})^{2.717}}  \ \ P_{lab} \geq 0.664 \ GeV/c
\end{equation}

\begin{center}
\fbox{\texorpdfstring{$\Sigma N \rightarrow \Lambda N$}{Lg}}
\end{center}
\begin{equation}
\sigma(p \Sigma^0 \rightarrow p \Lambda) =  \left\{
      \begin{aligned}
        & 100    \qquad \ &P_{lab} < 0.1  \ GeV/c \\
        & 8.23 \ P_{lab}^{-1.087}  &0.1 \ GeV/c \leq P_{lab}\\
      \end{aligned}
    \right.
\end{equation}

\begin{center}
\fbox{\texorpdfstring{$\Sigma N \rightarrow N' \Sigma'$}{Lg}}
\end{center}

\begin{equation}
  \left.
    \begin{aligned}
        \sigma(n \Sigma^0 \rightarrow p \Sigma^-)\\
        \sigma(n \Sigma^+ \rightarrow p \Sigma^0)\\
    \end{aligned}
  \right\} =
  \left\{
      \begin{aligned}
         &0   & P_{lab} < 162 \ MeV/c \\
         &13.79 \ P_{lab}^{-1.181} \ & P_{lab} \geq 162 \ MeV/c \\
      \end{aligned}
  \right.
\end{equation}

\begin{equation}
  \left.
    \begin{aligned}
        \sigma(p \Sigma^0 \rightarrow n \Sigma^+)\\
        \sigma(p \Sigma^- \rightarrow n \Sigma^0)\\
    \end{aligned}
  \right\} =
  \left\{
      \begin{aligned}
        & 200   & P_{lab} < 103.5 \ MeV/c \\
        & 13.79 \ P_{lab}^{-1.181} \  & P_{lab} \geq 103.5 \ MeV/c \\
      \end{aligned}
  \right.
\end{equation}

\begin{center}
\fbox{\texorpdfstring{$\overline{K} N \rightarrow \Lambda \pi$}{Lg}}
\end{center}
\begin{equation}
\sigma(p K^- \rightarrow \Lambda \pi^0) =
  \left\{
     \begin{aligned}
     &\begin{aligned}
        &40.24  && \ \ P_{lab} < 86.636 \ MeV/c \\
        &0.97 \ P_{lab}^{-1.523}  && \ 86.636 \ MeV/c \leq P_{lab} < 500 \ MeV/c \\
     \end{aligned}\\
     &\begin{aligned}
        \\
        & 1.23 \ P_{lab}^{-1.467} + 0.872\exp \left(-\frac{(P_{lab}-0.749)^2}{0.0045} \right)+ 2.337 \exp \left(-\frac{(P_{lab}-0.957)^2}{0.017} \right)\\
        & \qquad + 0.476 \exp \left(-\frac{(P_{lab}-1.434)^2}{0.136} \right)  \qquad \qquad \ 500 \ MeV/c \leq P_{lab} < 2 \ GeV/c \\
     \end{aligned}\\
     \\
     &\begin{aligned}
        &3  \ P_{lab}^{-2.57}  && \ 2 \ GeV/c \leq P_{lab} \\
     \end{aligned}
     \end{aligned}
  \right.
\end{equation}

\begin{center}
\fbox{\texorpdfstring{$\overline{K} N \rightarrow \Sigma \pi$}{Lg}}
\end{center}
\begin{equation}
\sigma(p K^- \rightarrow \Sigma^+ \pi^-) =
  \left\{
    \begin{aligned}
        & 70.166 \qquad \qquad \ P_{lab} < 100 \ MeV/c \\
        & 1.4 \ P_{lab}^{-1.7} + 1.88 \exp \left(-\frac{(P_{lab}-0.747)^2}{0.005} \right) & \ P_{lab} \geq 100 \ MeV/c \\
        &\qquad + 8 \exp \left(-\frac{(P_{lab}-0.4)^2}{0.002} \right)+ 0.8 \exp \left(-\frac{(P_{lab}-1.07)^2}{0.01} \right)&  \\
    \end{aligned}
  \right.
\end{equation}
\newpage
\begin{center}
\fbox{\texorpdfstring{$\overline{K} N \rightarrow \overline{K}' N'$}{Lg}}
\end{center}
\begin{equation}
\sigma(p K^- \rightarrow n \overline{K}^0) =
     \begin{dcases}
        0 & \ P_{lab} < 89.21 \ MeV/c \\
        0.4977 \frac{(P_{lab} - 0.08921)^{0.5581}}{P_{lab}^{2.704}} & \ 89.21 \ MeV/c \leq \ P_{lab} < 0.2 \ GeV/c\\
        2 \ P_{lab}^{-1.2} + \ 6.493 \exp \left(-0.5 \left( \frac{P_{lab}-0.3962}{0.02} \right)^2\right)  & \ 0.2 \ GeV/c \leq \ P_{lab} < 0.73 \ GeV/c\\
        2.3 \ P_{lab}^{-0.9} +  1.1 \exp \left(-0.5 \left( \frac{P_{lab}-0.82}{0.04} \right)^2\right) & \\
        \hspace{1cm}+ \ 5 \exp\left(-0.5 \left( \frac{P_{lab}-1.04}{0.1} \right)^2\right)  & 0.73 \ GeV/c \leq \ P_{lab} < 1.38 \ GeV/c \\
        \ 2.5 \ P_{lab}^{-1.68} + \ 0.7 \exp \left(-0.5 \left( \frac{P_{lab}-1.6}{0.2} \right)^2\right) & \\
        \hspace{1cm}+ \ 0.2 \exp \left(-0.5 \left( \frac{P_{lab}-2.3}{0.2} \right)^2\right) & \ 1.38 \ GeV/c \leq \ P_{lab}\\
     \end{dcases}
\end{equation}

The following cross section results from the balance detailed of the previous cross section.

\begin{equation}
\sigma(n \overline{K}^0 \rightarrow p K^-) =
      \begin{dcases}
        30  & \ P_{lab} < 100 \ MeV/c \\
        2 \ P_{lab}^{-1.2} + 6.493 \exp \left(-0.5 \left( \frac{P_{lab}-0.3962}{0.02} \right)^2\right)  & \ 0.1 \ GeV/c \leq \ P_{lab} < 0.73 \ GeV/c\\
        2.3 \ P_{lab}^{-0.9} + 1.1 \exp \left(-0.5 \left( \frac{P_{lab}-0.82}{0.04} \right)^2\right) & \\
        \qquad + 5 \exp \left(-0.5 \left( \frac{P_{lab}-1.04}{0.1} \right)^2\right) & \ 0.73 \ GeV/c \leq P_{lab} < 1.38 \ GeV/c\\
        2.5 \ P_{lab}^{-1.68} + 0.7 \exp \left(-0.5 \left( \frac{P_{lab}-1.6}{0.2} \right)^2\right) & \\
        \qquad + 0.2 \exp \left(-0.5 \left( \frac{P_{lab}-2.3}{0.2} \right)^2\right) & \ 1.38 \ GeV/c \leq \ P_{lab} \\
      \end{dcases}
\end{equation}

\begin{center}
\fbox{\texorpdfstring{$\overline{K} N \rightarrow N \overline{K} \pi$}{Lg}}
\end{center}
\begin{equation}
\sigma(p \overline{K}^0 \rightarrow p K^- \pi^+) =  101.3\frac{(P_{lab}-0.526)^{5.846}}{P_{lab}^{8.343}} \qquad P_{lab} \geq 526 \ MeV/c
\end{equation}

\begin{center}
\fbox{\texorpdfstring{$\overline{K} N \rightarrow \Sigma \pi \pi$}{Lg}}
\end{center}
\begin{equation}
\sigma(p K^- \rightarrow \Sigma^+ \pi^0 \pi^-) = 73.67 \frac{(P_{lab}-0.260)^{6.398}}{(P_{lab}+0.260)^{9.732}} + 0.21396 \exp \left(-\frac{(P_{lab}-0.4031)^2}{0.00115} \right)  \quad \ P_{lab} \geq 260 \ MeV/c
\end{equation}

\begin{center}
\fbox{\texorpdfstring{$\overline{K} N \rightarrow \Lambda \pi \pi$}{Lg}}
\end{center}
\begin{equation}
\sigma(p K^- \rightarrow \Lambda \pi^+ \pi^-) = 
   \left\{
     \begin{aligned}
        &6364 \frac{P_{lab}^{6.07}}{(P_{lab}+1)^{10.58}} + 2.158 \exp \left(-\frac{1}{2} \left(\frac{P_{lab}-0.395}{0.01984} \right)^2 \right)  &\ P_{lab} < 970 \ MeV/c \\
        & 46.3 \frac{P_{lab}^{0.62}}{(P_{lab}+1)^{3.565}}  &\ P_{lab}  \geq  \ 970\ MeV/c \\
      \end{aligned}
   \right.
\end{equation}

\newpage
\begin{center}
\fbox{\texorpdfstring{$\overline{K} N \rightarrow N \overline{K} \pi \pi$}{Lg}}
\end{center}
\begin{equation}
\sigma(p K^- \rightarrow p K^- \pi^+ \pi^-) = 26.8 \frac{(P_{lab}-0.85)^{4.9}}{P_{lab}^{6.34}}  \qquad \ P_{lab} \geq 850 \ MeV/c
\end{equation}

\begin{center}
\fbox{\texorpdfstring{$K N \rightarrow K' N' $}{Lg}}
\end{center}
\begin{align}
\sigma(n K^+ \rightarrow p K^0) & = 12.84 \frac{(P_{lab}-0.0774)^{18.19}}{(P_{lab})^{20.41}}  \qquad \ P_{lab} \geq 77.4 \ MeV/c \\
\sigma(p K^0 \rightarrow n K^+) & = 12.84 \frac{(P_{lab}+0.0774)^{18.19}}{(P_{lab}+0.1548)^{20.41}}   \qquad \ P_{lab} \geq 0 \ MeV/c
\end{align}
\begin{center}

\end{center}
\begin{center}
\fbox{\texorpdfstring{$K N \rightarrow N K \pi $}{Lg}}
\end{center}
\begin{equation}
\sigma(p K^0 \rightarrow p K^+ \pi^-) = 116.8 \frac{(P_{lab}-0.53)^{6.874}}{P_{lab}^{10.11}}  \qquad \ P_{lab} \geq 530 \ MeV/c \\
\end{equation}

\begin{center}
\fbox{\texorpdfstring{$K N \rightarrow N K \pi \pi$}{Lg}}
\end{center}
\begin{equation}
\sigma(p K^0 \rightarrow p K^+ \pi^0 \pi^-) =
  \left\{
     \begin{aligned}
        & 26.41 \frac{(P_{lab}-0.812)^{7.138}}{P_{lab}^{5.337}}  & \ 812 \ MeV/c \leq \ P_{lab}& < 1.744 \ GeV/c \\
        & 1572 \frac{(P_{lab}-0.812)^{9.069}}{P_{lab}^{12.44}}  & \ 1.744 \ GeV/c \leq \ P_{lab}& < 3.728 \ GeV/c \\
        & 60.23 \frac{(P_{lab}-0.812)^{5.084}}{P_{lab}^{6.72}}  & \ 3.728 \ GeV/c \leq \ P_{lab}.&\\
      \end{aligned}
    \right.
\end{equation}

\onecolumn
\section{The $\boldsymbol{\pi^+ p \rightarrow K^+ \Sigma^+}$ Legendre coefficients with pion momentum from $\boldsymbol{1282}$ up to $\boldsymbol{2473~MeV/c}$}
\label{Leg_Table}

In this appendix we summarize in a table the 9 first Legendre coefficients extracted from the differential cross sections published in \cite{piN7}. The reaction studied is $\pi^+ p \rightarrow K^+ \Sigma^+$ with pion momentum from $1282$ up to $2473~MeV/c$. This coefficients were determined using a ROOT minimization with a smoothness constraint.

\begin{table}[!ht]
\centering
\begin{tabular}{|c|ccccccccc|c|}
\hline
\begin{tabular}{c} $P_{lab}$ \\ $(MeV/c)$ \end{tabular} & $A_0$ & $A_1$ & $A_2$ & $A_3$ & $A_4$ & $A_5$ & $A_6$ & $A_7$ & $A_8$ & $\rchi$ $^2/NDF$ \\
\hline
1282 & 0.120 & -0.030 & -0.011 & 0.121 & -0.001 & -0.012 & -0.026 & 0.008 & -0.008 & 1.476 \\
1328 & 0.144 & -0.029 & -0.014 & 0.135 & 0.018 & 0.007 & -0.020 & -0.004 & -0.003 & 1.665 \\
1377 & 0.175 & -0.033 & 0.006 & 0.168 & 0.032 & 0.010 & -0.003 & 0.016 & -0.004 & 1.240 \\
1419 & 0.203 & -0.023 & 0.004 & 0.201 & 0.058 & 0.031 & -0.025 & -0.005 & -0.028 & 1.330 \\
1490 & 0.247 & -0.042 & 0.111 & 0.174 & 0.142 & -0.015 & 0.059 & -0.027 & -0.004 & 1.289 \\
1518 & 0.264 & -0.043 & 0.142 & 0.189 & 0.175 & -0.003 & 0.089 & -0.055 & -0.023 & 0.861 \\
1582 & 0.247 & -0.018 & 0.138 & 0.176 & 0.161 & 0.008 & 0.084 & -0.031 & -0.008 & 1.177 \\
1614 & 0.266 & -0.007 & 0.174 & 0.181 & 0.195 & 0.039 & 0.118 & -0.003 & -0.039 & 1.094 \\
1687 & 0.259 & 0.015 & 0.170 & 0.165 & 0.211 & 0.075 & 0.162 & -0.036 & -0.009 & 1.155 \\
1712 & 0.261 & 0.021 & 0.199 & 0.158 & 0.252 & 0.088 & 0.192 & -0.003 & -0.018 & 1.781 \\
1775 & 0.267 & 0.037 & 0.226 & 0.133 & 0.260 & 0.107 & 0.170 & -0.003 & -0.007 & 1.201 \\
1808 & 0.256 & 0.066 & 0.231 & 0.108 & 0.269 & 0.104 & 0.180 & -0.020 & 0.030 & 1.033 \\
1879 & 0.230 & 0.076 & 0.220 & 0.102 & 0.249 & 0.072 & 0.153 & -0.063 & 0.002 & 1.914 \\
1906 & 0.262 & 0.065 & 0.202 & 0.110 & 0.233 & 0.082 & 0.185 & -0.025 & -0.031 & 1.194 \\
1971 & 0.265 & 0.085 & 0.218 & 0.100 & 0.263 & 0.131 & 0.165 & -0.048 & -0.015 & 1.108 \\
1997 & 0.238 & 0.085 & 0.207 & 0.056 & 0.224 & 0.122 & 0 .154 & -0.009 & -0.027 & 0.981 \\
2067 & 0.259 & 0.103 & 0.186 & 0.081 & 0.203 & 0.181 & 0.148 & -0.008 & -0.063 & 1.011 \\
2099 & 0.246 & 0.158 & 0.183 & 0.112 & 0.200 & 0.200 & 0.114 & 0.001 & -0.085 & 0.779 \\
2152 & 0.242 & 0.121 & 0.224 & 0.064 & 0.209 & 0.188 & 0.174 & 0.041 & -0.086 & 1.339 \\
2197 & 0.248 & 0.101 & 0.230 & 0.051 & 0.218 & 0.223 & 0.211 & -0.013 & -0.058 & 1.491 \\
2241 & 0.252 & 0.121 & 0.246 & 0.061 & 0.186 & 0.199 & 0.161 & 0.044 & -0.070 & 1.129 \\
2291 & 0.254 & 0.154 & 0.235 & 0.125 & 0.170 & 0.269 & 0.208 & 0.092 & -0.071 & 1.324 \\
2344 & 0.264 & 0.144 & 0.279 & 0.110 & 0.242 & 0.254 & 0.215 & 0.087 & -0.040 & 0.911 \\
2379 & 0.245 & 0.172 & 0.246 & 0.114 & 0.206 & 0.278 & 0.237 & 0.133 & -0.040 & 1.239 \\
2437 & 0.262 & 0.167 & 0.315 & 0.106 & 0.286 & 0.249 & 0.281 & 0.150 & 0.016 & 1.308 \\
2473 & 0.281 & 0.158 & 0.347 & 0.095 & 0.344 & 0.230 & 0.345 & 0.083 & 0.088 & 1.306 \\
\hline
\end{tabular}
\caption{Legendre coefficients extracted from differential cross sections published in \cite{piN7}.}
\label{leg1}
\end{table}

\end{document}